\documentstyle[amssymb,prb,aps,multicol,epsfig]{revtex} 

\begin{document}
\title { A Spin Fluctuation Model for  $d-$wave  Superconductivity}
\author{Andrey V. Chubukov$^{1}$, David Pines$^{2}$, and
 J\"org Schmalian$^{3}$}

\address{$^1$ Department of Physics, University of Wisconsin,
 Madison, WI 53706}
\address{$^2$ Institute for Complex Adaptive Matter,
 University of California, T-Division, Los Alamos National Laboratory,
 Los Alamos, NM 87545 and 
Department of Physics, University of Illinois at Urbana-Champaign, Urbana, Il 61801}
\address{$^3$ Department of Physics and Astronomy  and Ames Laboratory, 
Iowa State University, Ames, IA 50011}  
\date{\today}
\maketitle
\vskip 2cm
\begin{abstract}
\leftskip 54.8pt \rightskip 54.8pt
We review the results of the spin-fermion model for correlated electron
materials that are sufficiently close to an antiferromagnatic instability that 
their staggered static magnetic susceptibility in the normal state is large 
compared to that found in a conventional Fermi liquid. We demonstrate that for such
materials magnetically-mediated superconductivity, brought about by
the exchange of spin fluctuations, is a viable alternative to
conventional phonon-mediated pairing, and leads to pairing in the
$d_{x^{2}-y^{2}}$ channel. If the dominant interaction between
quasiparticles is of electronic origin and, at energies much smaller
than the fermionic bandwidth, can be viewed as being due to the
emission and absorption of a collective, soft spin degree of
freedom, the low-energy physics of these materials is accurately
described by the spin-fermion model. The derived dynamic magnetic
susceptibility and quasiparticle interaction coincide with the the
phenomenonological  forms used to fit NMR experiments and  in
earlier Eliashberg calculations. In discussing normal state properties, 
the pairing instability and superconducting properties, we focus our  attention
on those materials that, like the cuprate, organic, and some heavy
electron superconductors, display quasi-two dimensional behavior.
In the absence of superconductivity, at sufficiently low temperatures
and energies, a nearly antiferromagnetic Fermi liquid
is unconventional, in that the characteristic energy above which a
Landau Fermi liquid  description is no longer valid is not the Fermi
energy, but is the much smaller  spin-fluctuation energy,$\omega_{\rm sf}$.
For energies (or temperatures) between $\omega_{\rm sf}$ and the Fermi energy, the
system behavior is quite different from that in a conventional Fermi liquid.
Importantly, it is universal in that it is governed by just two input
parameters -an effective spin-fermion interaction energy that sets
the overall energy scale, and a dimensionless spin-fermion coupling
constant that diverges at the antiferromagnetic quantum critical point.
We discuss the pairing instability cased by the spin-fluctuation exchange, and
"fingerprints" of a spin mediated pairing that are chiefly associated with the
emergence of the resonance peak in the spin response of a $d$-wave
 superconductor. We identify these fingerprints in spectroscopic experiments
 on cuprateb superconconductors. We conclude with a discussion of open questions
associated primarily with the nature of the pseudogap state found in
underdoped cuprates.
\end{abstract}

\newpage
\begin{multicols}{2}  
\narrowtext   

\tableofcontents

$ \ \ $
\vskip 2cm 

\section{Introduction and Overview}

The identification of the microscopic mechanisms responsible for
superconductivity and the nature of the superconducting pairing state \
continues to represent one of the most exciting theoretical challenges in
theoretical physics\cite{M2S}. In the so-called conventional superconductors, at
frequencies less than or comparable to the Debye frequency, the attractive
phonon-induced interaction between electrons wins out over the repulsive
screened Coulomb interaction\cite{BardeenPines} and brings about
superconductivity\cite{BCS}. The pairing of electrons in the superconducting
state is in an $s-$wave channel. The primacy of phonon-induced interaction
in conventional superconductors has been demonstrated with great clarity.
The phonon density of states, obtained by inelastic neutron scattering
experiments, and the spectrum of the bosons which mediate pairing, as
deduced from tunneling experiments, agree very well in systems like Pb\cite
{McMillan69,Scalapino69}. In addition to the isotope effect\cite
{isotope1,isotope2}, this comparison of two independent experiments is
generally considered to be a very reliable proof of a phonon-mediated
pairing state.

The analysis of the tunneling data relies heavily on the existence of strong
coupling effects in the quasiparticle density of states and assumes the
validity of the Eliashberg approach to superconductivity\cite{Eliashberg}.
Eliashberg theory for conventional superconductors is extremely robust due
to the decoupling of typical electron and phonon time scales caused by the
small ratio of the velocity of sound and the Fermi velocity. This smallness
also implies that the Debye frequency is much smaller than the Fermi energy,
and hence the quasiparticles that participate in the pairing are low-energy
quasiparticles, located in the near vicinity of the Fermi surface. Landau~ 
\cite{Landau} showed that the low energy properties of a normal Fermi liquid
are characterized by a small number of parameters and are independent of the
details of the underlying lattice Hamiltonian. Both, the normal and
superconducting states of a conventional superconductor may be viewed as
protected states of matter~\cite{LaughlinPines}, states whose generic low
energy properties, insensitive to microscopic details at large energy, are
determined by a higher organizing principle. In this view, in conventional
superconductors, the superconducting transition marks a transition from a
Landau Fermi liquid quantum protectorate to a BCS quantum protectorate. The
success of the BCS-Eliashberg theory for conventional phonon mediated
superconductors is almost unique for interacting many body systems.

It is well known however that the pairing state in the
Bardeen-Cooper-Schrieffer theory does not need to be caused by the
interaction between electrons and lattice vibrations. Generally there are
two distinct classes of theories of unconventional pairing. The first, and
more conservative approach is to replace phonons by another collective
bosonic excitation of the solid. This approach successfully describes the
physics of superfluid $^{3}$\textrm{He}\cite{Leggett_review}, where the
intermediate bosons are failed ferromagnetic spin fluctuations
(ferromagnetic paramagnons); a large value of the ferromagnetic correlation
length is not required in view of the Kohn-Luttinger effect~\cite{kl}. This
magnetically-mediated interaction causes pairing in a state with angular
momentum $l=1$ ($p$-wave pairing) and leads to a rich phase diagram and a
large class of new physical phenomena~\cite
{Leggett_review,Vollhardt90,Volovik}. Magnetically-mediated
superconductivity has been proposed for various organic and heavy fermion
superconductors by a number of authors\cite
{Emery83,Hirsch85,Miyake86,Scalapino,Bickers87,Millis88,Kotliar88}.

A second approach to unconventional pairing is more phenomenological, and is
based on the assumption that the superconducting condensation energy is not
determined by the attractive interaction mediated by some boson but rather by
the energy gain due to feedback effects associated with pairing. The latter
may, in principle, occur even for a purely repulsive pairing interaction. In
general, this approach assumes a non-Fermi-liquid behavior in the normal
state. Two examples are the inter-layer tunneling model of Refs.~\cite
{PWA,PWA2} and the mid-infrared model of Ref.~\onlinecite{AJL}. In both models,
the expectation value of the Hamiltonian is drastically different in the
superconducting and the normal states, and the energy gain due to pairing
apparently cannot be traced back to some boson-mediated attraction. 

The two approaches to unconventional pairing are not necessarily in
contradiction with each other. First, at strong coupling, the pairing
interaction mediated by a low-energy bosonic mode is highly retarded and is
a complex function of frequency, so it is not straightforward to determine
whether it is repulsive or attractive. Second the pairing obviously changes
the form of the fermionic self-energy and hence affects the kinetic energy.
And third, if the bosonic mode is itself made out of fermions, then the
propagator of this mode also changes when the system becomes
superconducting. This change affects the potential energy of the system.
Which of the two effects is larger depends on the details of the system
behavior, but in any case, there are clear similarities between a
strong-coupling theory which involves a pairing boson, and a scenario based
entirely on energy gain due to feedback from pairing.

\begin{figure}[tbp]
\epsfxsize=\columnwidth 
\begin{center}
\epsffile{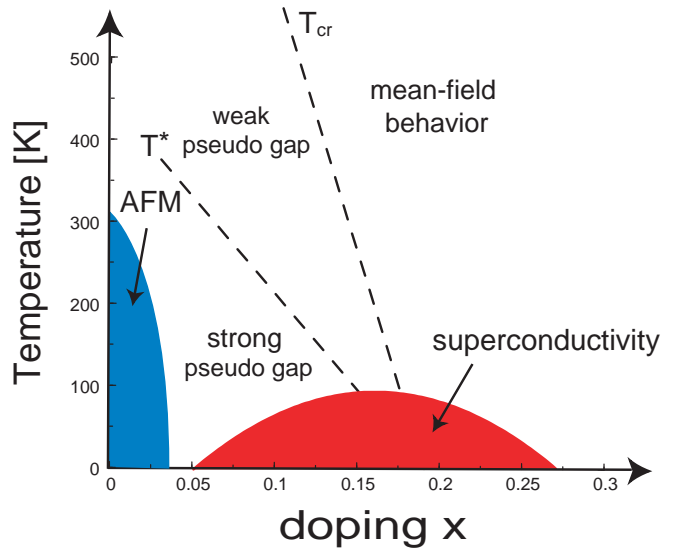}
\end{center}
\caption{Generic phase diagram of high temperature 
superconducting cuprates.
The thermodynamic phases (antiferromagnetic at low doping and
superconducting at higher doping) are depicted by the shaded regions. The
remaining lines are either phase transitions or crossovers, visible in a
variety of experiments.}
\label{PinesHouston}
\end{figure}

In what follows we adopt the first approach and investigate the role of
failed antiferromagnetic spin fluctuations (antiferromagnetic paramagnons)
as a possible cause for both anomalous normal state behavior and
unconventional i.e. non $s$-wave superconductivity. This approach is chiefly
motivated by the physics of \ the high temperature cuprate superconductors
which have been shown to exhibit \ both highly anomalous normal state
behavior and an unconventional pairing state with angular momentum $l=2$ ($d$%
-wave pairing) \cite{WVH93,TK94,Ott}. As may be seen in Fig.\ref
{PinesHouston}, the materials with the highest $T_{c}$ are located
reasonably close to an antiferromagnetic state and have been shown in
nuclear magnetic resonance and inelastic neutron scattering experiments to
exhibit significant antiferromagnetic correlations in the paramagnetic
state. \cite{reviewaf} We will show, in agreement with the calculations of
Monthoux \emph{et al.}\cite{Pines90,Monthoux91}, that in a quasi
two-dimensional material where those correlations are significant (e.g. a
spin correlation length larger than  a lattice constant) the normal
state behavior is anomalous while for Fermi surface parameters appropriate
for the cuprates, one always gets a d$_{x^{2}-y^{2}}$ superconducting
pairing state. We discuss other materials below, following a brief
historical overview of the developments in the spin-fluctuation approach
over the last decade. References to earlier works can be found in the papers
cited below. 

A $d_{x^{2}-y^{2}}$ pairing state in two dimensions due to the exchange of
near-antiferromagnetic spin fluctuations was found in the detailed Hubbard
model calculations of Bickers \emph{et al.}~\cite{Bickers87}. For parameters
\ believed to be relevant for cuprates in 1987, the superconducting
transition temperature was comparatively low ($<40\,\mathrm{K}$) under what
seemed to be optimal conditions. Furthermore, T$_{c}$ decreased as one
increased the planar hole concentration from a low level, in contrast to
experiment. These results, when taken together with the early
penetration-depth experiments that supported an s-wave pairing state, were
responsible for the fact that the magnetic mechanism and $d_{x^{2}-y^{2}}$
pairing had been abandoned by most of the high temperature superconductivity
community by the end of 1989 (Bedell \textit{et al.}\cite{Bedell90}).

At about this time, theoretical groups in Tokyo\cite{Moriya90,Moriya91} and
Urbana\cite{Pines90,Monthoux91} independently began developing a
semi-phenomenological, macroscopic theory of spin-mediated pairing. Both
groups assumed that the magnetic interaction between the planar
quasiparticles was responsible for the anomalous normal state properties and
found a superconducting transition to a $d_{x^{2}-y^{2}}$ pairing state at a
significantly higher temperature than those achieved using the Hubbard
model. Moriya \textit{et al.}\cite{Moriya90,Moriya91} used a self-consistent
renormalization group approach to characterize the dynamic spin
susceptibility. The resulting effective magnetic interaction between the
planar quasiparticles was then used to calculate T$_{c}$ and the normal
state resistivity. Monthoux \textit{et al.}~\cite{Monthoux91} did not
attempt a first-principles calculation of the planar quasiparticle
interaction. Rather, they turned to experiment and used quasiparticles whose
spectra was determined by fits to band structure calculations and angular
resolved photoemission spectroscopy (ARPES) experiments. The effective
magnetic interaction between these quasiparticles was assumed to be
proportional to a mean field dynamic spin susceptibility of the form
developed by Millis \textit{et al.}\cite{MMP} that had been shown to provide
an excellent description of NMR experiments on the YBa$_{2}$Cu$_{3}$O$_{7-y}$
system\cite{reviewaf,MMP}.

Both groups followed up their initial weak coupling calculations with strong
coupling (Eliashberg) calculations~\cite
{Ueda92,Monthoux92,Monthoux93,Monthoux94} that enabled them to take into
account lifetime effects brought about by the strong magnetic interaction.
These calculations showed that $d_{x^{2}-y^{2}}$ superconductivity at high T$%
_{c}$ is a robust phenomenon. Monthoux and Pines\cite{Monthoux93} also found
in a strong coupling calculation that they could obtain an approximately
correct magnitude and temperature dependence of the planar resistivity of
optimally doped YBa$_{2}$Cu$_{3}$O$_{7-y}$ using the same coupling constant
(and the same parameters to characterize the quasiparticle and spin
spectrum) that had yielded a T$_{c}$ of approximately $90\,\mathrm{K}$. They
concluded that they had established a ``proof of concept'' for a nearly
antiferromagnetic Fermi liquid (NAFL) description of the anomalous normal
state behavior and a spin fluctuation mechanism for high temperature
superconductivity. Referring back to Fig. 1, these calculations should apply
to the right of the $T_{\mathrm{cr}}$-line, where the normal state is an
unconventional Fermi liquid in which the characteristic energy above which
quasiparticles loose their Fermi liquid behavior is of order of the spin
fluctuation energy and low compared to the fermionic bandwidth.

Since their calculations unambiguously predicted a $d_{x^{2}-y^{2}}$ pairing
state, Monthoux and Pines challenged the experimental community to find
unambiguously the symmetry of the pairing state. 
At that time (1991-1992), only NMR Knight shift and $Cu$ spin-lattice
relaxation rate results supported $d_{x^{2}-y^{2}}$ pairing\cite
{Barrett90,Takigawa89,Monien90,Walstedt87,Warren87}. However within the next
year or so, the tide turned dramatically away from s-wave pairing, with
ARPES \cite{Shen93}, penetration depth\cite{Hardy93} and new NMR experiments 
\cite{Martindale93,Bulut93,Thelen93} on the oxygen spin-lattice relaxation
time and the anisotropy of the copper spin-lattice relaxation time all
supporting a $d_{x^{2}-y^{2}}$ state. The decisive experiments were the
direct phase-sensitive tests of pairing symmetry carried out by Van
Harlingen and his group in Urbana\cite{WVH93} as well as by Kirtley, Tsuei,
and their collaborators\cite{TK94}.

In subsequent work on the spin-fluctuation mechanism a microscopic,
Hamiltonian approach to the problem was developed\cite
{Chubukov95,Abanov01advphys}. It was shown that the low-energy physics of
spin-mediated pairing is fully captured by a model that describes the
interaction of low energy fermionic quasiparticles with their own collective
spin excitations (the spin-fermion model)~\cite{Chubukov95,Abanov01advphys}.
In particular, it was demonstrated that the phenomenological interaction
between quasiparticles 
could be derived in a controllable way, even at strong coupling, by
expanding either in the inverse number of hot spots in the Brillouin zone ($%
=8$ for the physical case), or in the inverse number of fermionic flavors.
We discuss this theory in detail in Section 4. As will be seen there, the
spin-fermion model contains only a small number of parameters. These
uniquely determine system behavior that is fully universal in the sense that
it does not depend on the behavior of the underlying electronic system at
energies comparable to the fermionic bandwidth. It is therefore possible to
verify its applicability by first using a few experimental results to
determine these parameters, and then comparing the predictions of the
resulting parameter-free theory with the larger subset of experimental
results obtained at temperatures and frequencies which are much smaller than
the fermionic bandwidth. A major prediction of the spin fermion model is
that the upper energy scale for the Fermi liquid behavior progressively
shifts down as the system approaches a quantum-critical point at $T=0$, and
there emerges a large intermediate range of frequencies where, on the one
hand, the system behavior is still a low-energy one and universal, and on
the other hand, it is quantum-critical and not a Fermi liquid.

%
%
%
\begin{figure}[t]
\epsfxsize=\columnwidth 
\begin{center}
\epsffile{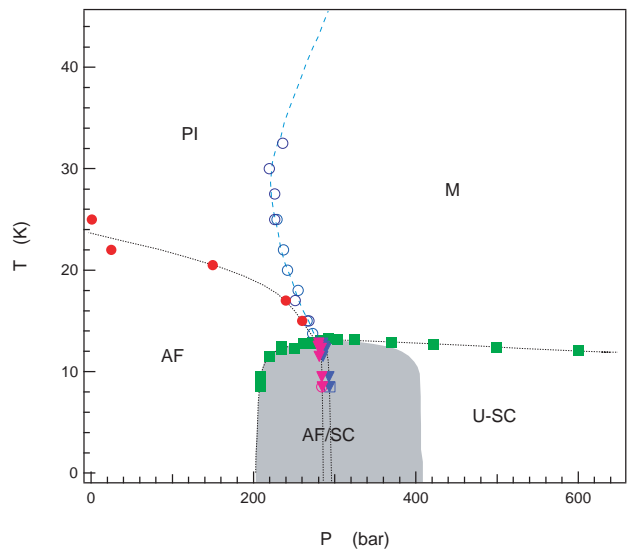}
\end{center}
\caption{The phase diagram of the layered organic superconductor $\protect%
\kappa $-(ET)$_{2}$Cu[N(CN)$_{2}$]Cl in the units of temperature and
pressure (from Ref.{\protect\onlinecite{Brown}}). PI refers to a paramagnetic
insulating regime, M to a metallic regime, AF to an antiferromagnetic
regime, and SC to a superconducting regime. In the region AF-SC,
superconductivity and antiferromagnetism co-exist. In the region U-SC, the
system is an unconventional superconductor}
\label{Fig_organics}
\end{figure}

Cuprate superconductors are not the only candidate materials for spin
fluctuation mediated pairing and non-Fermi liquid quantum protectorates. A
number of organic superconductors are anisotropic quasi-two-dimensional
materials that exhibit many of the anomalies typical of a system with an
unconventional pairing state. The phase diagram of a quasi two-dimensional
organic compound $\kappa $ -BEDT-TTF  is shown in Fig.~\ref{Fig_organics}.
One can see that, as in the case of the cuprates, the superconducting phase
is found in the vicinity of an antiferromagnetic phase. Several groups\cite
{Schmalian98_o,Kino,Kondo,Kuroki} have used spin-fluctuation theory to
predict the position of nodes of the superconducting order parameter of
these materials. An unconventional order-parameter with nodes of the gap is
indeed supported by NMR\cite{deSoto,Mayaff,Kanoda}, thermal conductivity\cite
{Bel}, millimeter transmission\cite{Singleton} and STM\cite{Arai}
experiments. However, the last two experiments seem to come to different
conclusions as far as the position of the nodes is concerned. Ref.\cite
{izawa} also finds nodes, but at a position that is not consistent with the
prediction of a spin fluctuation induced pairing state. Finally, penetration
depth experiments \cite{pen} and recent specific heat data\cite{steglich}
appear to support a conventional s-wave gap. Given those contradictory
experiments, whether quasi-two dimensional organic superconductors exhibit
an unconventional pairing state is, as of this writing, an open question. %

\begin{figure}[tbp]
\epsfxsize=\columnwidth 
\begin{center}
\epsffile{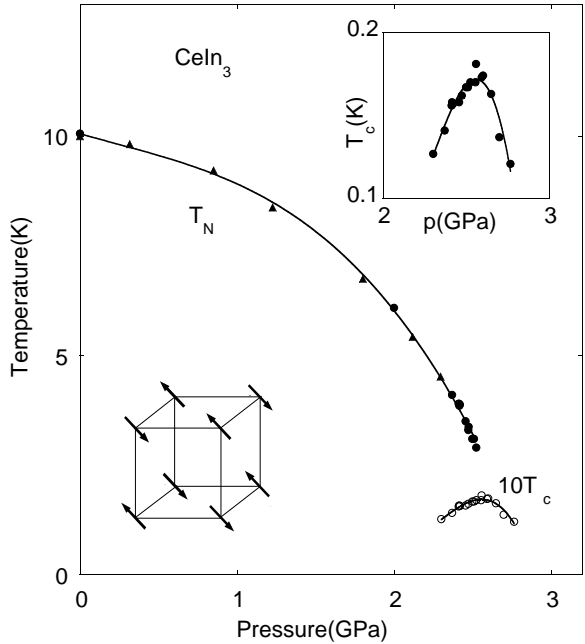}
\end{center}
\caption{The phase diagram of CeIn$_3$ in the units of temperature and
pressure (from Ref.{\protect\onlinecite{Mazur}}).}
\label{phased_cein3}
\end{figure}

\begin{figure}[t]
\epsfxsize=\columnwidth 
\begin{center}
\epsffile{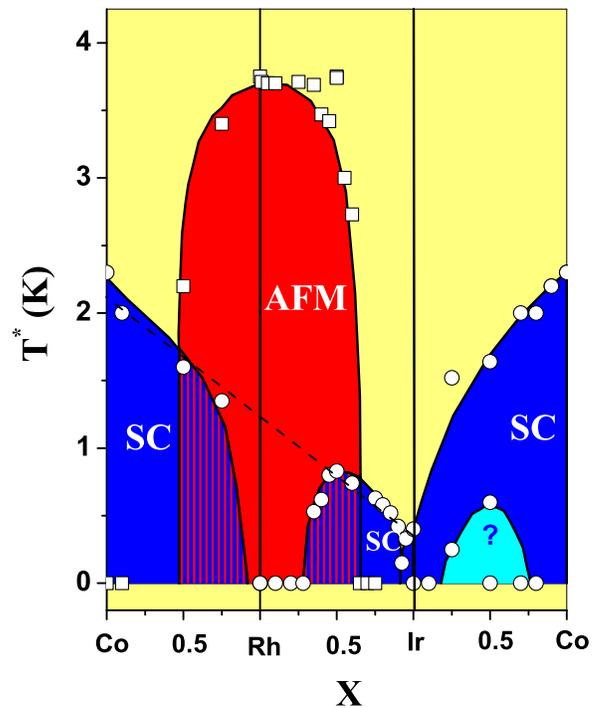}
\end{center}
\caption{The phase diagram of Ce$_{2}$X In$_{5}$ with X=Co, Rh, and Ir in
the units of temperature and doping (from Pagliuso \emph{et al.}{\ 
\protect\cite{Pagliuso00}}).}
\label{phased_115}
\end{figure}

Cerium-based heavy electron superconductors represent another class of
strongly correlated electron superconductors for which a spin-fluctuation
induced interaction between quasiparticles is a strong candidate for the
superconducting mechanism. Examples include $\mathrm{CeCu}_{2}\mathrm{Si}_{2}
$\cite{Steglich}, $\mathrm{CePd}_{2}\mathrm{Si}_{2}$, $\mathrm{CeIn}_{3}$%
\cite{Mazur} and the newly discovered 1-1-5 materials $\mathrm{CeXIn}_{5}$
with $\mathrm{X=Co}$, $\mathrm{Rh}$ and $\mathrm{Ir}$\cite{115first} or
mixtures thereof. As may be seen in the phase diagrams of Fig.\ref
{phased_cein3} and Fig.\ref{phased_115}, all these materials are close to
antiferromagnetism, with superconductivity occurring close to the critical
pressure or alloy concentration at which the magnetic ordering disappears.
Moreover, thermal conductivity measurements of  \textrm{Ce}$\mathrm{Co}$\textrm{In}$_{5}$, which becomes superconducting
at 2.4K at ambient pressures -  the highest known value of $T_{c}$
for a heavy fermion based system - strongly support a superconducting gap with nodes along the $(\pm \pi,\pm \pi)$ directions, as found  in a d$_{x^2-y^2}$ pairing state\cite{izawa115}.
  Another exciting aspect of these systems is
that by changing the relative compositions of Ir and Rh in 1-1-5 materials 
\textrm{Ce}$\mathrm{Rh}_{1-x}\mathrm{Ir}_{x}$\textrm{In}$_{5}$, one can move
the system from an antiferromagnetic to a superconducting state at ambient
pressure.

Another widely studied material in which pairing is possibly due to spin
fluctuation exchange is 
\textrm{Sr}$_{2}\mathrm{RuO}_{4}$\cite{Maeno94,Maeno01,Rice95,MonthouxLon}, where
 NMR Knight shift 
experiments\cite{Ishidasrru} and spin-polarized neutron scattering measurements\cite{Duffysrru} reveal that the spin susceptibility is unchanged upon entering the superconducting state, consistent with spin-triplet 
superconductivity. 

In summary, the cuprates, the 1-1-5 heavy fermion materials and the layered
organic superconductors are strongly correlated materials that exhibit
unconventional normal state and superconducting behavior, while the
superconducting phases are located in the vicinity of magnetic instabilities
in their corresponding phase diagrams. It is then quite natural to assume
that in all three cases magnetic interactions play a dominant role in the
pairing. 

The presence of antiferromagnetic and superconducting regions in the phase
diagram raises the question of whether antiferromagnetism and
superconductivity should be treated on equal footing in a spin fluctuation
approach. If they should, the theoretical analysis would be complex.
Fortunately, this is not the case, at least as long as the characteristic
energy scales for the magnetic interactions are smaller than the fermionic
bandwidth. The point is that superconductivity is generally a low-energy
phenomenon associated with fermions in the near vicinity of the Fermi
surface. On the other hand, antiferromagnetism originates in fermions with
energies comparable to the bandwidth. Perhaps the easiest way to see this is
to formally compute the static spin susceptibility in the random phase approximation  (RPA).
An RPA analysis yields 
$\chi ^{-1}({\bf{q}})\propto 1-g_{\mathrm{eff}}(%
{\bf{q}})\Pi ({\bf{q}})$ where $g_{\mathrm{eff}}({\bf{q}})$ is some
effective interaction, and $\Pi ({\bf{q}})$ is the static spin
polarization operator (a particle-hole bubble with Pauli matrices in the
vertices). For an antiferromagnetic instability we need $g_{\mathrm{eff}}(%
{\bf{Q}})\Pi ({\bf{Q}})=1$. One can easily make sure, by evaluating $\Pi
({\bf{Q}})$ for free fermions, that the momentum/frequency integration in
the particle-hole bubble is dominated by the upper energy limit that is the
fermionic bandwidth. This implies that whether or not a system orders
antiferromagnetically is primarily determined by high-energy fermions that
are located far away from the Fermi surface, and hence the antiferromagnetic
correlation length, that measures the proximity of a material to a nearby
antiferromagnetic region in the phase diagram, should not be calculated but
rather be taken as an input for any low-energy analysis. We discuss the
practical meaning of this separation of energies in Section 4.

A more subtle but important issue is whether the dynamical part of the spin
susceptibility should be considered simply as an input for a low-energy
model (as in the case for phonons), or whether the spin dynamics is produced
by the same electrons that are responsible for the superconductivity and
hence needs to be determined consistently within the low-energy theory. The
first issue one has to consider here is whether a one-band description is
valid, i.e., whether localized electrons remain quenched near the
antiferromagnetic instability and form a single large Fermi surface together
with the conduction electrons to which they are strongly coupled\cite
{Hertz,Millis93}, or \ whether the magnetic instability is accompanied by
the un-quenching of local moments. In the latter case, the volume of the
Fermi surface changes discontinuously at the magnetic transition and could
e.g., cause a jump in the Hall coefficient\cite{Coleman}. The quenching
versus un-quenching issue is currently a subject of intensive debate in
heavy fermion materials~\cite{Coleman,si}. In cuprates the quenching versus
un-quenching issue does not seem to play a role; it is widely accepted that
the formation of Zhang-Rice singlets\cite{ZhangRice} gives rise to a single
electronic degree of freedom. Similarly, in organic materials, the charge
transport in the metallic and superconducting parts of the phase diagram is
due to the same missing electrons in otherwise closed filled molecular
orbital states. Whether or not the spin dynamics originates in low-energy
fermions then reduces to the geometry of a single, large Fermi surface. For
a Fermi surface with hot spots, connected by the wave vector at which the
spin fluctuation spectrum peaks, the low-energy spin dynamics is dominated
by a process in which a collective spin excitation decays into a
particle-hole pair. By virtue of energy conservation, this process involves
fermions with \ frequencies comparable to the frequency of a spin
excitation. Consequently, the spin dynamics is not an input. If, however,
the Fermi surface does not contain hot spots, spin damping is forbidden at
low-energies and spin fluctuations are magnon-like propagating excitations.
It is easy to show that in the latter situation, the full form of the spin
propagator comes from particle-hole excitations at energies comparable to
the bandwidth and therefore should be considered as an input for the
low-energy theory.

In this chapter we consider in detail the scenario in which the Fermi
surface contains hot spots and the spin damping by quasiparticles is
allowed. Our approach to the cuprates is largely justified by the results of
extensive ARPES and NMR and neutron measurements that indicate that the
Fermi surface possesses hot spots, and that spin excitations are overdamped
in the normal state.

Whether or not spin fluctuations are overdamped is also of significant
conceptual importance for spin mediated pairing, since this mechanism
requires that quasiparticles be strongly coupled to the collective spin
excitation mode. At first glance, the undamped (magnon) form of the spin
propagator appears more favorable for spin-mediated pairing than the
overdamped form. Indeed, if one assumes that the spin-mediated interaction
is just proportional to the spin susceptibility, the magnon-like form is
preferable. By the Goldstone theorem, in the antiferromagnetically ordered
state, the transverse spin susceptibility $\chi ({\bf{q}})$ (that yields
an attraction in the $d_{x^{2}-y^{2}}$ channel) even diverges as ${\bf{q}}$
approaches the antiferromagnetic momentum ${\bf{Q}}$, hence the $d-$wave
attraction appears to be the strongest. This reasoning, however, is
incorrect. Schrieffer and \ his collaborators \cite{SWZ,Schrieffer95} and
others\cite{ChubFrenk,ChubMorr}
 have shown that the Goldstone modes of an ordered
antiferromagnet cannot give rise to a strong $d-$wave pairing because the
full spin mediated interaction is the product of the spin susceptibility and
the square of the fully renormalized coupling constant between fermions and
magnons. The latter vanishes in the ordered SDW state at ${\bf{q}}={\bf{%
Q}}$ and this effect exactly compensates the divergence of the static
susceptibility. The vanishing of the effective coupling is a consequence of
the Adler principle which states that true Goldstone modes always decouple
from other excitations in a system \cite{Adler}.

Schrieffer later argued~\cite{Schrieffer95} that the near cancellation
between the enhancement of the spin susceptibility and the reduction of the
effective magnon-fermion interaction persists in the paramagnetic state as
long as spin fluctuations remain propagating excitations. This would
substantially reduce (although not eliminate~\cite{sushkov}) the
spin-mediated $d$-wave attraction. This argument is however inapplicable to
overdamped spin fluctuations. These are \textit{not} Goldstone modes
although they become gapless at the magnetic instability. Goldstone modes
appear only in the ordered state at the smallest ${\bf{q-Q}}$ values~\cite
{ChubMorr}. For near-gapless, but overdamped spin excitations, the Adler
principle does not work. After all, the damping itself is due to the strong
coupling of the collective mode to fermions. Consequently, the spin-fermion
vertex does not vanish at the magnetic transition and hence cannot cancel
out the enhancement of the $d$-wave interaction due to the increase of the
spin susceptibility near ${\bf{Q}}$. Thus, overdamped spin fluctuations
are better for spin-mediated pairing than magnon-like excitations.

Another aspect of the fact that spin dynamics is made out of low-energy
fermions is that the retarded interaction which causes the pairing changes
when fermions acquire a superconducting gap. This feedback from
quasiparticle pairing on the form of the pairing interaction distinguishes
pairing mediated by overdamped spin fluctuations from conventional phonon
induced pairing. In the latter the bosonic propagator is an input and is
only very weakly affected by the opening of the gap in the quasiparticle
spectrum. We will discuss in detail how  feedback forces one to go
beyond an approach in which one solely replaces a phonon by a spin
fluctuation, and requires that one consistently calculates the spin dynamics
at low energies. While doing this is a theoretical challenge, the approach
is appealing since it reduces the number of unknown parameters in the
problem. In particular, we will see that in the superconducting state, the
propagator of spin fluctuations acquires the same form as for optical
phonons, but the collective mode that is the analog of the phonon frequency
is fully determined by the superconducting gap and the normal state spin
damping. This gives rise to new, unique ''fingerprints'' of spin mediated
pairing, whose presence can be checked experimentally.

What is the role of dimensionality? As noted above, many of the candidates
for spin-mediated pairing are strongly anisotropic, quasi-two dimensional
systems. This not only holds for the cuprates, but also for a large class of
organic superconductors. Also, heavy fermion superconductors such as $%
\mathrm{CeCoIn}_{5}$ display a considerable spatial anisotropy. On the other
hand, $\mathrm{CeIn}_{3}$ and to a lesser extent $\mathrm{CeCu}_{2}\mathrm{Si%
}_{2}$ do not display appreciable quasi two-dimensionality in their
electronic properties. The dimensionality of the electronic system is
important to the spin fluctuation model for both normal state and
superconducting behavior. We will see that the dynamics of the fermions in
the normal state is very differently affected by antiferromagnetic spin
fluctuations in two and in three dimensional systems. While in the latter
case only small (logarithmic) corrections to the ideal Fermi gas behavior
occur \ in the vicinity of hot spots, we shall see that in 2d systems, the
strong interaction between fermions and spins gives rise to non-Fermi
liquid, diffusive behavior of low energy fermions as the quantum critical
point is approached. The importance of dimensionality for superconductivity
has been emphasized by Monthoux and Lonzarich\cite{MonthouxLon} who have
shown that  it exerts a considerable influence on the superconducting transition
temperature. They pointed out that in three dimensions one cannot avoid
repulsive contributions to the pairing interaction in choosing a pairing
state with nodes, so that the same spin-mediated quasiparticle interaction is
far less effective in bringing about superconductivity in three dimensions
than in two.

\begin{figure}[tbp]
\epsfxsize=\columnwidth 
\begin{center}
\epsffile{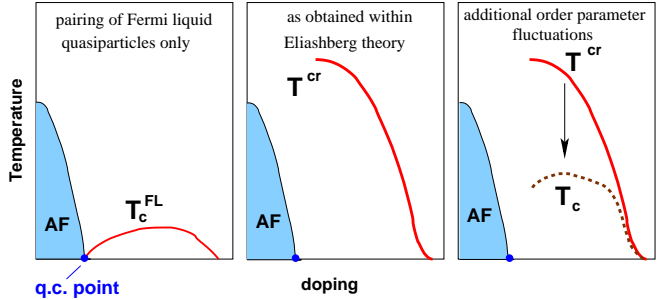}
\end{center}
\caption{The candidate phase diagrams in the units of temperature and doping
of a one-band electronic system near an antiferromagnetic quantum critical
point. Left panel - the phase diagram for the hypothetical situation when
only coherent, Fermi liquid quasiparticles contribute to the pairing (the
McMillan theory applied to spin fluctuations). The antiferromagnetic and
superconducting regions are completely decoupled. Central panel - the
solution of the coupled set of the Eliashberg equations for the onset of
spin-mediated pairing instability~\protect\cite{ACF}. The solution shows
that at strong coupling, the the pairing instability is predominantly
produced by incoherent fermions, and the instability temperature remains
finite at $\protect\xi =\infty $. Right panel, the proposed phase diagram
based on the solution of the Eliashberg equations below the pairing
instability and general arguments about superconducting fluctuations 
\protect\cite{Abanov01epl}.}
\label{fig_tc}
\end{figure}

Since the non-Fermi-liquid behavior of fermionic quasiparticles extends down
to progressively lower frequencies as one approaches the magnetic transition
at $T=0$, one can inquire whether pairing near this quantum-critical point
is caused by the fermions at the lowest energies that are still coherent, or
comes from those at higher energies (that are still smaller than the
bandwidth) that display non-Fermi-liquid behavior. If only coherent fermions
are involved in the pairing, then, according to McMillan's extension of the
BCS theory \cite{McMillan}, the resulting superconducting transition
temperature, $T_{c}^{\mathrm{FL}}$, is comparable to the upper energy cutoff
of the Fermi liquid regime, and thus will be of the order of the spin
fluctuation energy. This energy vanishes at the critical point, and
therefore magnetic criticality is unaffected by pairing (see the left panel
in Fig.\ref{fig_tc}). If, however, ``non-Fermi-liquid'' fermions can give
rise to a pairing instability, then the onset temperature of this
instability in the particle-particle channel, that we will identify with $%
T_{cr}$ in the phase diagram of Fig.\ref{PinesHouston}, generally scales
with the upper cutoff energy for the quantum-critical behavior and remains
finite at criticality\cite{ACF,Andy_triplet,Abanov01epl}. In this situation, the quantum critical point is
necessarily surrounded by a dome beneath which pairing correlations cannot
be neglected\ as shown in the central panel of Fig.\ref{fig_tc}. The
critical behavior inside and outside the dome is different, and the
``primary'' critical behavior (which gives rise to pairing) can only be
detected outside the dome. We will demonstrate below that the incoherent
pairing temperature saturates at a finite value when the magnetic
correlation length diverges. Furthermore, for parameters relevant to
cuprates at low doping, this temperature is of order of the magnetic
exchange interaction $J$, i.e., it is not small.

A related issue is whether the pairing instability at $T_{cr}$ implies the
onset of true superconductivity (i.e., $T_{cr}\equiv T_{c}$), or whether it
marks the onset of pseudogap behavior. In the latter scenario, for which we
will see there is considerable experimental support, fluctuations prevent a
superfluid density $\rho _{s}$ from developing a nonzero value until one
reaches a much smaller $T_{c}$, and the paired incoherent fermions do not
participate in superconductivity. The phase between $T_{cr}$ and $T_{c}$
would then be a new state of matter, the pseudogap state. Abanov \emph{et al.%
}\cite{Abanov01epl} conjectured that a 
pseudogap regime is a universal feature of
the spin fluctuation scenario, as below $T_{cr}$, quasiparticles that 
are paired into singlets still 
remain incoherent and cannot carry a supercurrent.
 True superconductivity
is reached only at much smaller $T_{c}$ where the systems recovers coherent,
Fermi-liquid behavior (see the right panel in Fig.\ref{fig_tc}). %
%
%
%

The term pseudogap was introduced by Friedel\cite{Friedel88} to describe
the fact that in the underdoped regime of the cuprates, the planar
quasiparticles begin to develop a gap-like structure well above $T_{c}$.
This behavior was first seen in Knight shift measurements of the uniform
spin susceptibility\cite{Alloul}, and later detected in almost all measured
properties of underdoped cuprates. At present, the physics of the pseudogap
phase in the underdoped cuprate superconductors is not yet fully understood
and its origin continues to be an open question. We believe that the
``magnetic scenario'' for the pseudogap provides a reasonable explanation,
but many details still need to be worked out. Some researchers on the other
hand have suggested that the pseudogap phase emerges due to strong
fluctuations of the phase of the superconducting order parameter.\cite
{Emery95,Franz98,Kwon98,Ioffe99,Franz01}. Others suggest that the pseudogap
phase may actually be a new phase of matter with a hidden order parameter
associated with bond currents~\cite{schulz89,affleck_marston} 
(most recently, this idea has been explored in detail by Chakravarty \emph{%
et al.} \cite{Chakravartyddw}). A somewhat more general phenomenological
possibility discussed by several researchers is that there exists an
additional quantum critical point of yet unknown origin slightly above
optimal doping\cite{MFL,VarmaQCP,Castellani97,VZS00}
 (a number of experiments suggest that this point is at doping
concentration $x\approx 0.19$). The pseudogap and Fermi liquid phases are
assumed to be to the left and to the right of this new quantum critical
point, respectively.

Our main goal is to discuss in detail the ``primary'' quantum-critical
behavior within the magnetic scenario and how it gives rise to pairing at $%
T_{cr}$. A detailed theory of the pseudogap state of high temperature
superconductors is beyond the scope of this Chapter. However, in the
interest of providing a base line against which to compare both experiment
and future theoretical developments, we summarize the predictions of the
spin-fermion model for the pseudogap in Section 6 and discuss other
alternatives.

To spell out the expected regions of applicability to the superconducting
cuprates ( in doping and temperature) of the spin-fluctuation theory without
pseudogap physics involved, we return to the candidate generic phase diagram
in Fig.\ref{PinesHouston}. The two lines, $T_{cr}$ and $T^{\ast }$ determine
distinct regimes of physical behavior. Above $T_{cr}$, pseudogap physics
plays no role; the theory of a nearly antiferromagnetic Fermi liquid (NAFL)
presented in this Chapter should be applicable for both the normal state and
the superconducting state. Since $T_{cr}$ crosses $T_{c}$ near the optimal
doping concentration, the theory with no pseudogap involved is roughly
applicable at and above optimal doping (from an experimental perspective,
optimally doped materials do show some pseudogap behavior, but only over a
very limited temperature regime). For the overdoped and nearly optimally
doped cuprates the transition is then from a nearly antiferromagnetic Fermi
liquid to a BCS superconductor with $d_{x^{2}-y^{2}}$-pairing symmetry. We
will argue in Section 6 that there is a great deal of experimental evidence
that at and above optimal doping the normal state is indeed a NAFL, and that
the pairing is of magnetic origin. It is also likely that the theory can
also be extended into a so-called ``weak pseudogap'' regime between $T_{cr}$
and $T^{\ast }$\cite{Schmalian98,Abanov01epl}, but we will not discuss this
issue here. 

In Section 2 we introduce and motivate the spin fermion model that we use to
study spin fluctuation induced pairing. We discuss the weak coupling
approach to the pairing problem and the symmetry of the magnetically
mediated pairing state. In Section 3 we review the main results and
arguments used to justify Eliashberg theory for conventional phonon
superconductors. In particular, we discuss the physical origin of the Migdal
theorem that allows a controlled approach to phonon-induced pairing. In
Section 4 we then analyze in detail the strong coupling theory for the
spin-fermion model. We first discuss the normal state properties of this
model and calculate the low frequency dynamics of quasiparticles and spin
fluctuations. We next consider spin-fluctuation induced superconductivity.
We show that for magnetically-mediated superconductivity one can again
analyze the pairing problem in controlled calculations that on the level of
the equations involved resemble the Eliashberg equations for
 electron-phonon superconductivity\cite{Ioffe_MillisUspech,Abanov01advphys}. We demonstrate that the actual physical origin of
the applicability of a generalized Eliashberg approach for spin mediated
pairing is qualitatively different from the phonon case, and is associated
with the overdamped nature of the spin excitations. We solve the resulting
equations in certain limits and investigate the role of quantum critical
pairing. In Section 5 we present a general discussion of some of the
observable fingerprints of spin fluctuation induced superconductivity, and
in Section 6 we compare our results with experiments and discuss to what
extent the fingerprints of spin mediated pairing have already been seen in
optimally doped cuprate superconductors. In our concluding Section 7 we
summarize our results and comment on several topics that are of interest for
a further understanding of spin mediated pairing, including the extent to
which our theory can be extended to address the physics of the pseudogap
state in underdoped cuprates.

\section{Spin-fermion model}

\subsection{Physical motivation of the spin fermion model}

We first discuss the formal strategy one has to follow to \emph{derive} an
effective low-energy model from a microscopic Hubbard-type Hamiltonian with
a four fermion interaction: 
\begin{eqnarray}
{\mathcal{H}}&=&\sum_{{\bf{k}},\alpha }{\bf{\ }}\varepsilon _{{\bf{k}}%
}\psi _{{\bf{k}},\alpha }^{\dagger }\psi _{{\bf{k}},\alpha }
\label{hubb1} \\
&+& \sum_{%
{\bf{k_{i}}},\alpha _{i}}U_{{\bf{k}}_{1},{\bf{k}}_{2},{\bf{k}}_{3},%
{\bf{k}}_{4}}^{\alpha _{1},\alpha _{2},\alpha _{3},\alpha _{4}}~\psi _{%
{\bf{k}}_{1},\alpha _{1}}^{\dagger }\psi _{{\bf{k}}_{2},\alpha
_{2}}^{\dagger }\psi _{{\bf{k}}_{3},\alpha _{3}}\psi _{{\bf{k}}_{4}%
,\alpha _{4}}  \nonumber
\end{eqnarray}
Here $U_{{\bf{k}}_{1},{\bf{k}}_{2},{\bf{k}}_{3},{\bf{k}}%
_{4}}^{\alpha _{1},\alpha _{2},\alpha _{3},\alpha _{4}}$ is the four-fermion
interaction, $\psi _{{\bf{k}},\alpha }^{\dagger }$ is the creation
operator for fermions with spin $\alpha $ and momentum ${\bf{k}}$, and $%
\varepsilon _{{\bf{k}}}$ is the band-structure dispersion. For a one band
Hubbard model with local Coulomb interaction, 
\begin{eqnarray}
U_{{\bf{k}}_{1},{\bf{k}}_{2},{\bf{k}}_{3},{\bf{k}}_{4}}^{\alpha
_{1},\alpha _{2},\alpha _{3},\alpha _{4}}&=&U\delta _{{\bf{%
k}_{1}+k_{2}-k_{3}-k_{4}}} \\
& & \times \left( \delta _{\alpha _{1}\alpha _{4}}\delta
_{\alpha _{2}\alpha _{3}}-\delta _{\alpha _{1}\alpha _{3}}\delta _{\alpha
_{2}\alpha _{4}}\right) .\nonumber
\end{eqnarray}
In a perturbation theory for Eq.\ref{hubb1} involving $U$ and the fermion
band width, the contributions from large and small fermionic energies are
mixed. However, near a magnetic instability much of the non-trivial physics
is associated (at any $U$) with the system behavior at low energies. To
single out this low-energy sector, one can borrow a strategy from field
theory: introduce a characteristic energy cut off, $\Lambda $, and generate
an effective low energy model by eliminating all degrees of freedom above $%
\Lambda $ in the hope that some of the system properties will be universally
determined by the low-energy sector and as such will not depend sensitively
on the actual choice of $\Lambda $. Eliminating these high energy degrees of
freedom is the central theoretical difficulty in the field of strongly
correlated electron systems. In our case, except for a poorly controlled RPA
analysis, there are no known ways to perform such a renormalization
procedure. Furthermore, the separation between high energies and
low-energies can be rigorously justified only if the interaction is smaller
than the fermionic bandwidth. For larger interactions, the critical
exponents very likely will remain the same, but the pre-factors will depend
on system properties far from the Fermi surface.

Still, it is possible to assume that the separation of scales is possible,
i.e., that the interaction is smaller than the bandwidth and to pursue the
consequences of that assumption. This is the strategy we adopt. By itself,
this does not guarantee that there exists a universal physics confined to
low energies. This we will have to prove. This also does not mean that the
system is in the weak coupling regime, as near the antiferromagnetic
transition we will find a strong, near-divergent contribution to the
fermionic self-energy that comes from low frequencies. What the separation
of scales actually implies (to the extent that we find universal, low-energy
physics) is that Mott physics does not play a major role. In particular, in
our analysis the Fermi surface in the normal state remains large, and its
volume satisfies Luttinger theorem. How well this approximation is satisfied
depends on doping for a given material and also varies from one material to
another. Most of our experimental comparisons will be made with the
cuprates. In cuprates, the Hubbard $U$ in the effective one-band model for $%
\mathrm{CuO}_{2}$ unit (a charge transfer gap) is estimated to be between 1
and 2 \textrm{eV}. The bandwidth, measured by ARPES and resonant Raman
experiments, roughly has the same value. This suggests that lattice effects, beyond a universal
low energy theory,
do, indeed, play some role. At half-filling, lattice effects are crucial as
evidenced by the fact that half-filled materials are both Mott insulators
and antiferromagnets with local (nearest-neighbor) spin correlations. Doping
a Mott insulator almost certainly initially produces a small Fermi surface
(hole or electron  pockets). This small Fermi surface evolves as doping increases and
eventually transforms into a large, ``Luttinger'' Fermi surface. How this
evolution actually occurs is still a subject of debate. From our
perspective, it is essential that at and above optimal doping, all ARPES
data indicate that the Fermi surface is large. Correspondingly, magneto-oscillation
experiments in BEDT-TTF based organic superconductors also show that the Fermi 
surface of these materials is large.  We believe that in this
situation, lattice effects change the system behavior quantitatively but not
qualitatively, and the neglect of lattice effects is justified. We emphasize
however that our analysis certainly needs to be modified to incorporate Mott
physics close to  half-filling.

Several aspects of our approach have a close similarity to the fluctuation
exchange approximation (FLEX), which in case of a single band Hubbard model
corresponds to a self consistent summation of bubble 
and ladder diagrams\cite{MonthScal,Bickers}.
Specifically, the emergence of a sharp gapless resonance mode in the spin
excitation spectrum of a d-wave superconductor, 
the feedback of this mode on
the fermions and the anomalous normal state behavior of low energy fermions
close to an antiferromagnetic instability are very similar in both
approaches\cite{TW1,TW2,GR1,GR2}. On the other hand, the FLEX approach attempts to determine the
static spin response and thus the actual position of the quantum critical
point in terms of the bare parameters of the model such as the local or
additional nonlocal Coulomb repulsions as well as the band structure $%
\varepsilon _{{\bf{k}}}$. As discussed above, the static spin response,
characterized by the correlation length $\xi $, strongly depends on the
behavior of fermions with large energy. Details of the underlying
microscopic model which are hard to specify uniquely as well as uncontrolled
approximations in the treatment of the high energy behavior strongly affect
the static spin response within the FLEX  approach, making it hard to
discriminate model dependent aspects from universal behavior. It is this
latter aspect which is resolved in our approach which concentrates
exclusively on the universal low energy physics for a given $\xi $.

What should be the form of the low-energy action? Clearly, it should involve
fermions which live near the Fermi surface. It also should involve
collective spin bosonic degrees of freedom with momenta near ${\bf Q}$, as these
excitations become gapless at the magnetic transition. The most
straightforward way to obtain this action is to introduce a spin-1 bose
field ${\bf{S}}$ and decouple the four-fermion interaction using the
Hubbard-Stratonovich procedure~\cite{ref_to_HS,ref_to_HS2}. This yields 
\begin{eqnarray}
{\mathcal{H}} &=&\sum_{{\bf{k}},\alpha } \ \varepsilon _{{\bf{k}}}
\psi _{{\bf{k}},\alpha }^{\dagger }\psi _{{\bf{k}},\alpha }+
\sum_{{\bf{q}}}U\left( {\bf{q}}\right) {\bf{S}}_{{\bf{q}}}\cdot {\bf{S}}%
_{-{\bf{q}}}  \nonumber \\
&&\ +\sum_{{\bf{k}},{\bf{q}},\alpha ,\beta }U\left( {\bf{q}}\right)
\psi _{{\bf{k+q}},\alpha }^{\dagger }
\sigma _{\alpha \beta }\psi _{{\bf{k}},\beta }\cdot {\bf{S}}_{-{\bf{q}}}
\end{eqnarray}
where the $\sigma _{\alpha \beta }$ are Pauli matrices and we assumed that
the four fermion interaction only makes a contribution in the spin channel
with momentum transfer ${\bf{q}}$. Integrating formally over energies
larger than $\Lambda $ we obtain the effective action in the form (see e.g., 
\onlinecite{CastroNeto00}) 
\begin{eqnarray}
{\mathcal{S}} &=&-\int_{k}^{\Lambda }G_{0}^{-1}\left( k\right) \psi _{k,\alpha
}^{\dagger }\psi _{k,\alpha }+\frac{1}{2}\int_{q}^{\Lambda }\chi
_{0}^{-1}\left( q\right) \ {\bf{S}}_{q}\cdot {\bf{S}}_{-q}  \nonumber \\
&&+g\int_{k,q}^{\Lambda }\psi _{k+q,\alpha }^{\dagger }\sigma _{\alpha \beta
}\psi _{k,\beta }\cdot {\bf{S}}_{-q}+O(S^{4}).\   \label{startac}
\end{eqnarray}
The last term is a symbolic notation for all terms with higher powers of $S$%
. Fortunately, in dimensions $d\geq 2$ these higher order terms are
irrelevant (marginal for $d=2$) and can therefore be neglected~\cite
{Hertz,Millis93}.

The integration over $k$ and $q$ in \ref{startac} is over $2+1$
dimensional vectors $q=\left( {\bf{q}}, i\omega _{m}\right) $ with
Matsubara frequency $\omega _{m}$. In explicit form, the integrals read 
\begin{equation}
\int_{q}^{\Lambda }...=\int_{\left| {\bf{q-Q}}\right| <\Lambda }\frac{d^{d}%
{\bf{q}}}{\left( 2\pi \right) ^{d}}\ T\sum_{m}...
\end{equation}
in the boson case, and 
\begin{equation}
\int_{k}^{\Lambda }...=\int_{\left| {\bf{k-k}}_{F}\right| <\Lambda }\frac{%
d^{d}{\bf{k}}}{\left( 2\pi \right) ^{d}}T\sum_{m}...
\end{equation}
in the fermion case. Further, $g$ is the effective coupling constant, $%
G_{0}\left( k\right) $ is the bare low-energy fermion propagator, and $%
\chi _{0}\left( q\right) $ is the bare low-energy collective spin boson
propagator. As we have emphasized, a controlled derivation of $g$, $%
G_{0}(k)$, and $\chi _{0}(q)$ is impossible. We therefore will not try to
calculate $g$, etc. Rather, we use the fact that antiferromagnetism
predominantly comes from high-energy fermions and further assume that the
integration over high energies does not produce singularities in both
bosonic and fermionic propagators. Then, quite generally, $G_{0}(k)$, and $%
\chi _{0}(q)$ should have Fermi-liquid and Ornstein-Zernike forms,
respectively 
\begin{equation}
G_{0}\left( k\right) =\frac{z_{0}}{i\omega_m -\epsilon _{{\bf{k}}}},
\label{gf_f}
\end{equation}
\begin{equation}
\chi _{0}\left( q\right) =\frac{\alpha }{\xi _{0}^{-2}+\left( {\bf{q-Q}}%
\right) ^{2}+\omega_m ^{2}/c^{2}}.  \label{gf_b}
\end{equation}
Here, $\xi _{0}$ is the bare value of the spin correlation length, and the
other notations are self-explanatory. The actual $\xi $ generally differs
from the bare one because the low energy fermions that damp the spin
fluctuation modes might change as well their static properties.  $%
\xi $ acquires an additional temperature dependence due to spin-spin interactions. We
return to this question later. It is essential that the bare spin propagator
does not contain a term linear in $\omega $. The latter will only appear
when we consider the interaction \textit{within} the low-energy model.

We also assume that (i) the momentum dependence of the effective coupling $g$
is non-singular and can be neglected (recall that we are only interested in
a narrow range of bosonic momenta near ${\bf{Q}}$), (ii) the low-energy
fermionic dispersion can be linearized in ${\bf{k}}-{\bf{k}}_{F}$: $%
\epsilon _{{\bf{k}}}={\bf{v}}_{F}\cdot ({\bf{k}}-{\bf{k}}_{F})$, and
(iii) that the Fermi velocity is non-singular near hot spots and to first
approximation its magnitude can be approximated by a constant. In doing this
we neglect effects due to a van Hove singularity in the density of states.
Finally, the exact values of $z_{0}$ and $\alpha $ are not relevant, as both
can be absorbed into an effective coupling with dimension of an energy 
\begin{equation}
{\bar{g}}=g^{2}z_{0}^{2}\alpha ,  \label{g_def}
\end{equation}
while  $v_{F}$ and $\xi $ will always appear only in the combination $%
v_{F}\xi ^{-1}$.

We see therefore that the
 input parameters in Eq.~\ref{startac} are the effective coupling
energy ${\bar{g}}$, the typical quasiparticle energy $v_{F}\xi ^{-1}$, and
the upper cutoff $\Lambda $. An additional parameter is the angle $\phi _{0}$
between the Fermi velocities at the two hot spots separated by ${\bf{Q}}$,
but this angle does not enter the theory in any significant manner as long
as it is different from $0$ or $\pi $. When hot spots are located near $%
(0,\pi )$ and $(\pi ,0)$ points, as in optimally doped cuprates, $\phi _{0}$
is close to $\pi /2$, the value  we use in what follows.

As we have emphasized, we will demonstrate that the low-energy properties of
the model are universal and do not depend on $\Lambda $, which then can be
set to infinity. Out of the two parameters that are left, one can construct
a
 doping dependent dimensionless ratio 
\begin{equation}
\lambda =\frac{3}{4\pi }~\frac{{\bar{g}}}{v_{\mathrm{F}}\xi ^{-1}},
\label{lambda}
\end{equation}
which will turn out to be the effective dimensionless coupling constant of
the problem (the factor $3/4\pi $ is introduced for further convenience).
The fact that $\lambda $ scales with $\xi $ immediately implies that close
enough to a magnetic transition $\lambda >1$, i.e., the system will
necessarily be in a strong coupling limit. Besides $\lambda $ the only other
free parameter of the theory is an overall energy scale, i.e., ${\bar{g}}$
(or alternatively the quasiparticle energy $v_{\mathrm{F}}\xi ^{-1}$). All
physical quantities that we discuss will be expressed in terms of these two
parameters only.

Eqs.\ref{startac}-\ref{gf_b} determine the structure of the perturbation
theory of the model. The interaction between fermions and collective spin
excitations yields self energy corrections to both bosonic and fermionic
propagators. We will show below that at strong coupling, the fermionic
self-energy strongly depends on frequency and also displays some dependence
on the momentum along the Fermi surface. However, its dependence on the
momentum transverse to the Fermi surface can be neglected together with
vertex corrections. The fermionic and bosonic propagators are then given by
the Gor'kov expressions, which for generality we present in the
superconducting state, 
\begin{equation}
G_{{\bf{k}}}(i\omega )=\frac{i\omega +\Sigma _{{\bf{k}}}(i\omega
)+\varepsilon _{{\bf{k}}}}{(\left( i\omega +\Sigma _{{\bf{k}}}(i\omega
)\right) ^{2}-\Phi _{{\bf{k}}}^{2}(i\omega )-\varepsilon _{{\bf{k}}}^{2}},
\label{GGorkov}
\end{equation}
\begin{equation}
F_{{\bf{k}}}(i\omega )=-\frac{\Phi _{{\bf{k}}}(i\omega )}{\left(i \omega
+\Sigma _{{\bf{k}}}(i\omega )\right) ^{2}-\Phi _{{\bf{k}}}^{2}(i\omega
)-\varepsilon _{{\bf{k}}}^{2}},  \label{FGorkov}
\end{equation}
\begin{equation}
\chi _{{\bf{q}}}\left( i\omega \right) =\frac{\alpha \xi ^{2}}{1+\xi
^{2}\left( {\bf{q-Q}}\right) ^{2}-\Pi _{{\bf{Q}}}\left( i\omega \right) }%
.  \label{chitot}
\end{equation}
Here $\Sigma _{{\bf{k}}}(i\omega )$ and $\Pi _{{\bf{Q}}}(i\omega )$ are
fermionic and bosonic self-energies (${\bf k}$ stands for the component along the Fermi
surface), and $F_{{\bf{k}}}(i\omega )$ and $\Phi _{{\bf{k}}}(i\omega )$
are the anomalous Green's function and the anomalous self-energy,
respectively. In the next sections we  compute
 $\Sigma _{{\bf{k}}}(i\omega )$, $\Pi _{{\bf{Q}}}(i\omega )$
 and $\Phi _{{\bf{k}}}(i\omega )
$.

One can also motivate the spin fermion model by following the approach
developed by Landau in his theory of Fermi liquids\cite{Landau}, i.e., by
assuming that the influence of the other fermionic quasiparticles on a given
quasiparticle can be described in terms of a set of molecular fields\cite
{Leggett_review,Monthoux93}. In the present case the dominant molecular
field is an exchange field produced by the Coulomb interaction $U({\bf{q}}%
) $. However, one has to consider this field as dynamic, not static. The
corresponding part of the action contains 
\begin{equation}
{\cal S}=-\int_{k}^{\Lambda }G_{0}^{-1}\left( k\right) \psi _{k,\alpha
}^{\dagger }\psi _{k,\alpha }+\frac{1}{2}\int_{q}^{\Lambda }\ \ {\bf{H}}%
_{q}^{\mathrm{int}}\cdot {\bf{s}}_{-q}
\end{equation}
with fermionic spin density ${\bf{s}}_{-q}=\int_{k}^{\Lambda }\psi
_{k+q,\alpha }^{\dagger }\sigma _{\alpha \beta }\psi _{k,\beta }$. As in the
Landau theory of Fermi liquids, we assume that the molecular field ${\bf{H%
}}_{q}^{\mathrm{int}}$ is given by the linear response function, 
\begin{equation}
{\bf{H}}_{q}^{\mathrm{int}}=g^{2}\ \chi _{0}\left( q\right) {\bf{s}}_{-q}
\end{equation}
This expression is valid as long as one is not in a region so close to a
magnetic instability that nonlinear magnetic effects play an important role.
Formally, this expression can be obtained from Eq.\ref{startac} by
integrating out the collective degrees of freedom ${\bf{S}}$. A relation
between this purely fermionic approach and the bosonic spin susceptibility $%
\chi \left( {\bf{q}}\right) $ of Eq. \ref{chitot} can be established by
evaluating the reducible four point vertex in the spin channel in the lowest
order of perturbation theory. We find 
\begin{equation}
\Gamma _{\alpha \beta, \gamma \delta }(k,k^{\prime },q)=-V_{\mathrm{eff}}(q)~%
{\bf{\sigma }}_{\alpha \beta }{\bf{\sigma }}_{\gamma \delta }
\label{ffvertex}
\end{equation}
where the effective quasiparticle interaction is proportional to the
renormalized spin propagator of Eq.~\ref{chitot}: 
\begin{equation}
V_{\mathrm{eff}}(q)=g^{2}\chi (q)=\frac{g^2 \alpha \xi^2}{1+\xi^2({\bf q-Q})^2-\Pi_{\bf Q}(\omega)}.  \label{sfint}
\end{equation}

\subsection{Weak coupling approach to the pairing instability}

One of the most appealing aspects of the spin-fluctuation theory is that it
inevitably yields an attraction in the $d_{x^{2}-y^{2}}$ channel. As with
any unconventional pairing, this attraction is the result of a specific
momentum dependence of the interaction, not of its overall sign which for
spin fluctuation exchange is positive, i.e., repulsive in the $s$-wave
channel. This differentiates spin induced pairing from the pairing mediated
by phonons. In the latter case, the effective interaction between fermions
is negative up to a Debye frequency. In configuration space the
spin-fluctuation interaction, Eq.\ref{sfint}, can easily seen to be
repulsive at the origin and alternate between attraction and repulsion as
one goes to nearest neighbors. It is always repulsive along the diagonals,
 and this is why a $d_{x^{2}-y^{2}}$ state, in which 
 the nodes of the gap are along the
diagonals, is the energetically preferred pairing state due to 
 spin-fluctuation exchange \cite{MonthouxJCPS}.

Suppose first that  the spin-fermion coupling is small enough such 
 that conventional
perturbation theory is valid. To second order in the spin-fermion coupling,
the spin mediated interaction has the following form, Eq.~\ref{ffvertex}.
The total antisymmetry of the interaction implies that the orbital part is
symmetric when the spin part is antisymmetric and vice versa. The orbital
part of the interaction at low frequencies has the same sign as the
propagator of optical phonons at $\omega <\omega _{D}$. However, the spin
part involves a convolution of the Pauli matrices, and is different from
phononic $\delta _{\alpha \beta }\delta _{{\gamma \delta }}-\delta _{\alpha
\delta }\delta _{\beta \gamma }$. Using ${\bf{\sigma} }_{\alpha \beta }%
{\bf{\sigma }}_{\gamma \delta }=2\delta _{\alpha \delta }\delta _{\beta
\gamma }-\delta _{\alpha \beta }\delta _{\gamma \delta }$ we find 
\begin{equation}
\Gamma _{\alpha \gamma ,\beta \delta }(q )=- g^{2}\chi (q )~(T_{\alpha
\gamma ,\beta \delta }-3S_{\alpha \gamma ,\beta \delta })
\end{equation}
where $T_{\alpha \gamma ,\beta \delta }=(\delta _{\alpha \beta }\delta _{{%
\gamma \delta }}+\delta _{\alpha \delta }\delta _{\beta \gamma })/2$ and $%
S_{\alpha \gamma ,\beta \delta }=(\delta _{\alpha \beta }\delta _{{\gamma
\delta }}-\delta _{\alpha \delta }\delta _{\beta \gamma })/2$ are triplet
and singlet spin configurations, respectively. We see that there is an extra
minus sign in the singlet channel. This obviously implies that in
distinction to phonons, the isotropic $s-$wave component of the interaction
is repulsive, i.e., isotropic $s-$wave pairing due to spin fluctuation
exchange is impossible. There are two other possibilities: unconventional
singlet pairing for which we will need a partial component of $\chi (q)$ to
be negative, or triplet pairing, for which the corresponding partial
component of $\chi (q)$ should be positive.

\begin{figure}
\epsfxsize=2.4in 
\epsfysize=2.2in
\begin{center}
\epsffile{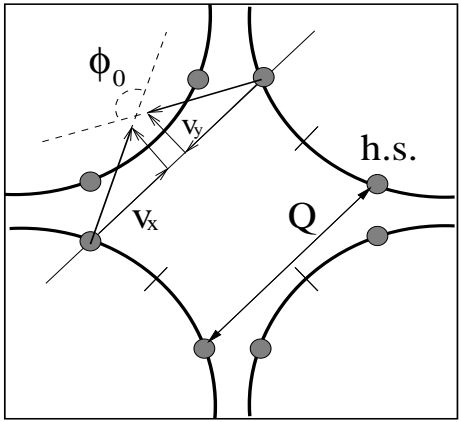}
\end{center}
\caption{A representative Brillouin zone and Fermi surface for a
near-optimally doped cuprate superconductor. The hot spots are connected by the
antiferromagnetic wave vector ${\bf{Q}}=(\protect\pi ,\protect\pi )$. For
a weakly incommensurate magnetic response, the number of hot spots doubles
but simultaneously the intensity of the magnetic response at momenta
connecting pairs of hot spots drops by a factor of two. The net effect then
remains the same as for the commensurate response.
 The figure is taken from~\protect\onlinecite{Abanov01advphys}. }
\label{FigBZ}
\end{figure}

It turns out that for a spin susceptibility peaked at or near the
antiferromagnetic momentum ${\bf{Q}}$, the triplet components are all negative, and triplet pairing is impossible. We therefore focus on singlet pairing. The
linearized equation for the pairing vertex $F_{{\bf{k}}}(i\omega _{m})$ in
the singlet channel is 
\begin{equation}
F(k)=-3g\int_{k^{\prime }}^{\Lambda }F(k')\chi(k-k')G_0(k')G_0(-k').
\label{lineq}
\end{equation}
If $\chi(k-k')=\chi_{\bf{k-k}'} (i\omega _{m}-i\omega _{m}^{\prime })$ was
independent of momentum, a solution of this equation would not exist. In our
situation, however, the interaction predominantly occurs between fermions
that are separated in momentum space by ${\bf{Q}}=(\pi ,\pi )$ (see Fig.%
\ref{FigBZ}).

Since both ${\bf{k}}$ and ${\bf{k}}+{\bf{Q}}$ should be near the Fermi
surface, the pairing predominantly involves fermions near hot spots 
${\bf{k}}_{\rm hs}$ and ${\bf{k}}_{\rm hs}+{\bf{Q}}$. As the hot spots are well
separated in momentum space, we can change the sign in the r.h.s. of Eq.\ref
{lineq} by requiring that the pairing gap changes sign between ${\bf{k}}%
_{\rm hs}$ and ${\bf{k}}_{\rm hs}+{\bf{Q}}$. Indeed, substituting 
$F_{{\bf{k+Q}}}(i\omega _{m})=-F_{{\bf{k}}}(i\omega _{m})$ into Eq.\ref{lineq}, we find
that both sides of this equation have the same sign. The pairing problem
then becomes almost identical to that for phonons, and the gap equation has
a solution at some $T_{c}$.

Altogether, there are $N=8$ hot spots in the Brillouin zone; these form 4
pairs separated by ${\bf{Q}}$, between which the gap should change sign.
This still does not specify the relative sign of the gap between adjacent
hot spots. However, if the hot spots are close to $(0,\pi )$ and symmetry
related points, as in optimally doped cuprates, the gap is unlikely to
change sign between them. This leaves as the only possibility a gap that
vanishes along diagonal directions $k_{x}=\pm k_{y}$, i.e., it obeys $%
d_{x^{2}-y^{2}}$ symmetry.

We emphasize that although we found that the pairing gap necessarily should
have $d_{x^{2}-y^{2}}$ symmetry, its momentum dependence is generally
more complex than simply $\cos k_{x}-\cos k_{y}$. To understand this, we
consider how one can generally analyze Eq.\ref{lineq}~\cite{ACF}. A standard
strategy is to expand both the pairing vertex and the interaction in the
eigenfunctions of the representations of the $D_{4h}$ symmetry group of the
square lattice. There are four singlet representations of $D_{4h}$ labeled $%
A_{1g},B_{1g},B_{2g}$ and $A_{2g}$. The basic eigenfunctions in each
representation are $1$ in $A_{1g}$, $\cos k_{x}-\cos k_{y}$ in $B_{1g}$, $%
\sin k_{x}\sin k_{y}$ in $B_{2g}$ and $(\cos k_{x}-\cos k_{y})\sin k_{x}\sin
k_{y}$ in $A_{2g}$. Other eigenfunctions in each representation are obtained
by combining the basic eigenfunction with the full set of eigenfunctions
with full $D_{4h}$ symmetry. Obviously, the eigenfunctions with $%
d_{x^{2}-y^{2}}$ symmetry belong to $B_{1g}$ channel. The orthogonal
functions in the set can be chosen as $d_{n}({\bf{k}})=\cos nk_{x}-\cos
nk_{y}$.

One can easily make sure that the pairing problem close to T$_{c}$ decouples
between different representations. As $\xi \rightarrow \infty $, the $B_{1g}$
components of the spin susceptibility diverge while other components remain
finite. Obviously, the pairing is in the $B_{1g}$ channel. At moderate $\xi $%
, all components are generally of the same order, but numerical calculations
show that $B_{1g}$ components continue to be the largest. We therefore
neglect other channels and focus only on $B_{1g}$ pairing. Still, there are
an infinite number of $B_{1g}$ eigenfunctions, and the pairing problem does
not decouple between them. Indeed, expanding $F_{{\bf{k}}}(\omega _{m})$
and $\chi ({\bf{k}},\omega _{m})$ in $d_{n}({\bf{k}})$ as $F_{{\bf{k}}%
}=\sum_{p}F_{p}d_{p}({\bf{k}})$, $\chi ({\bf{k-k}}^{\prime },i\omega
-i\omega ^{\prime })=\sum_{p}\chi _{p}(i\omega -i\omega ^{\prime })d_{p}(%
{\bf{k}})d_{p}({\bf{k}}^{\prime })$ and substituting the result into (%
\ref{lineq}) we obtain~\cite{ACF} 
\begin{equation}
F_{n}(i\omega _{m})=-3g^{2}T\sum_{p,m'}
F_{p}(i\omega _{m'})~\chi _{n}(i\omega _{m-m'})\Gamma _{n,p}  \label{a1}
\end{equation}
where 
\begin{equation}
\Gamma _{n,p}=\int \frac{d^{2}k^{\prime }}{4\pi ^{2}}~\frac{d_{n}({\bf{k}}%
^{\prime })d_{p}({\bf{k}}^{\prime })}{(\omega _{m}^{\prime
})^{2}+(\epsilon _{{\bf{k}}}^{\prime })^{2}}  \label{comp}
\end{equation}
We see that all $\Gamma _{n,p}$ are nonzero for arbitrary $n$ and $p$ as the
products $d_{n}(k)d_{p}(k)$ and 
$G_0(k)G_0(-k)=1/((\omega _{m})^{2}+(\epsilon _{{\bf{k}}})^{2})$ are symmetric under $D_{4h}$%
. As a result, Eq.~\ref{a1} couples together an infinite number of partial
components $F_{n}$. As long as the partial components of the susceptibility
are dominated by momenta relatively far from ${\bf{Q}}$, all $\chi _{n}$
and hence $F_{n}$ are of the same order. In some particular lattice models,
e.g. in the Hubbard model with nearest neighbor interaction, $U=4t$ and $\xi
\sim 2$, an RPA analysis shows that $\chi _{1}$ is numerically significantly
larger than all other $\chi _{n}$~\cite{Monthoux93}. In this situation, $F_{%
{\bf{k}}}\approx F_{1}d_{1}({\bf{k}})$, i.e., the momentum dependence of
the gap should closely resemble the $\cos k_{x}-\cos k_{y}$ form. In real
space, this implies that the pairing (even though non-local in time) is
local and involves fermions from the nearest neighbor sites on the lattice.
In models with different high-energy physics, other $\chi _{n}$ may
dominate. Clearly, the momentum dependence of the gap is not universal as
long as the pairing is local in real space.

Universality is recovered when the susceptibility becomes strongly peaked at 
${\bf{Q}}$. In this limit (in which we can also rigorously justify
restricting our attention to the $B_{1g}$ channel~\cite{ACF}), all partial
components $\chi_n$ scale as $\log \xi$ and differ only in sub-leading
non-logarithmic terms. A straightforward trigonometric exercise shows that
for near-equal $\chi_n $, the pairing gap is very different from $\cos k_x -
\cos k_y$: it is reduced near the nodes and enhanced near the maxima at hot
spots with the ratio of the gap maximum and the slope near the nodes
increasing as $\xi^2$. In this situation the pairing problem is confined to
hot spots, and the pairing symmetry is inevitably $d_{x^2 -y^2}$.

\section{Summary of strong coupling theory for electron phonon pairing}

For phonon-mediated superconductivity, Eliashberg theory offers a successful
approach to study the system behavior at strong coupling. It is justified by
Migdal's theorem which states that the vertex corrections, $\delta g/g$ and $%
(1/v_{F})d\Sigma ({\bf{k}},i\omega )/dk$, are small due to the smallness
of the ratio of sound velocity and Fermi velocity stemming from the
smallness of the ratio of the electronic and ionic masses. Eliashberg has
demonstrated that for $\Sigma ({\bf{k}},i\omega )\approx \Sigma (i\omega )$
and $g_{\rm tot}=g+\delta g\simeq g$, the phonon-mediated pairing
problem can be solved exactly. A review of Eliashberg theory for \ the
electron-phonon problem is given in Ref.\onlinecite{RMP-Carbotte}. As we already
pointed out, for spin fluctuation induced pairing there is no Migdal theorem
simply because the spin propagator is made out of electrons and hence a
typical spin velocity is of the same order as the Fermi velocity. It is
therefore natural to inquire whether this implies that an Eliashberg-type
treatment is inapplicable to the spin problem. To properly address this
issue, we review the case of electron- phonon interactions and examine why
the smallness of the mass ratio affects $d\Sigma ({\bf{k}},i\omega )/d{\bf k}$
but not $d\Sigma ({\bf{k}},i\omega )/d\omega $.

The electron-phonon interaction is generally rather involved and has been
discussed in detail in the literature\cite{Scalapino69}. For our purposes,
it is sufficient to consider the simplest Fr\"{o}hlich-type model of the
electron-phonon interaction in which low-energy electrons are coupled to
optical phonons by a momentum, frequency and polarization independent
interaction $g_{\mathrm{ep}}$. Despite its simplicity, this model captures
the key physics of phonon-mediated pairing.

The propagator of an optical phonon has the form 
\begin{equation}
D({\bf{q}},i\omega _{m})=\frac{\omega _{0}}{\omega _{0}^{2}+\omega
_{m}^{2}+(v_{s} {\bf{q}})^{2}}.  \label{ph1}
\end{equation}
Here $\omega _{0}$ is a typical phonon frequency, which is of the order of
the Debye frequency, and $v_{s}$ is the sound velocity. Both $\omega _{0}$
and $v_{s}$ are input parameters, unrelated to fermions. The ratio $%
v_{s}/v_{F}$ scales as $(m/M)^{1/2}$ where $M$ is the ionic mass, and $m$ is
the electron mass. In practice, $v_{s}/v_{F}\sim 10^{-2}$~\onlinecite{Scalapino69}%
. This is a real physically motivated small parameter for the electron- phonon problem.

Far from structural instabilities, $\omega _{0}\sim v_{s}a$ where a is the
distance between ions. To make the analogy with spin fluctuations more
transparent, we will consider below the case when $\omega _{0}\ll v_{s}a$,
i.e., the correlation length for optical phonons is large. This last
assumption allows us to simplify the calculations but will not change their
conclusions.

In the presence of two different velocities, it is not \textit{a priori}
obvious what is the best choice of a dimensionless expansion parameter in
the theory. Two candidate dimensionless parameters are 
\begin{equation}
\lambda _{\mathrm{ep}}=g_{\mathrm{ep}}^{2}/(4\pi (v_{s}v_{F}))
\end{equation}
(the factor $4\pi $ is chosen for further convenience), and 
\begin{equation}
{\tilde{\lambda}}_{\mathrm{ep}}=\lambda _{\mathrm{ep}}v_{s}/v_{F}\sim (g_{%
\mathrm{ep}}/v_{F})^{2}.
\end{equation}
For simplicity, we set the lattice constant $a=1$. Obviously, $\lambda _{%
\mathrm{ep}}\gg {\tilde{\lambda}}_{\mathrm{ep}}$. In practice, ${\tilde{%
\lambda}}_{\mathrm{ep}}\ll 1$ for all reasonable $g_{\mathrm{ep}}$ while $%
\lambda _{\mathrm{ep}}$ can be either small or large. 

Which of the two dimensionless couplings appears in perturbation theory?
Consider first the lowest-order diagram for the spin-fermion vertex at zero
external frequency and zero bosonic momentum with all fermionic momenta at
the Fermi surface. Choosing $T=0$ and the $x$ axis along the direction of $%
{\bf{v}}_{F}$ of an external fermion, we obtain the vertex correction in
the form: 
\begin{equation}
\frac{\delta g_{\mathrm{ep}}}{g_{\mathrm{ep}}}=\frac{g_{\mathrm{ep}}^{2}}
{8\pi ^{3}}
\int \frac{ d\omega _{m}d^{2}q   }{(i\omega _{m}-v_{F}q_{x})^{2}}
\frac{\omega _{0}}{\omega _{0}^{2}+\omega
_{m}^{2}+(v_{s}{\bf{q}})^{2}}.
\label{ph2}
\end{equation}
The evaluation of the momentum and frequency integrals is straightforward.
On performing the integration we obtain for $\omega _{0}\ll v_{F}$, 
\begin{equation}
\frac{\delta g_{\mathrm{ep}}}{g_{\mathrm{ep}}}={\tilde{\lambda}}_{\mathrm{ep}%
}=\lambda _{\mathrm{ep}}\frac{v_{s}}{v_{F}}.  \label{ph3}
\end{equation}
We see that the vertex correction scales as ${\tilde{\lambda}}_{\mathrm{ep}}$
and is small. For $\omega _{0}\sim v_{s}$, the expression is more complex
but still is of order ${\tilde{\lambda}}$. One can easily check that
higher-order diagrams yield higher powers of ${\tilde{\lambda}}_{ep}$, i.e.,
are also small. This result, first obtained by Migdal\cite{migdal}, is
Migdal's theorem.

The physics behind this result is transparent and can be directly deduced
from an analysis of the integrand in Eq.\ref{ph2}. We see that the product
of the two fermionic Green's functions yields a double pole at $%
q_{x}=i\omega _{m}/v_{F}\sim \omega _{0}/v_{F}$. The integral over $q_{x}$
is then finite only due to the presence of the pole in the bosonic
propagator which is located at a much larger momentum $q_{x}\sim i\omega
_{0}/v_{s}$. This implies that the electrons that contribute to the vertex
correction are oscillating not at their own fast frequencies but at much
slower phonon frequencies. At vanishing $v_{s}/v_{F}$, fermions can be
considered as static objects with a typical $q_{x}\sim q_{y}\sim \omega
_{0}/v_{s}$. The two dimensional momentum integration over the product of
two fermionic propagators then yields $O(1/v_{F}^{2})$. Simultaneously, the
frequency integral over the bosonic propagator yields $D(q\sim \omega
_{0}/v_{s},0)=O(1)$. Combining the two results, we obtain $\delta g_{%
\mathrm{ep}}/g_{\mathrm{ep}}=O({\tilde{\lambda}})$ as in  Eq.~\ref{ph3}.

\begin{figure}[tbp]
\epsfxsize=2.4in 
\epsfysize=1.0in
\begin{center}
\epsffile{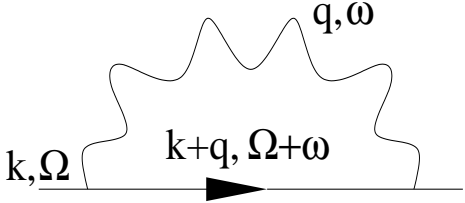}
\end{center}
\caption{A Feynman diagram for the electronic self energy due to electron
phonon interaction in the Eliashberg theory.}
\label{phonon-self-energy}
\end{figure}

We next consider the fermionic self-energy. Evaluating the diagram in Fig.%
\ref{phonon-self-energy} and subtracting the contribution at $\omega =0,~%
{\bf{k}}={\bf{k}}_{F}$, we obtain 
\begin{equation}
\Sigma ({\bf{k}},i\omega _{m})=
(i\omega _{m}-\epsilon _{{\bf{k}}})I({\bf{k}},i\omega _{m})  \label{ph4}
\end{equation}
where 
\begin{eqnarray}
I({\bf{k}},i\omega _{m}) &=&g_{\mathrm{ep}}^{2}\int \frac{d\Omega
_{m}d^{2}q}{8\pi ^{3}}\frac{\omega _{0}}{\omega _{0}^{2}+\Omega
_{m}^{2}+(v_{s}{\bf{q}})^{2}}  \label{ph5} \\
&&\times \frac{1}{i\Omega _{m}-v_{F}q_{x}}\frac{1}{i(\omega _{m}+\Omega
_{m})-\epsilon _{k}-v_{F}q_{x}}  \nonumber
\end{eqnarray}
We again have chosen the $q_{x}$ axis along the Fermi velocity 
${\bf{v}}_{F}$.

Suppose for a moment that $I({\bf{k}},i\omega _{m})$ is non-singular in
the limit ${{\bf k} \rightarrow 0}$ and $\omega _{m}\rightarrow 0$. Then the
wave function renormalization and the velocity renormalization are both
expressed via $I(0,0)$. This quantity can be evaluated in exactly the same
way as the vertex correction. Performing the integration, we find 
\begin{equation}
I(0,0)=-{\tilde{\lambda}}_{\mathrm{ep}}  \label{ph6}
\end{equation}
Substituting this into Eq.\ref{ph4} we find $v_{F}^{-1}d\Sigma /dk=-d\Sigma
/d(i\omega) ={\tilde{\lambda}}_{\mathrm{ep}}$. As ${\tilde{\lambda}}_{\rm ep}\ll 1$,
this result, if true, would imply that both derivatives of the self-energy
are small, i.e., the system is fully in the weak coupling regime. The
physical interpretation of this result would parallel that for the vertex
correction: fast electrons vibrate at slow phonon velocities, and this
well-out-of-resonance vibration cannot substantially affect electronic
properties.

However, this result is incorrect - $d\Sigma /d\omega $ is of order $\lambda
_{\mathrm{ep}}$, not ${\tilde{\lambda}}_{ep}$. 
The answer lies in the singular behavior of $I({\bf{k}},i\omega _{m})$ in
the limit of vanishing ${\bf{k}}$ and $\omega _{m}$. To see this, we need
to compute $I({\bf{k}},i\omega )$ in Eq.\ref{ph5} at small but finite $%
\omega $ and $\epsilon _{{\bf{k}}}$. An explicit computation yields 
\begin{equation}
I({\bf{k}},i\omega _{m})=-{\tilde{\lambda}}_{ep}+\lambda _{ep}\frac{%
i\omega _{m}}{i\omega _{m}-\epsilon _{{\bf{k}}}},  \label{ph7}
\end{equation}
demonstrating that  $I(0,0)$
and the limit of $I({\bf{k}},i\omega )$ at $\omega ,\epsilon _{{\bf{k}}%
}\rightarrow 0$ do not coincide.
The additional contribution in Eq.~\ref{ph7} comes from the regularization of the double pole in the
integrand in Eq.\ref{ph5} and is in fact mathematically similar to the
chiral anomaly in quantum chromodynamics\cite{Jackiw}.

It is clear from Eq.\ref{ph7} that $I({\bf{k}},i\omega _{m})$ has singular
low frequency and small momentum limits. Substituting Eq.\ref{ph7} into Eq.%
\ref{ph4}, we find that the actual self-energy is 
\begin{equation}
\Sigma ({\bf{k}},i\omega _{m})=i\lambda _{ep}~\omega _{m}-{\tilde{\lambda}}%
_{ep}(i\omega _{m}-\epsilon _{{\bf{k}}}).  \label{ph8}
\end{equation}
We see that $d\Sigma /d\omega $ scales with $\lambda _{ep}$ whereas $%
v_{F}^{-1}d\Sigma /dk$ is proportional to ${\tilde{\lambda}}_{\mathrm{ep}}$.
At strong coupling, $\lambda _{ep}\geq 1$, this self-energy gives rise to a
strong renormalization of both the quasiparticle mass and the quasiparticle
wave function. Still, vertex corrections and the renormalization of the
Fermi velocity scale with ${\tilde{\lambda}}_{ep}$ and are small

To understand the physical origin of the distinction between the frequency
and momentum dependence of the fermionic self-energy, it is essential to
realize that the second, $O(\lambda _{ep})$, term in Eq. \ref{ph7} is not
caused by real electron-phonon scattering. Rather, a careful examination of
the structure of the expression for $I({\bf{k}},\omega _{m})$ shows that
this term accounts for the pole in the fermionic particle hole polarization
bubble at small momentum and frequency. This pole is known to describe a
zero sound bosonic collective excitation -- a vibration of the Fermi surface
in which it changes its form but preserves its volume.\cite{LL,PinesNoz}
This implies that the fermionic self-energy is actually caused by coupling
of fermions to their own zero-sound collective modes, while phonons just
mediate this coupling.

The neglect of vertex corrections and $v_{F}^{-1}d\Sigma /dk$ leads to the
well known Eliashberg equations for $\Sigma _{m}=\Sigma (i\omega _{m})$ and
(in the superconducting state) the pairing self energy $\Phi _{m}=\Phi
\left( i\omega _{m}\right) $ that also depends only on frequency. For
phonon-mediated superconductors, these equations have the form 
\begin{eqnarray}
\ \Sigma _{m} &=&\ \frac{\lambda _{ep}T}{2}\sum_{n}\frac{\left( \omega
_{n}+\Sigma _{n}\right) D\left( i\omega _{m-n}\right) }{\sqrt{\Phi
_{n}^{2}+\left( \omega _{n}+\Sigma _{n}\right) ^{2}}}~\   \label{si1} \\
\Phi _{m} &=&~\frac{\lambda _{ep}T}{2}\sum_{n}\frac{\Phi _{n}D\left( i\omega
_{m-n}\right) }{\sqrt{\Phi _{n}^{2}+\left( \omega _{n}+\Sigma _{n}\right)
^{2}}}~\ .
\end{eqnarray}
This set of equations is solved for a given $D\left( i\omega _{m}\right) $
which by itself 
 is only very weakly affected by fermions. In the normal state and at $%
\omega \ll \omega _{0}$, the self-energy Eq.\ref{si1} reduces to the first
term in Eq.\ref{ph8}.

We have reviewed the derivation of these well known equations to underline
the physical origin of the applicability of the Eliashberg approach. We see
that the electron-phonon interaction gives rise to two physically distinct
classes of interaction processes that contribute very differently to the
fermionic self-energy and vertex corrections. In the first class, fast
electrons are forced to vibrate at slow phonon frequencies (i.e., phonons
are in resonance, but electrons are far from resonance). This gives rise to
both vertex and self-energy corrections, but these are small ($\propto
v_{s}/v_{F}$) and are neglected in the Eliashberg theory. In the second
class, phonons mediate an effective coupling between fermions and their
zero-sound collective modes. This process involves fermions with energies
near resonance and phonons far away from resonance, and yields a strong
renormalization of the fermionic propagator. The applicability of the
Eliashberg approach is then based on the fact that one has to include the
influence of phonon-mediated scattering on zero-sound vibrations, but can
neglect the direct scattering of electrons by phonons.

These insights into the origin of the applicability of the Eliashberg theory
will now be used to justify a generalized Eliashberg approach for spin
mediated pairing.

\section{Strong coupling approach to spin-fermion interaction}

As shown in the previous section, for electron phonon interactions the
smallness of vertex and velocity renormalizations is caused by the small
ratio of the Bose velocity and the Fermi velocity. The
spin problem is qualitatively different. The bare spin-fluctuation
propagator, Eq.\ref{gf_b}, describes propagating magnons whose velocity $c$
is expected to be of the same order as the Fermi velocity. There is then no
a priori reason to neglect vertex and velocity renormalizations.
Fortunately, this argument is not correct for the following reasons: First,
 as we just found 
in case of electron-phonon interaction,  the fermionic self-energy in the Eliashberg
 theory is insensitive to the ratio of sound and Fermi velocities. The small ratio of $v_{s}/v_{F}$ is
only necessary to eliminate regular terms in the self-energy, which are due
to real scattering by phonons. For these terms to be small it is sufficient that bosons are 
slow modes compared to fermions. Second,  the dynamics of the spin fluctuations 
is drastically modified, compared to the ballistic behavior of Eq.~\ref{gf_b}.
A  strong spin-fermion interaction
at low-energies changes the bosonic dynamics from propagating to diffusive.
Diffusive modes have a different relationship between typical momenta and energies compared to ballistic ones, 
making them slow modes compared to electrons. Consequently, regular terms in the
fermionic self-energy again become smaller than singular ones.

 We will show
that in dimensions between $d=2$ and $d=3$, $d\Sigma \left( \omega \right)
/d\omega $ scales as $\lambda ^{3-d}$ where, we recall, $\lambda \propto \xi 
$ is the dimensionless spin-fermion coupling, while $v_{F}^{-1}d\Sigma /dk$
remains non-singular at $\xi =\infty $ in $d>2$ and only logarithmically
increases for $d=2$. This implies that an Eliashberg-type approach near a
magnetic instability is fully justified for $d>2$ and is ``almost''
justified for $d=2$. In the latter case (which is the most interesting
because of the cuprates), we will have to invoke an extra approximation (an
extension to large $N$) to be able to perform calculations in a controllable
way.

Our strategy is the following. We first establish that one can indeed
develop a controlled approach to the spin fluctuation problem in the normal
state, see also Ref.\onlinecite{Ioffe_MillisUspech}. Next we apply this theory to the pairing problem and show that there
is indeed  $d-$wave superconductivity caused by antiferromagnetic spin
fluctuations. We discuss the value of $T_{c}$ near optimal doping and the
properties of the superconducting state, particularly the new effects
associated with the feedback from superconductivity on the bosonic
propagator, since these distinguish between spin-mediated and phonon-mediated
pairings.

\subsection{ Normal state behavior}

For our normal state analysis we follow Ref.~\onlinecite{Abanov01advphys} and
perform computations assuming that the Eliashberg theory is valid, analyze
the strong coupling results, and then show that vertex corrections and $%
v_{F}^{-1}d\Sigma _{{\bf{k}}}\left( \omega =0\right) /d{\bf{k}}$ are
relatively small at strong coupling.

We begin by obtaining the full form of the dynamical spin susceptibility as
it should undergo qualitative changes due to interactions with fermions. The
self-energy for the spin susceptibility (that is the spin polarization
operator) is given by the convolution of the two fermionic propagators with
the momentum difference near ${\bf{Q}}$ and the Pauli matrices in the
vertices. Collecting all combinatorial factors, we obtain: 
\begin{eqnarray}
\Pi _{{\bf{q}}} \left( i\omega_{m}\right) &=&
-8{\bar{g}}\xi^2 \int
 \frac{d^{2}kd\Omega _{m}}{(2\pi )^{3}} G_{{\bf{k}}}\left(i \Omega_{m}\right) \nonumber \\
& & \times ~G_{{\bf{k+q}}}
\left( i\omega _{m} +i\Omega_{m}\right)
  \label{pi_def}
\end{eqnarray}
Here $G_{{\bf{k}}}\left( i\omega _{m}\right) $ is a full fermionic
propagator 
\begin{equation}
G_{{\bf{k}}}\left( i\omega _{m}\right) =\frac{1}{i\omega _{m}+\Sigma
\left( i\omega _{m}\right) -\varepsilon _{\bf k }}
\end{equation}
and $\Sigma \left( i\omega _{m}\right) $ is the self energy which remains to
be determined. In principle, even in Eliashberg theory, this self-energy
depends on the momentum component along the Fermi surface. However, for 
computations of $\Pi _{{\bf{Q}}}\left( i\omega _{m}\right) $ this
dependence can be safely neglected;  both fermions in the fermion
polarization bubble should be close to the Fermi surface, and hence the
momentum integration is necessarily confined to a narrow region around hot
spots.

The $2+1$ dimensional integral over momentum and frequency in Eq.~\ref{pi_def}
 is ultraviolet divergent; this implies that its dominant piece
comes from highest energies which we have chosen to be $O(\Lambda )$. This
contribution, however, should not be counted as it is already incorporated
into the bare susceptibility. In addition, $\Pi _{{\bf{Q}}}\left( i\omega
_{m}\right) $ contains a universal piece which remains finite even if we
extend the momentum and frequency integration to infinity. The most
straightforward way to obtain this contribution is to first integrate over
momentum and then over frequency, i.e. perform computations assuming an
infinite momentum cutoff but keeping energies still finite. One can easily
make sure that for this order of limits, the high-energy contribution is
absent due to a cancellation of infinities, and the full $\Pi _{{\bf{q}}%
}\left( i\omega _{m}\right) $ coincides with the universal part. The
integration is straightforward and on performing it we obtain 
\begin{equation}
\Pi _{{\bf{Q}}}\left( i\omega _{m}\right) =-\frac{\left| \omega
_{m}\right| }{\omega _{\mathrm{sf}}}  \label{Pins}
\end{equation}
where the characteristic energy scale is the spin fluctuation frequency, 
\begin{equation}
\omega _{\mathrm{sf}}=\frac{\pi }{4}\frac{(v_{F}\xi ^{-1})^{2}}{{\bar{g}}%
\sin {\phi _{0}}}.
\end{equation}
We recall that $\phi _{0}$ is the angle between the Fermi velocities at the
two hot spots separated by ${\bf q}\sim {\bf  Q}$. 
In optimally doped cuprates, $\phi _{0}\approx \pi /2$ 
and weakly depends on ${\bf{q}}$ as long as one
is far from momenta connecting Fermi points along \ the zone diagonal. This
implies that we can safely set $\phi _{0}=\pi /2$ and obtain, on making use of Eq.\ref{lambda},
\begin{equation}
\omega _{\mathrm{sf}}=\frac{\pi }{4}\frac{v_{F}^{2}\xi ^{-2}}{{\bar{g}}}=%
\frac{3}{16}\frac{v_{F}\xi ^{-1}}{\lambda }=\frac{9}{64 \pi} {\bar g} \lambda^{-2}.
\end{equation}

An analytical continuation of Eq.\ref{Pins} to the real frequency axis
yields $\Pi _{{\bf{Q}}}\left( \omega \right) =i|\omega |/\omega _{%
\mathrm{sf}}$. As we anticipated, the coupling to low-energy fermions gives
rise to a finite decay rate for a spin fluctuation. As typical spin
frequencies are of order $v_{F}\xi ^{-1}\sim c \xi ^{-1}$, at strong
coupling, $\lambda \geq 1$, the induced damping term $\omega /\omega
_{sf}\sim \lambda ~\omega /(v_{F}\xi ^{-1})$ is large compared to the
frequency dependent $\omega ^{2}/(c \xi ^{-1})^{2}$ term in the bare
susceptibility, i.e., \ the interaction with low energy fermions changes the
form of the spin dynamics from a propagating one to a relaxational one.
Neglecting the bare frequency term, we obtain 
\begin{equation}
\chi _{{\bf{q}}}\left( \omega +i0^{+}\right) =\frac{\alpha \xi ^{2}}{1+\xi
^{2}\left( {\bf{q-Q}}\right) ^{2}-i\omega /\omega _{\mathrm{sf}}},
\label{MMP}
\end{equation}
The same purely relaxational 
dynamic spin susceptibility  was introduced phenomenologically by
Millis, Monien and Pines \cite{MMP} to describe the spin dynamics observed in
NMR experiments on the cuprates.

We emphasize that the polarization operator, Eq.\ref{Pins}, is independent
of the actual form of the fermionic self energy $\Sigma _{{\bf{k}}_{%
\mathrm{hs}}}\left( \omega \right) $. The explanation of why a ${\bf{k}}-$%
independent fermionic self-energy does not affect the polarization bubble at
a finite momentum transfer was given by Kadanoff \cite{kadanoff} who
analyzed a similar problem for phonons. He pointed out that the
linearization of the fermionic energy around $k_{F}$ is equivalent to
imposing an approximate Migdal sum rule on the spectral function
 $A({\bf{k}},\omega )=(1/\pi )~\mathrm{Im}G({\bf{k}},\omega )$ 
\begin{equation}
\int d\epsilon _{{\bf{k}}}A({\bf{k}},\omega )=1,  \label{A}
\end{equation}
where $\epsilon _{{\bf{k}}}$, the bare fermionic dispersion relation,
 $={\bf{v}}_{{\bf{k}}}\cdot \left( {\bf{k-k}}_{\mathrm{F}%
}\right) $ in our case. Expressing $G({\bf{k}},\omega )$ in terms of $A(%
{\bf{k}},\omega )$ via the Kramers-Kronig relation and making use of Eq.%
\ref{A}, one finds 
\begin{equation}
\Pi ({\bf{q}},\omega )=-\frac{i\omega }{\omega _{\mathrm{sf}}}%
\int_{-\infty }^{\infty }d\omega ^{\prime }\frac{df(\omega ^{\prime })}{%
d\omega ^{\prime }}+{\mathcal{O}}(\omega ^{2}),
\end{equation}
where $f(\omega ^{\prime })$ is the Fermi function. Eq.\ref{Pins} then
follows from the fact that $f(\omega ^{\prime })$ is $1$ at $\omega ^{\prime
}=-\infty $ and $0$ at $\omega ^{\prime }=+\infty $.

We next determine the fermionic self energy $\Sigma _{{\bf{k}}}\left(
i\omega \right) $. In Eliashberg theory it is given by 
\begin{equation}
\Sigma _{{\bf{k}}}\left( i\omega _{m}\right) =-3g^{2}\int \frac{d^{2}%
{\bf{q}}d\varepsilon }{(2\pi )^{3}}~\chi _{{\bf{q}}}\left( i\varepsilon
\right) G_{{\bf{k}}+{\bf{q}}}\left( i\omega _{m}+i\varepsilon \right).
\label{SEns_a}
\end{equation}
Consider first the self-energy at a hot spot. Introducing $i\widetilde{%
\Sigma }_{n}=i\omega _{n}+\Sigma _{{\bf{k}}_{\mathrm{hs}}}\left( i\omega
_{n}\right) $ and  ${\bf{q-Q}}={\bf{\tilde{q}}}$, we obtain from Eq.\ref
{SEns_a} 
\begin{eqnarray}
\Sigma _{{\bf{k}}_{\mathrm{hs}}}\left( i\omega _{m}\right) &=&\frac{-3{\bar{g}}}
{\left( 2\pi \right) ^{3}}~\int \frac{d\omega _{m'}d^{2}{\tilde{q}}}
{{\bf{\tilde{q}}}^{2}+\xi ^{-2}\left( 1+\ \frac{\left| \omega
_{m}\right| }{\omega _{\mathrm{sf}}}\right) } \nonumber \\
& & \times \frac{1}{i \widetilde{%
\Sigma }_{m+m'} -v_{F}\tilde{q}}%
\end{eqnarray}
with ${\bf{\tilde{q}}}^{2}={\tilde{q}}_{\parallel }^{2}+{\tilde{q}}_{\perp
}^{2}$. The integration over the component ${\tilde{q}}_{\parallel }$,
parallel to the velocity ${\bf{v}}_{{\bf{k}}_{\mathrm{hs}}}$ at 
${\bf{k}}_{\mathrm{hs}}$, is elementary and yields 
\begin{equation}
\Sigma _{{\bf{k}}_{\mathrm{hs}}}\left( i\omega _{n}\right) =\frac{3%
\overline{g}}{8\pi ^{2}v_{F}^{2}}\int d\omega _{m}^{\prime }~d{\tilde{q}}%
_{\perp }\frac{1}{\sqrt{{\tilde{q}}_{\perp }^{2}+q_{0}^{2}\ }\left( {\tilde{q%
}}_{\perp }-iq_{E}\right) }  \label{one_more}
\end{equation}
where $q_{0}=\xi ^{-1}\left( 1+\ \frac{\left| \omega _{m}\right| }{\omega _{%
\mathrm{sf}}}\right) ^{1/2}$ and $q_{E}=\widetilde{\Sigma }_{m+m'}/v_{F}$.

We next assume, and then verify, that a typical ${\tilde{q}}_{\perp }\ll
q_{0}$. $q_{\perp }^{2}$ in the first term can then be ignored, and the
momentum integration yields 
\begin{eqnarray}
\int d{\tilde{q}}_{\perp }\frac{1}{\sqrt{{\tilde{q}}_{\perp }^{2}+q_{0}^{2}}}%
\frac{1}{\tilde{q}_{\perp }-iq_{E}} &\approx &\frac{1}{q_{0}}\int_{-\infty
}^{\infty }d{\tilde{q}}_{\perp }\frac{1}{\left( {\tilde{q}}_{\perp } -i
q_{E}\right) }  \nonumber \\
&=&i\pi \mathrm{sign} q_E  \nonumber \\
& =& i\pi \mathrm{sign} (\omega _{m}+\omega
_{m}^{\prime }).  \label{apprx}
\end{eqnarray}
On substituting this result into Eq.\ref{one_more} and splitting the
frequency integral into several pieces, depending on the sign of $\omega
_{m}+\omega _{m}^{\prime }$, we obtain 
\begin{equation}
\Sigma _{{\bf{k}}_{\mathrm{hs}}}\left( i\omega _{m}\right) =i\lambda
~\int_{-\omega _{m}}^{0}d\omega _{m}^{\prime }\frac{\mathrm{\ }1}{\sqrt{%
1+|\omega _{m}^{\prime }|/\omega _{\mathrm{sf}}}},
\end{equation}
where $\lambda $ is given by Eq.\ref{lambda}. The remaining frequency
integration is elementary and yields 
\begin{equation}
\Sigma _{{\bf{k}}}\left( i\omega _{m}\right) =i\lambda ~\frac{2\omega _{m}%
}{1+\sqrt{1+\frac{\left| \omega _{m}\right| }{\omega _{\mathrm{sf}}}}}.
\label{SEns}
\end{equation}

We now return to the approximation we made above, that a typical ${\tilde{q}}%
_{\perp }\ll q_{0}$. Since the typical values for ${\tilde{q}}_{\perp }$,
i.e. those which dominate the integral in Eq.\ref{apprx}, are of order $q_{E}
$, the approximation is valid if $q_{0}\gg q_{E}$. Further, a typical
internal frequency $\omega _{m}^{\prime }$ is of order $\omega _{m}$, hence $%
q_{E}\sim |\omega _{m}+\Sigma (\omega _{m})|/v_{F}$ and $q_{0}\sim (\omega
_{sf}\xi ^{2})^{-1}(|\omega _{m}|+\omega _{\mathrm{sf}})^{1/2}\sim {\bar{g}}%
^{1/2}(|\omega _{m}|+\omega _{\mathrm{sf}})^{1/2}/v_{F}$. At weak coupling, $%
\lambda \ll 1$, the condition $q_{0}\gg q_{E}$ implies that $\omega \ll {%
\bar{g}}/\lambda $, i.e., at energies smaller that ${\bar{g}}$, the
approximation is well justified. At strong coupling and $\omega \ll \omega _{%
\mathrm{sf}}$, $q_{E}\approx \lambda \omega /v_{F}$, $q_{0}\sim \lambda
\omega _{\mathrm{sf}}/v_{F}$, and the criterion $q_{0}\gg q_{E}$ is
 again satisfied. Finally, at strong coupling and $\omega >\omega _{sf}$, 
$q_{E}\sim q_{0}(\sim {\bar{g}}\omega _{m})^{1/2}/v_{F}$. In this limit, the
approximation we made is qualitatively but not quantitatively correct. In
order to develop a well controlled theoretical framework for this limit,
Abanov \emph{et al.}\cite{ac,ACF} developed a controllable $1/N$ approach by
extending the model to a large number of hot spots, $N$, in the Brillouin
zone ($N=8$ in the physical case). This allows an expansion in terms of $1/N$%
. Alternatively, one can extend the model to a large number of fermion
flavors, $M$, and expand in $1/M$. For large $N$, $q_{0}/q_{E}\sim N^{2}$,
i.e. the approximation $q_{0}\gg q_{E}$ is justified. Another appealing feature
of this $\frac{1}{N}$ expansion is that within it, vertex corrections and
the dependence of the self-energy of the momentum component transverse to
the Fermi surface are also small in $\frac{1}{N}$ and can be computed
systematically together with the corrections to the frequency dependent part 
 of the self energy \cite{Abanov01advphys}. 
To keep our discussion focused on the key results, we
will not discuss further the details of the $1/N$ approach. Rather, we just
emphasize that (i) Eq.\ref{SEns} is quantitatively correct even if $%
q_{0}\sim q_{E}$ and (ii) numerically the difference between approximate and
more involved ``exact'' results for the fermionic $\Sigma _{{\bf{k}}_{%
\mathrm{hs}}}(\omega )$ is only few percent~\cite{Abanov01advphys}.

We next analyze the functional form  of 
 Eq.\ref{SEns}. We see
that $\Sigma _{{\bf{k}}_{\mathrm{hs}}}(i\omega )$ scales with $\lambda $
and at strong coupling exceeds the bare $i\omega $ term in the inverse
fermion propagator $G_{{\bf{k}}_{\mathrm{hs}}}^{-1}(i\omega )=i\omega
-\Sigma _{{\bf{k}}_{\mathrm{hs}}}(i\omega ).$ As was the case with phonons,
it originates in the scattering on zero-sound vibrations of the electronic
subsystem, while spin-fluctuations mediate the interaction between electrons
and their zero-sound fluctuations. Further, $\Sigma _{{\bf{k}}_{\mathrm{hs}%
}}(i\omega )$ evolves at the same typical energy $\omega _{\mathrm{sf}}$ as
the bosonic self-energy. This interconnection between bosonic and fermionic
propagators is one of the key ``fingerprints'' of the spin-fermion model.

\begin{figure}[tbp]
\epsfxsize=2.8in 
\epsfysize=2.2in
\begin{center}
\epsffile{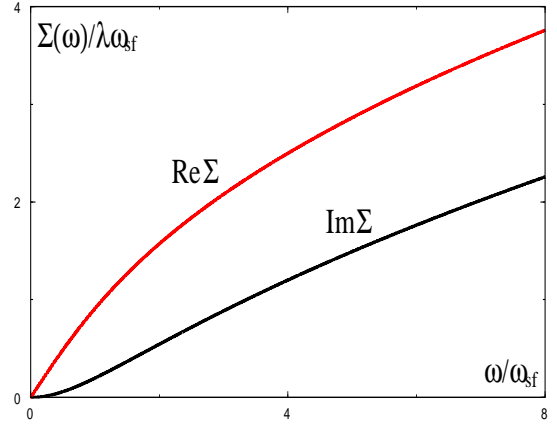}
\end{center}
\caption{The frequency dependence of the normal state electron self energy
due to spin fluctuation-fermion interaction for a quasi particle at a hot
spot (from Ref.~\protect\onlinecite{Abanov01advphys}). }
\label{selfenergyns}
\end{figure}
\begin{figure}[tbp]
\epsfxsize=2.8in 
\epsfysize=2.2in
\begin{center}
\epsffile{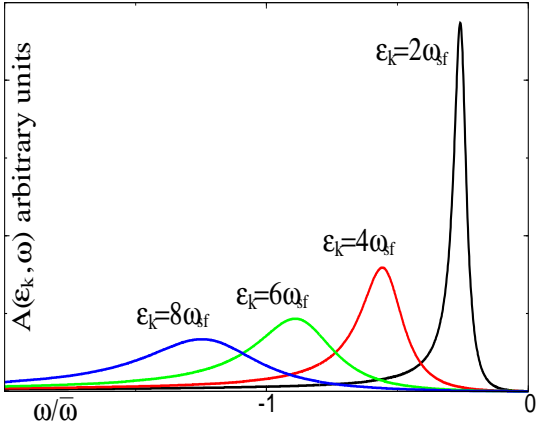}
\end{center}
\caption{The normal state spectral function. Note the absence of a
quasiparticle peak. This is the consequence of the proximity to an
antiferromagnetic quantum critical point. The figure is taken from Ref~ 
\protect\onlinecite{Abanov01advphys}. }
\label{Figure_NFL}
\end{figure}

In Fig.\ref{Figure_NFL} we show the behavior of the quasiparticle spectral
function $A(\epsilon _{{\bf{k}}},\omega )$ at various $\epsilon _{{\bf{k}}}$. 
We see that the spectral weight of the quasiparticle peak rapidly
decreases as $\epsilon _{{\bf{k}}}$ becomes larger than $\omega _{\mathrm{%
sf}}$.

For small frequencies, $\omega \ll \omega _{\mathrm{sf}}$, the spin
susceptibility can be approximated by its static form. For the fermionic
self-energy we find after analytical continuation 
\begin{equation}
\Sigma _{{\bf{{k}}}_{\mathrm{hs}}}\left( \omega \right) =\lambda \left(
\omega +\frac{i\omega \left| \omega \right| }{4\omega _{\mathrm{sf}}}\right)
; \ \ \ (\omega \ll \omega _{\mathrm{sf}}).  \label{si_lo}
\end{equation}
We see that the quasiparticle damping term, although quadratic in $\omega $
as it should be in a Fermi liquid, scales inversely with $\omega _{\mathrm{sf%
}}$, not with the Fermi energy as in conventional metals. As $\omega _{%
\mathrm{sf}}$ vanishes at the critical point, the width of the Fermi liquid
region, where damping is small compared to $\omega $, progressively shrinks
as $\xi $ increases. The quasiparticle renormalization factor $d\Sigma _{%
{\bf{k}}_{\mathrm{hs}}}(\omega )/d{\omega }|_{\omega =0}=\lambda \propto
\xi $ increases as the system approaches the magnetic quantum critical
point. The quasiparticle $z-$factor simultaneously decreases as 
\begin{equation}
z_{{\bf{k}}_{\mathrm{hs}}}=\frac{1}{1 + \left. \frac{\partial \Sigma _{%
{\bf{k}}_{\mathrm{hs}}}\left( \omega \right) }{\partial \omega }\right|
_{\omega =0}}=\frac{1}{1+\lambda }
\end{equation}
and vanishes at criticality.

At frequencies above $\omega _{\mathrm{sf}}$, the imaginary part of the
fermionic self-energy \ resembles a linear function of $\omega $ over a
substantial frequency range up to about $8\omega _{\mathrm{sf}}$, and then
eventually crosses over to 
\begin{equation}
\Sigma _{{\bf{k}}_{\mathrm{hs}}}\left( i\omega \right) =i\mathrm{sign}%
\omega ~\left( \overline{\omega }|\omega |\right) ^{1/2}  \label{sqrtNFL}
\end{equation}
where 
\begin{equation}
\overline{\omega }=4 \lambda ^{2}\omega _{\mathrm{sf}}=\frac{9}{16\pi }{\bar{g}%
}.
\end{equation}
Thus ${\overline{\omega }}$ remains finite as $\xi \rightarrow \infty $. At
the critical point $\omega _{\mathrm{sf}}\rightarrow 0$, the self energy
displays the non-Fermi-liquid behavior \ of Eq.\ref{sqrtNFL} down to $\omega
=0$. A plot of the fermionic self-energy is presented in Fig.\ref
{selfenergyns}. The intermediate quasi-linear regime is clearly visible.
Note also that the deviations from  Fermi liquid behavior starts already
at small $\omega \sim \omega _{\mathrm{sf}}/2$.

We now consider how well the Eliashberg approximation is satisfied, i.e.,
whether vertex corrections and the momentum dependent piece in the fermionic
self-energy are relatively small. To do this one needs to evaluate $\Sigma _{%
{\bf{k}}}(\omega =0)$ at ${\bf{k}}\neq {\bf{k}}_{\mathrm{hs}}$ and the
vertex correction $\delta g/g$. The details of the derivation can be found
in Ref.\onlinecite{Abanov01advphys}. We have 
\begin{equation}
\Sigma _{{\bf{k}}}\left( \omega =0\right) =\left[\frac{3}{4\pi }\log \lambda \right]
~\varepsilon _{{\bf{k+Q}}}
\end{equation}
and 
\begin{equation}
\frac{\delta g}{g}=\frac{Q(\phi _{0})}{8}\log \lambda ,
\end{equation}
where $Q(\phi _{0})=2\phi _{0}/\pi $. For $\phi _{0}\approx \pi /2$, $%
Q\approx 1$. We see that these two corrections depend only logarithmically
on the coupling and at large $\lambda $ are parametrically small compared to
the frequency dependence of \ the self-energy. Furthermore, at large $N$,
both $\delta g/g$ and $\Sigma _{{\bf{k}}}$ contain an extra factor of $1/N 
$, i.e., scale as $(1/N)\log \lambda $. We see therefore that 
$(1/N) \log \lambda $
is the analog of the second coupling constant ${\tilde{\lambda}}_{ep}$ for
the phonon case. Just as for phonons, the applicability of the Eliashberg
theory is related to the fact that this second coupling constant is much
smaller than the primary coupling $\lambda $. In our case, this requires
that $\lambda \gg (\log \lambda )/N$. We however emphasize that the
smallness of the two ``couplings'' is not the result of the smallness of the
velocity ratio but the consequence of the proximity to a critical point.
Also, in distinction to phonons, the second coupling still diverges at the
critical point and therefore corrections to Eliashberg theory cannot be
neglected close to the antiferromagnetic transition. These corrections,
however, have been analyzed within a renormalization group approach \ in
Ref. \onlinecite{ac2,Abanov01advphys} and found to be of minor relevance 
 at intermediate coupling  discussed here.

It is also instructive to explicitly compute the fermionic self-energy in the
same way as we did in Section 3 for  systems with electron-phonon interaction 
and verify that
the reason for the dominance of the frequency dependence of the self energy
is common to that for the phonon case. To see this, we introduce, by analogy
with Eq.~\ref{ph4} 
\begin{equation}
\Sigma ({\bf{k}},i\omega _{m})=(i\omega _{m}-\epsilon _{{\bf{k+Q}}})I(%
{\bf{k}},i\omega _{m}).  \label{ph4sf}
\end{equation}
In the present case, $I({\bf{k}},i\omega _{m})$ is singular as ${\bf{k}}%
\rightarrow {\bf{k}}_{\mathrm{hs}}$ and $\omega _{m}\rightarrow 0$, and $I(%
{\bf{k}}= {\bf{k}}_{\mathrm{hs}},i\omega _{m}\rightarrow 0)\neq
I({\bf k} \rightarrow {\bf{k}}_{\mathrm{hs}},0)$~\cite{Abanov01advphys}. Evaluating $I({\bf{%
k}}_{\mathrm{hs}},i\omega _{m})$ in the same way as in the phonon case, we
find at $\omega \ll \omega _{sf}$ 
\begin{equation}
I({\bf{k}}_{\mathrm{hs}},i\omega _{m})=I_{reg}+ \lambda~ 
\frac{i \omega _{m}}{%
i\omega _{m}-\epsilon _{{\bf k+Q}}},
\end{equation}
where $I_{reg}=O(\log \lambda )$. This form is the same as that found for
phonons (see Eq.~\ref{ph7}). The analogy implies that the dominant, $%
O(\lambda )$, contribution to the fermionic self-energy comes from
magnetically mediated interactions between fermions and their zero sound
excitations, whereas the actual spin-fermion scattering process in which
fermions at forced to vibrate at typical spin frequencies yields a smaller 
$ O(\log \lambda )$ contribution.

Finally, away from hot spots but still at the Fermi surface, the fermionic
self-energy is given by the same expression, Eq.~\ref{SEns}, as at a hot
spot, but with a momentum dependent coupling constant $\lambda _{{\bf{k}}}$
and energy scale $\omega _{\mathrm{sf}}({\bf{k}})$ which obey 
\begin{eqnarray}
\lambda _{{\bf{k}}} &=&\lambda /(1+\delta k\xi ),  \nonumber \\
~\omega _{\mathrm{sf}}({\bf{k}}) &=&\omega _{\mathrm{sf}}(1+\delta k\xi
)^{2},  \label{lao}
\end{eqnarray}
where $\delta k=\left| {\bf{k}}_{\mathrm{F}}{\bf{-k}}_{\mathrm{hs}%
}\right| $ is the momentum deviation from a hot spot along the Fermi
surface. We see that the effective coupling decreases upon deviation from a
hot spot, while the upper energy scale for the Fermi liquid behavior
increases. The increase of $\omega _{\mathrm{sf}}$, however, is
counterbalanced by the fact that $\omega _{\mathrm{sf}}\propto \sin \phi _{0}
$, and $\phi _{0}$, which we had set to be $\approx \pi /2$, increases away
from a hot spot~\cite{rob2}.

We see from Eq.\ref{lao} that the width of the region where $\Sigma _{%
{\bf{k}}}(\omega )$ is independent of ${\bf{k}}$ (i.e., the ``size'' of
a hot spot) depends on frequency. At the lowest frequencies, $\omega <\omega
_{\mathrm{sf}}({\bf{k}})$, $\Sigma _{{\bf{k}}}(\omega )=\lambda _{%
{\bf{k}}}\omega $, and the hot spot physics is confined to a region of
width $\xi ^{-1}$ which progressively shrinks as $\xi $ increases. However,
at frequencies above $\omega _{\mathrm{sf}}({\bf{k}})$,
 $\Sigma _{{\bf{k}}}(\omega )\sim \lambda _{{\bf{k}}}(\omega 
\omega _{\mathrm{sf}}({\bf{k}})^{1/2}=\lambda (\omega \omega _{\mathrm{sf}})^{1/2}$ is independent of $%
{\bf{k}}$. Accordingly, physical processes that happen on these scales are
isotropic (apart from the dependence on $\phi _{0}$). In this sense 
 the whole Fermi surface acts as one big ``hot spot''.

One can easily perform the above analysis in dimensions larger than $d=2$.
One finds that the quasiparticle spectral weight behaves \ for large $%
\lambda $ and the lowest energies $\omega <\omega _{\mathrm{sf}}$ as 
\begin{equation}
z_{{\bf{k}}_{\mathrm{hs}}}\propto \lambda ^{d-3}
\end{equation}
for $2<d<3$, and vanishes logarithmically for $d=3$. Correspondingly, in the
quantum critical regime we find for the self energy along ''hot lines'' 
\begin{equation}
\Sigma _{{\bf{k}}_{\mathrm{hs}}}\left( i\omega \right) \propto i\omega
\left| \omega \right| ^{\frac{d-3}{2}},
\end{equation}
an expression that transforms into $\Sigma _{{\bf{k}}_{\mathrm{hs}}}\left(
i\omega \right) \propto i\omega \log \left( \left| \omega \right| \right) $
for $d=3$. Above $d=3$ no non-Fermi liquid effects result from the proximity
to the quantum critical point. This demonstrates that many of the effects
caused by the incoherent nature of the fermions are peculiar to 2d and are
considerably less pronounced in three dimensional systems.

\subsection{The $d_{x^{2}-y^{2}}$ pairing instability temperature}

\begin{figure}[tbp]
\epsfxsize=2.8in 
\epsfysize=0.7in
\begin{center}
\epsffile{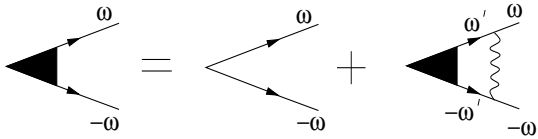}
\end{center}
\caption{Diagrammatic representation for the pairing vertex\protect\cite{ACF}%
. The solid and wavy lines are fermionic and spin fluctuation propagators,
respectively. }
\label{pairing}
\end{figure}

We now consider the development of the pairing instability in the
spin-fermion model. We follow Ref.~\onlinecite{ACF}. It is customary \ in an
analysis of the pairing problem to introduce an infinitesimally small
particle-particle vertex $\Phi _{{\bf{k,-k}}}^{(0)}(\omega ,-\omega
)\equiv \Phi _{{\bf{k}}}^{(0)}(\omega )$ and study its renormalization by
the pairing interaction. The corresponding diagrams are presented in Fig.~\ref{pairing}.
 The temperature at which the renormalized vertex diverges, i.e.,
when the equation for the full $\Phi _{{\bf{k}}}(\omega )$ has a
nontrivial solution at vanishing $\Phi _{{\bf{k}}}^{(0)}(\omega )$, marks
the onset of pairing. 

As noted above, the spin-mediated pairing interaction gives rise to $%
d_{x^{2}-y^{2}}$ superconductivity. We argued above that near the magnetic
instability, the gap is maximum near the hot spots. One can check (see \onlinecite{ACF}) that the pairing problem is confined to the hot regions
in the sense that the momentum integration never extends to $|{\bf{k}}-%
{\bf{k}}_{\mathrm{hs}}|$ where the momentum dependence of the self-energy
or infinitesimal pairing vertex becomes relevant. We can then assume that
the pairing vertex is flat near the hot spots. The underlying $d-$wave
symmetry then implies that the gap changes sign between two hot regions
separated by the antiferromagnetic ${\bf{Q}}$. A separation into hot and
cold regions is indeed valid only if typical 
$|{\bf{k}}-{\bf{k}}_{\mathrm{hs}%
}|\leq k_{F}$. We will see below 
that the  $d-$wave pairing problem is confined to
frequencies of order ${\bar{g}}$. For these frequencies, the width of the
hot region is \textit{finite} and is constrained only by the requirement
that $\omega _{\mathrm{sf}}({\bf{k}})<{\bar{g}}$,
 a condition that implies (after a more accurate account of
the overall factors) that the effective coupling is smaller than the
fermionic bandwidth. As discussed in the Introduction, the latter is \ a
necessary condition for the separation between high and low energies, 
and we assume it to hold. We comment below on what happens if 
typical $|{\bf{k}}-{\bf{k}}_{\mathrm{hs}%
}| > k_{F}$, i.e., hot and cold regions cannot be separated.

The value of the transition temperature depends sensitively on the behavior
of fermions that are paired by the spin-mediated interaction. Our analysis
of the normal state has shown that the character of the fermionic degrees of
freedom changes at energies of order $\omega _{\mathrm{sf}}$. For energies
smaller than $\omega _{\mathrm{sf}}$, fermions display Fermi liquid
behavior, while at higher energies they display behavior that
 is different from that in a Fermi liquid. In the BCS theory of
superconductivity only Fermi liquid degrees of freedom contribute to the
pairing. Let us suppose that this also holds in the present case. Then the
pairing problem would be qualitatively similar to that of BCS, since for
frequencies smaller than $\omega _{\mathrm{sf}}$, the spin susceptibility
that mediates pairing can be approximated by its static value. The
linearized equation for the pairing vertex then has the form 
\begin{equation}
\Phi \left( \omega \right) =\frac{\lambda }{1+\lambda }~\int_{T_{c}^{\mathrm{%
FL}}}^{\omega _{\rm sf}}d\omega ^{\prime }\frac{\Phi \left( \omega ^{\prime
}\right) }{\omega ^{\prime }} , \label{gaplin01}
\end{equation}
where the $1+\lambda $ factor in the denominator is the result of  mass
renormalization in the Fermi liquid regime ($\Sigma (\omega )\approx \lambda
\omega $). The solution of this equation~\cite{McMillan} yields 
\begin{equation}
T_{\mathrm{c}}^{\mathrm{FL}}\sim \omega _{\mathrm{sf}}\exp \left( -\frac{%
1+\lambda }{\lambda }\right) .  \label{mcm}
\end{equation}
At weak coupling, this is just the BCS result. At strong coupling, the mass
renormalization compensates the coupling constant, and $T_{\mathrm{c}}^{%
\mathrm{FL}}$ saturates at $T_{\mathrm{c}}^{\mathrm{FL}}\sim \omega _{%
\mathrm{sf}}$. This result, if correct, would imply that the pairing
fluctuations become progressively less relevant as one approaches the
quantum critical point $\xi ^{-1}\rightarrow 0$ (see the left panel in Fig.%
\ref{fig_tc}). At a first glance, this is what happens,because pairing of
non Fermi-liquid degrees of freedom seems hard to accomplish. Indeed, at
frequencies larger than $\omega _{\mathrm{sf}}$, the pairing interaction
decreases as $(1+|\omega |/\omega _{\mathrm{sf}})^{-1/2}$, and this
apparently makes the frequency integral ultraviolet convergent, i.e., the
``logarithmic'' pairing problem does not appear to extend above $\omega _{%
\mathrm{sf}}$. The flaw in this argument is that when the interaction
decreases, the mass renormalization produced by the \textit{same}
interaction also decreases, and the large overall $\lambda $ is no longer
compensated by $1+\Sigma (\omega )/\omega $. Indeed, for $\omega \gg \omega
_{\mathrm{sf}}$, $\Sigma (\omega )=({\bar{\omega}}\omega )^{1/2}$, where, we
recall, ${\bar{\omega}}=(9/16\pi ){\bar{g}}$, and the mass renormalization
is $1+(\bar{\omega}/\omega )^{1/2}=1+2\lambda (\omega _{\mathrm{sf}}/\omega
)^{1/2}\ll \lambda $. Furthermore, we see that at frequencies between $%
\omega _{\mathrm{sf}}$ and ${\bar{\omega}}$, the effective mass and
effective interaction both scale as $(\omega )^{-1/2}$. The product of the
two then scales as $1/\omega $, i.e. the ``logarithmic'' pairing problem
extends to frequencies of the order of ${\bar{\omega}}$ which, we recall,
remains finite at $\xi =\infty $.

\begin{figure}[tbp]
\epsfxsize=\columnwidth 
\begin{center}
\epsffile{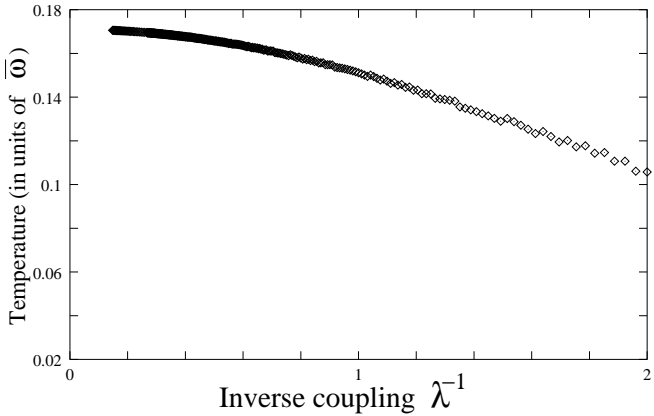}
\end{center}
\caption{The results for the instability temperature $T_{\mathrm{cr}}$
obtained from the solution of the linearized Eliashberg equations for
different values of the coupling constant $\protect\lambda $. The figure is
taken from Ref.~\protect\onlinecite{ACF}.}
\label{Tcrvslambda}
\end{figure}

By itself, this effect does not guarantee that the pairing instability
temperature is of order ${\bar \omega}$ 
 as the pairing interaction depends on
the transferred frequency $\omega -\omega ^{\prime }$, and  the
linearized equation for the pairing vertex becomes an integral equation in
frequency. In particular, for $\xi =\infty $, and hence $\omega _{\mathrm{sf}%
}=0$, we need to solve 
\begin{equation}
\Phi \left( \omega \right) =\frac{1}{4}\int_{T_{\mathrm{cr}}}^{\bar{\omega}%
}d\omega ^{\prime }\Phi \left( \omega ^{\prime }\right) \frac{1}{\sqrt{%
\omega ^{\prime }}}\frac{1}{\sqrt{\left| \omega -\omega ^{\prime }\right| }}.
\label{eq_ga}
\end{equation}
Observe that this equation does not have any adjustable parameter and is
therefore fully universal when $T$ is expressed in units of ${\bar{\omega}}$%
. Eq.~\ref{eq_ga} has been analyzed in detail by Finkel'stein, Abanov and
one of us\cite{ACF}. They found that it does have a nontrivial solution at 
\begin{equation}
T_{\mathrm{cr}}\sim 0.2{\bar{\omega}}.  \label{tcqc}
\end{equation}
They also analyzed the pairing problem at a finite $\lambda $ and found that
incoherent fermions dominate pairing down to  a surprisingly small $\lambda
\sim 0.5$. The McMillan like formula, Eq.\ref{mcm}, becomes valid only at
smaller $\lambda $. A  plot of $T_{\mathrm{cr}}$ versus $\lambda $ is
presented in Fig.\ref{Tcrvslambda}. 

Several comments are in order here. First, at these values of the coupling
 constant (and doping) $T_{\mathrm{cr}}$
does not coincide with the onset temperature for superconductivity,  $T_{c}$,
 but rather represents the onset
temperature for  pseudogap behavior; the actual $%
T_{c}$ is lower, as discussed below. Second, we have neglected
 fluctuation effects due to the quasi-two dimensionality. The latter
are expected to yield Kosterlitz-Thouless physics~\cite{KostTh}. Third, Eq.%
\ref{tcqc}, is only valid when ${\bar{g}}<W$ where we recall $W\sim v_{F}$
is the fermionic bandwidth. In the opposite limit ${\bar{g}}>W$, lattice
effects are important and controlled analytical calculations are difficult
to perform. One can, however, easily estimate that in this limit $T_{\mathrm{%
cr}}\propto v_{F}^{2}/{\bar{g}}\propto \omega _{sf}\xi ^{2}$. This estimate
coincides with the result of Monthoux and Pines who extracted 
$T_{\mathrm{cr}} \propto \omega_{sf} \xi^2 \sim v^2_F/{\bar g}$ 
 from their numerical analysis\cite
{Monthoux94}. Since $v_{F}$ is proportional to the hopping matrix 
element, and ${\bar g}$ scales with the Hubbard $U$ at large $\xi$,
 it follows  that $T_{\mathrm{cr}}\propto J$, where $J$ is the
magnetic exchange integral of the corresponding Heisenberg model which
describes antiferromagnetism at half-filling.

Eq.~\ref{tcqc} demonstrates that the $d$-wave
 pairing instability temperature of a two dimensional system
 at an antiferromagnetic quantum critical point is finite.
 It is interesting to note that the same holds for $p$-wave pairing at a
 ferromagnetic quantum critical point as shown by Roussev and
 Millis\cite{Andy_triplet}.

\subsection{Superconducting state}

We next extend the Eliashberg theory to the spin- fluctuation induced
superconducting state. The discussion in this section follows Ref. \cite
{Abanov01epl}. We derive a generalized set of Eliashberg equations for the
fermionic self-energy and the gap function that include an additional
coupled equation for the spin polarization operator. The latter, as
discussed in the Introduction, is produced by low-energy fermions and has to
be determined self consistently.

As discussed above, the infinitesimal pairing vertex and the fermionic
self-energy in the normal state depend weakly on momentum in the hot region $%
|{\bf k}-{\bf k}_{hs}|\leq {\bar{g}}/W$ where $W\propto v_{F}$ is the fermionic
bandwidth. We will see that in the superconducting state, the momentum
integration is also confined to hot regions. We can then safely neglect the
weak momentum dependence of both $\Sigma (i\omega )$ and $\Phi (i\omega )$, as
we did above in calculating  $T_{\mathrm{cr}}$. Subtle effects due to
 this weak momentum dependence will be considered in the next section.
We will not attempt to discuss the behavior of the gap near the nodes. The
latter is central for the interpretation of the experimental data at the
lowest temperatures and frequencies, but not at energies comparable or
larger than the maximum pairing gap. From our perspective it is not
essential for the spin fluctuation induced pairing state.

\subsubsection{ Generalized Eliashberg equations}

The derivation of \ the Eliashberg equations is straightforward. In the
superconducting state, the normal and anomalous fermionic Green's functions $%
G_{{\bf{k}}}\left( i\omega _{n}\right) $ and $F_{{\bf{k}}}\left(i \omega
_{n}\right) $ and the dynamical spin susceptibility are given by Eqs.\ref
{GGorkov} -  \ref{chitot}. It is convenient to rewrite $G_{%
{\bf{k}}}\left( i\omega _{n}\right) $ and $F_{{\bf{k}}}\left(i \omega
_{n}\right) $ as 
\begin{eqnarray}
G_{{\bf{k}}}\left(i \omega _{n}\right) &=&-\frac{\epsilon _{{\bf{k}}}+i%
\widetilde{\Sigma }_{n}}{\epsilon _{{\bf{k}}}^{2}+\widetilde{\Sigma }%
_{n}^{2}+\Phi ^{2}(i\omega _{n})},  \nonumber \\
F_{{\bf{k}}}\left(i \omega _{n}\right) &=&i\frac{\Phi (i\omega _{n})}{%
\epsilon _{{\bf{k}}}^{2}+\widetilde{\Sigma }_{n}^{2}+\Phi ^{2}(i\omega _{n})%
}  \nonumber \\
F_{{\bf{k}}+{\bf{Q}}}\left( i\omega _{n}\right) &=&-i\frac{\Phi (i\omega
_{n})}{\epsilon _{{\bf{k+Q}}}^{2}+\widetilde{\Sigma }_{n}^{2}+\Phi
^{2}(i\omega _{n})}
\end{eqnarray}
where 
$i\widetilde{\Sigma }_{n}=i\omega _{n} 
+\Sigma _{{\bf k}_{\rm hs}}\left( i\omega _{n}\right) $
 (in real frequencies,  $\widetilde{\Sigma } (\omega) =\omega  +
 \Sigma _{{\bf k}_{\rm hs}} (\omega)$).  Without losing
generality we can set $\epsilon _{{\bf{k}}}=v_{x}{\tilde{k}}_{x}+v_{y}{%
\tilde{k}}_{y}$ and $\epsilon _{{\bf{k+Q}}}=-v_{x}{\tilde{k}}_{x}+v_{y}{%
\tilde{k}}_{y}$ where ${\tilde{k}}=k-k_{hs}$. The sign change between 
$F_{{\bf{k}}}$ and $F_{{\bf{k}}+{\bf{Q}}}$ is the result of $%
d_{x^{2}-y^{2}}$ symmetry. The spin susceptibility, we recall, is given by 
\begin{equation}
\chi_{\bf{q}}\left(i\omega_m\right) =\frac{\alpha \xi ^{2}}{1+\xi ^{2}\left( 
{\bf{q-Q}}\right) ^{2}-\Pi _{{\bf{Q}}}\left(i \omega_m \right) }.
\label{chitot_1}
\end{equation}
 
We need to obtain the equations for the fermionic self-energy $\Sigma _{%
{\bf{k}}_{\mathrm{hs}}}(i\omega _{m})$, the anomalous vertex $\Phi _{%
{\bf{k}}_{\mathrm{hs}}}(i\omega _{m})$, and the spin polarization operator $%
\Pi _{{\bf{Q}}}(i\omega _{m})$.
The spin polarization operator is obtained in the same way as in the normal
state, but now there are two particle-hole bubbles: one is the convolution
of $G_{{\bf{k}}}G_{{\bf{k+Q}}}$ and the other is the convolution
 of $F_{{\bf{k}}}F_{{\bf{k+Q}}}$. We have 
\begin{eqnarray}
\Pi _{{\bf{Q}}}\left( i\omega _{n}\right) &=&-8{\bar{g}}\xi^2 T\sum_{m}\int \frac{%
d^{2}k}{\left( 2\pi \right) ^{2}}\left[G_{{\bf{k}}+{\bf{Q}}}\left(i \omega _{n+m}\right)  \right.   \\
&&\times\left.G_{{\bf{k}}}\left( i\omega
_{m}\right) -  F_{{\bf{k}}+{\bf{Q}}%
}\left(i \omega _{n+m}\right)F_{{\bf{k}}}\left(i \omega _{m}\right) \right] \nonumber
\end{eqnarray}
(the negative sign between the two terms originates in the summation over
the spin components). The momentum integration can be performed explicitly
and yields 
\begin{eqnarray}
\Pi _{{\bf{Q}}}\left(i \omega _{n}\right) &=&
-\frac{4{\bar{g}} \xi^2 }{v_{F}^{2}}
T\sum_{m}\left[ 1-g\left(i \omega _{m}\right) 
g\left( i\omega _{n+m}\right)\right. 
\nonumber \\
& & \left. -f\left( i\omega _{m}\right) f\left(i \omega _{n+m}\right) \right]  \label{El1}
\end{eqnarray}
where 
\begin{equation}
g\left( i\omega _{m}\right) =~\frac{\widetilde{\Sigma }_{m}}{\sqrt{\widetilde{%
\Sigma }_{m}^{2}+\Phi ^{2}(i\omega _{m})}}
\end{equation}
and 
\begin{equation}
f\left(i \omega _{m}\right) =\frac{\Phi (i\omega _{m})}{\sqrt{\tilde{\Sigma}%
_{m}^{2}+\Phi ^{2}(i\omega _{m})}}
\end{equation}
The first term in Eq.\ref{El1} is the result of the regularization of the
ultraviolet singularity. The additional sign change between $gg$ and $ff$
terms in Eq.\ref{El1} is due to the $d-$wave form of $F_{{\bf{k}}}$.

Eq.\ref{El1} takes into account the change of the low energy spin dynamics
in the pairing state. In the case of electron-phonon interaction a
corresponding change of the phonon dynamics exists as well, causing a shift
of the phonon frequency and line width below T$_{c}$. While this effect is
only a minor correction to the phononic dynamics and is often neglected~\cite
{mahan}, it leads to a dramatic change of the spin dynamics in our case.

The other two equations are formally the same as for phonon-mediated
superconductors. The fermionic self-energy $\Sigma (\omega _{n})$ is given
by 
\begin{equation}
\Sigma \left(i \omega _{n}\right) =3g^{2}T\sum_{m}\int \frac{d^{2}q}{\left(
2\pi \right) ^{2}}\chi \left( {\bf{q}},i\omega _{m}\right) G_{{\bf{k}_{\rm hs}}+%
{\bf{q}}}\left( \omega _{n+m}\right)
\end{equation}
Performing the momentum integration along the same lines as in the normal
state calculations we find 
\begin{equation}
\Sigma \left( i\omega _{n}\right) =\frac{3 g^2}{2 v_F }T\sum_{m}D\left(
i\omega _{m}\right) g\left(i \omega _{n}+i\omega _{m}\right)  \label{Elsi}
\end{equation}
where $D$ is the effective bosonic propagator that is obtained by
integrating the dynamical spin susceptibility over the momentum component
along the Fermi surface and setting other momentum components to ${\bf{Q}}$
(the last step is equivalent to the approximation we discussed below Eq.\ref
{apprx}). We have 
\begin{eqnarray}
D\left( i\omega _{m}\right) &=&\left. \int \frac{dq_{\parallel }}{2\pi }\chi
\left( {\bf{q}},i\omega _{m}\right) \right| _{q_{\perp }=Q}  \nonumber \\
&=&\frac{\alpha \xi }{2 \sqrt{1-\Pi _{{\bf{Q}}}\left( i\omega _{m}\right) }}
\end{eqnarray}
An analogous equation is obtained for the anomalous vertex 
\begin{equation}
\Phi (i\omega _{n})=\frac{3 g^2 }{2 v_F }T\sum_{m}D\left( i\omega _{m}\right)
f\left(i \omega _{n}+i\omega _{m}\right) .  \label{Elph}
\end{equation}
Eqs.~\ref{El1}, \ref{Elsi}, and \ref{Elph} constitute the full set of
Eliashberg equations for the spin-mediated superconductivity. Alternatively
to $\Phi (\omega )$ and $\Sigma (\omega )$ we can also introduce 
\begin{equation}
Z(\omega )=1+\frac{\Sigma (\omega )}{\omega }~~\Delta (\omega )=\frac{\Phi
(\omega )}{Z(\omega )}
\end{equation}
The complex function $\Delta (\omega )$ reduces to the superconducting gap $%
\Delta $ in  BCS theory where we also have $Z(\omega )=1$. In Eliashberg
theory, the superconducting gap, defined as a frequency where the density of
states has a peak, is the solution of $\Delta (\omega =\Delta )=\omega $.

We again emphasize that the Eliashberg equations are valid for fermionic momenta
which deviate from hot spots by  less than ${\bar g}/v_F$. For these momenta,
the pairing vertex can be approximated by a ${\bf{k}}$-independent
function which however changes sign between two hot regions separated by $%
{\bf{Q}}$. For larger deviations, the anomalous vertex rapidly goes down
and eventually vanishes along zone diagonals.

\subsubsection{Solution of the Eliashberg equations:}

We discuss the general structure of the solutions of the set of Eliashberg
equations, and then present the results of their numerical solution. First,
we see from Eq.~\ref{El1} that, as in the normal state, $\Pi (\omega =0)=0$
for any $\Sigma (\omega )$ and $\Phi (\omega )$. This physically implies
that the development of the gap does not change the magnetic correlation
length. This result becomes evident if one notices that $d$-wave pairing
involves fermions from opposite sub-lattices. Second, the opening of the
superconducting gap changes the low frequency spin dynamics. Now
quasiparticles near hot spots are gapped, and a spin fluctuation can decay
into a particle-hole pair only when it can pull two particles out of the
condensate of Cooper pairs. This implies that the decay into particle-hole
excitations is only possible if the external frequency is larger than $%
2\Delta $. At smaller frequencies, we should have $\Pi ^{\prime \prime
}(\omega )=0$ at $T=0$~\cite{Ding,ac}. This result indeed readily follows
from Eq.~\ref{El1}. The Kramers-Kronig relation $\Pi ^{\prime }(\omega
)=(2/\pi )\int_{0}^{\infty }\Pi ^{\prime \prime }(x)/(x^{2}-\omega ^{2})$ \
then implies that because of a drop in $\Pi ^{\prime \prime }(\omega )$, the
spin polarization operator in a superconductor acquires a real part, which
at low $\omega $ is quadratic in frequency and is of order $\omega
^{2}/(\Delta \omega _{\rm sf})$. Substituting this result into Eq.\ref{chitot},
we find that at low energies, spin excitations in a $d$-wave superconductor
are propagating, gapped magnon-like quasiparticles 
\begin{equation}
\chi ({\bf{q}},\omega )\propto \frac{\Delta _{s}^{2}}{\Delta
_{s}^{2}+c_{s}^{2}({\bf{q}}-{\bf{Q}})^{2}-\omega ^{2}}.  \label{chil}
\end{equation}
where 
\begin{equation}
\Delta _{s}\sim (\Delta \omega _{\mathrm{sf}})^{1/2}
\end{equation}
and $c_{s}^{2}\sim v_{\mathrm{F}}^{2}\Delta /{\bar{g}}$. The re-emergence of
 propagating spin dynamics implies that the dynamical spin susceptibility
acquires a resonance peak which at ${\bf{q=Q}}$ is located at $\omega
=\Delta _{s}$.

Eq.\ref{chil} is indeed meaningful only if $\Delta _{s}\leq \Delta $, i.e., $%
\omega _{\mathrm{sf}}\leq \Delta $. Otherwise the use of the quadratic form
for $\Pi (\omega )$ is not justified. To find out how $\Delta $ depends on the
coupling constant, one needs to carefully analyze the full set of 
Eqs.\ref{El1}-\ref{Elph}. 
This analysis is rather involved~\cite{ACF,Abanov01epl}, and is
not directly related to the goal of this Chapter. We skip the details and 
quote the result. It turns out that at strong coupling, $\lambda \geq 1$%
, i.e. for optimally and underdoped cuprates, the condition $\Delta >\omega
_{\mathrm{sf}}$ is satisfied:  the gap scales with ${\bar{\omega}}$ and
saturates at $\Delta \approx 0.35{\bar{\omega}}=0.06{\bar{g}}$ at $\lambda
\rightarrow \infty $, while $\omega _{\rm sf}\propto \lambda ^{-2}\rightarrow 0$%
. In this situation, the spin excitations in a superconductor are
propagating, particle-like modes with a gap $\Delta _{s}$. However, in
distinction to phonons, these propagating magnons get their identity from a
strong coupling feedback effect in the superconducting state.

\begin{figure}[tbp]
\epsfxsize=\columnwidth 
\begin{center}
\epsffile{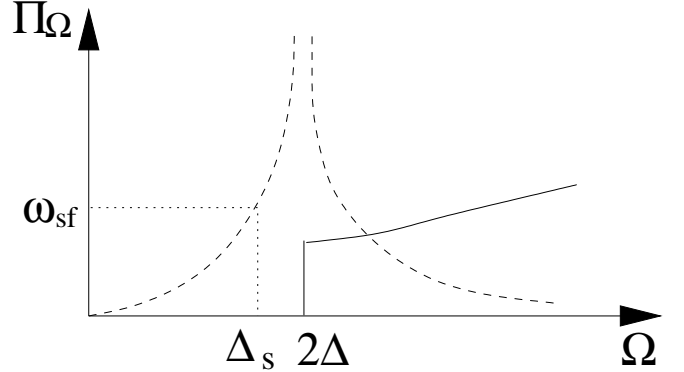}
\end{center}
\caption{Schematic behavior of the real (dashed line) and imaginary (solid
line) parts of the particle hole bubble in the superconducting state. Due to
the discontinuous behavior of $\Pi ^{\prime \prime }\left( \protect\omega
\right) $ at $\protect\omega =2\Delta $, the real part $\Pi ^{\prime }(%
\protect\omega )$ is logarithmically divergent at $2\Delta $. For small $%
\protect\omega $, the real part behaves like $\protect\omega ^{2}/\Delta $.
The figure is taken from ~\protect\onlinecite{acs_finger}.}
\label{Figure3}
\end{figure}

At weak coupling, the superconducting problem is of the  BCS type, and $\Delta \
\ll \omega _{\mathrm{sf}}$. This result is intuitively obvious as $\omega _{%
\mathrm{sf}}$ plays the role of the Debye frequency in the sense that the
bosonic mode that mediates pairing decreases at frequencies above $\omega _{%
\mathrm{sf}}$. For $\Delta \ll \omega _{sf}$, $\chi ({\bf{Q}},\omega )$
does not have a pole at frequencies where $\Pi (\omega )\propto \omega ^{2}$%
. Still, a pole in $\chi ^{\prime \prime }({\bf{Q}},\omega )$ does exist even at
weak coupling~\cite{Ding,Norman_1,weakcoul_neutr,oleg,ac,lee_brink}. To see
this, note that at $\omega \approx 2\Delta $ one can simultaneously set both
fermionic frequencies in the bubble to be close to $\Delta $, and make both
propagators singular due to the vanishing of $\sqrt{{\tilde{\Sigma}}%
^{2}-\Phi ^{2}}$ where, we recall, ${\tilde{\Sigma}}=\omega +\Sigma (\omega
) $. Substituting ${\tilde{\Sigma}}^{2}(\omega )-\Phi ^{2}(\omega )\propto
\omega -\Delta $ into Eq.~\ref{El1} and using the spectral representation, we
obtain for $\omega =2\Delta +\epsilon $ 
\begin{equation}
\Pi ^{\prime \prime }(\omega )\propto \int_{0}^{\epsilon }\frac{dx}{%
(x(\epsilon -x))^{1/2}}.  \label{ya}
\end{equation}
Evaluating the integral, we find that $\Pi ^{\prime \prime }$ undergoes a
finite jump at $\omega =2\Delta $. By the Kramers-Kronig relation, this jump
gives rise to a logarithmic singularity in $\Pi ^{\prime }(\omega )$ at $%
\omega =2\Delta $: 
\begin{equation}
\Pi ^{\prime }(\omega )=\frac{2}{\pi }~\int_{2\Delta }^{\infty }dx\frac{\Pi
^{\prime \prime }(x)}{x^{2}-\omega ^{2}}\propto \Delta \log \frac{2\Delta }{%
|\omega -2\Delta |}.
\end{equation}
The behavior of $\Pi ^{\prime }(\omega )$ and $\Pi ^{\prime \prime }(\omega
) $ is schematically shown in Fig.\ref{Figure3}. The fact that $\Pi ^{\prime
}(\omega )$ diverges logarithmically at $2\Delta $ implies that no matter
how small $\Delta /\omega _{\rm sf}$ is, $\chi ({\bf{Q}},\omega )$ has a pole
at $\Delta _{s}<2\Delta $, when $\Pi ^{\prime \prime }(\omega )$ is still
zero. Simple estimates show that for weak coupling, where $\omega _{\mathrm{%
sf}}\gg \Delta $, the singularity occurs at $\Delta _{s}=2\Delta (1-Z_{s})$
where $Z_{s}\propto e^{-\omega _{\rm sf}/(2\Delta )}$ is also the spectral
weight of the resonance peak in this limit. 

We see therefore that the resonance in the spin susceptibility exists both
at weak and at strong coupling. At strong coupling, the resonance frequency
is $\Delta _{s}\sim \Delta /\lambda \ll \Delta$, i.e., the resonance occurs
in the frequency range where spin excitations behave as propagating
magnon-like excitations. At weak coupling, the resonance occurs very near $%
2\Delta $ due to the logarithmic singularity in $\Pi ^{\prime }(\omega )$.
In practice, however, the resonance at weak coupling can hardly be observed
because the residue of the peak in the spin susceptibility $Z_{s}$ is
exponentially small.

\begin{figure}[tbp]
\epsfxsize=2.8in 
\epsfysize=2.2in
\begin{center}
\epsffile{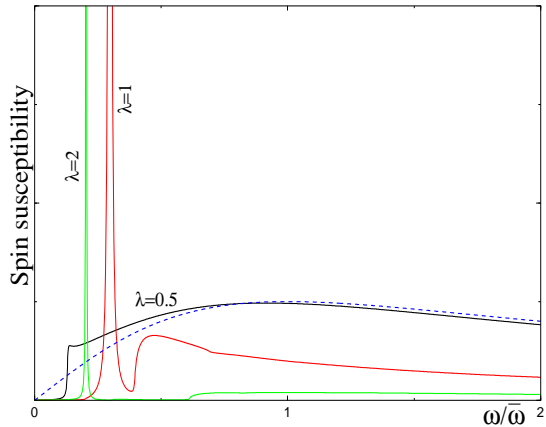}
\end{center}
\caption{Imaginary part of the dynamical spin susceptibility in the
superconducting state at $T\ll T_{c}$ obtained from the solution of the set
of three Eliashberg equations for coupling constants $\protect\lambda =0.5$, 
$\protect\lambda =1$, and $\protect\lambda =2$. The figure is taken from ~ 
\protect\onlinecite{acs_finger}.}
\label{Figure4}
\end{figure}

Fig.\ref{Figure4} shows the results for $\chi ({\bf{Q}},\omega )$ obtained
from the full solution of the set of three coupled equations at $T\approx 0$
\ and three different coupling constants\cite{Abanov01epl}. For $\lambda
\geq 1$, the spin susceptibility has a sharp peak at $\omega =\Delta _{s}$.
The peak gets sharper when it moves away from $2\Delta $. At the same time,
for $\lambda =0.5$, corresponding to weak coupling, the peak is very weak
and is washed out by a small thermal damping. In this case, $\chi ^{\prime
\prime }$ only displays a discontinuity at $2\Delta $.

We next show that the resonance peak does not exist for $s-$wave
superconductors\cite{oleg_1}. In the latter case, the spin polarization
operator is given by almost the same expression as in Eq.~\ref{El1}, but
with a different sign of the $ff$-term; recall that the original sign in Eq.(%
\ref{El1}) originated from the fact that the two fermions in the spin
polarization bubble differ in momentum by ${\bf{Q}}$, and the $d$-wave gap
changes sign under ${\bf{k}}\rightarrow {\bf{k}}+{\bf{Q}}$. One can
immediately check that for a different sign of the anomalous term, $\Pi
^{\prime \prime }$ is continuous at $2\Delta $. Accordingly, $\Pi ^{\prime
}(\omega )$ does not diverge at $2\Delta $, and hence there is no resonance
at weak coupling. Still, however, one could expect the resonance at strong
coupling as at small frequencies $\Pi ^{\prime }(\omega )$ is quadratic in $%
\omega $ by virtue of the existence of the threshold for $\Pi ^{\prime
\prime }$. It turns out, however, that for $s$-wave pairing the resonance is
precluded by the fact that $\Pi (\omega =0)$ changes between the normal and
the superconducting states. 

One can make sure that in an $s-$wave
 superconductor, $\Pi (\omega =0)<0$,
and that this negative term overshadows the positive $\omega ^{2}$ term in $%
\Pi (\omega )$ in such a way that for all frequencies below $2\Delta $, $\Pi
(\omega )<0$ and hence the resonance simply does not exist. That $\Pi
(\omega =0)<0$ in $s-$wave superconductors can be easily explained: a
negative $\Pi (0)$ implies that the spin correlation length decreases as the
system becomes superconducting. This is exactly what one should expect as $s$%
-wave pairing involves fermions both from different sub-lattices as well as
from the same sub-lattice. The pairing of fermions from the same sub-lattice
into a spin-singlet state obviously reduces the antiferromagnetic
correlation length.
\begin{figure}[tbp]
\epsfxsize=2.8in 
\epsfysize=3.8in
\begin{center}
\epsffile{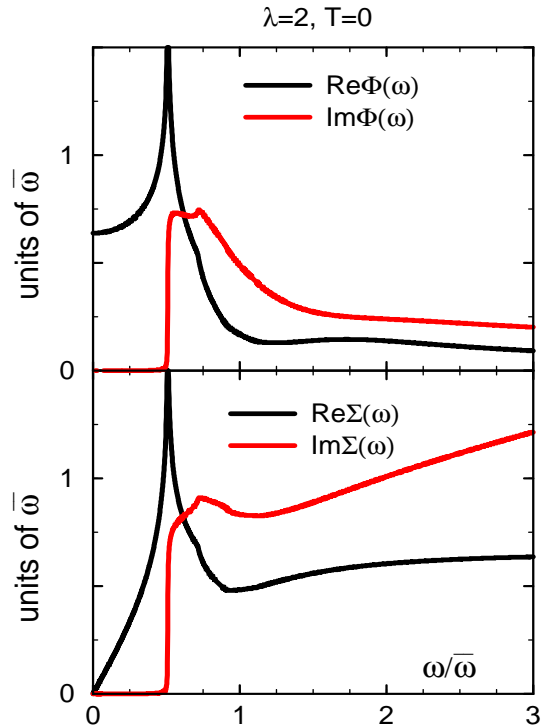}
\end{center}
\caption{The real and imaginary parts of the fermionic self-energy $\Sigma (%
\protect\omega )$ and the pairing vertex $\Phi (\protect\omega )$ for $%
\protect\lambda =2$ and the lowest $T$. The results are from Ref.~ 
\protect\onlinecite{artem}. }
\label{sigel1}
\end{figure}

We also comment on the dispersion of the resonance peak. In Eq.~\ref{chil}
we assumed that $\Delta _{s}$ is a constant. In fact, $\Delta _{s}$ depends
on ${\bf{q}}$ since for any given ${\bf{q}}$, $\Delta _{s}^{2}\propto
\Delta ({\bf{q}})$ where $\Delta ({\bf{q}})$ is a $d-$ wave gap at the
points at the Fermi surface which are connected by ${\bf q}$. In particular, $%
\Delta _{s}$ should vanish at ${\bf q}={\bf Q}_{\rm min}$ which connects the nodal points.
This effect accounts for the ``negative'' dispersion of the resonance peak 
\cite{Norman_1,oleg_1}. The latter certainly overshadows the positive dispersion
due to $({\bf{q-Q}})^{2}$ term for ${\bf q}$ close to ${\bf Q}_{\rm min}$ and may do so even
for ${\bf{q}}$ near ${\bf{Q}}$ if the correlation length is not large
enough. This effect is, however, not a part of the quantum-critical
description (it should become progressively less relevant for ${\bf{q}}%
\neq {\bf{Q}}_{min}$ when $\xi $ increases), and we ignore it in the
subsequent analysis. Note, however, 
that the negative dispersion of the
peak implies that the peak exists only \ for a small range of momenta
 between ${\bf{Q}}$ and ${\bf{Q}}_{\mathrm{min}}$. 
In optimally doped cuprates, $%
{\bf{Q}}_{\mathrm{min}}\approx (0.8\pi ,0.8\pi )$~\cite
{Fedorov99,Kaminski00}, and the momentum range for the peak does not exceed $%
6\%$ of the Brillouin zone. The actual ${\bf{q}}$ region where the peak is
observable is even smaller as the intensity of the peak also decreases when $%
{\bf{q}}$ approaches ${\bf{Q}}_{\mathrm{min}}$. The smallness of the $%
{\bf{q}}-$range for the peak accounts for small overall spectral intensity 
$I_{0}=\int S({\bf{q}},\omega )d^{2}qd\omega /(2\pi )^{3}$ that turns out
to be substantially smaller than $S(S+1)/3=1/4$. Still, at ${\bf{Q}}$, the
intensity of the peak is not small (experimentally, $\int S({\bf{Q}}%
,\omega )d\omega \sim 1.5$ in optimally doped YBCO~\cite{neutrons,dai}), and
we have verified that for the frequencies that we consider below,\ the
typical ${\bf{q-Q}}$ that account for the feedback on the fermions are
well within the ${\bf{q}}$ range between ${\bf{Q}}$ and ${\bf{Q}}_{%
\mathrm{min}}$. In other words, the small overall intensity of the resonance
peak does not preclude strong feedback effects from the resonance peak on
fermionic variables.
\begin{figure}[tbp]
\epsfxsize=2.8in 
\epsfysize=3.2in
\begin{center}
\epsffile{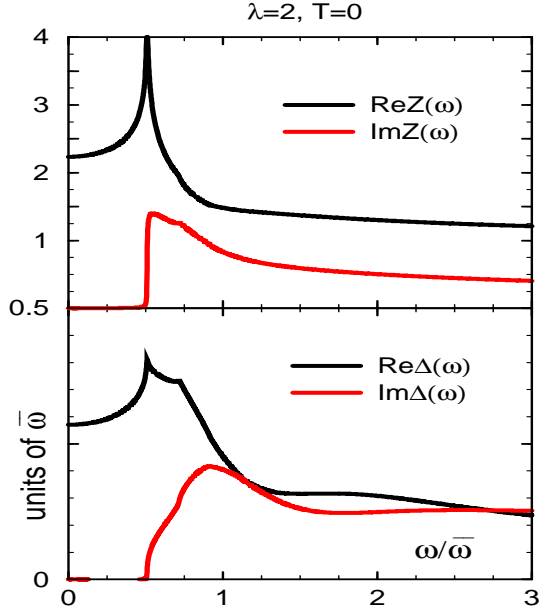}
\end{center}
\caption{The real and imaginary parts of the effective gap $\Delta (\protect%
\omega)$ and the quasiparticle renormalization factor $Z(\protect\omega)$
for $\protect\lambda =2$ and the lowest $T$. The results are from Ref.~ 
\protect\onlinecite{artem}. }
\label{sigel2}
\end{figure}

 For completeness, 
in Figs.\ref{sigel1} and \ref{sigel2} we present results for the fermionic
self-energy and the pairing vertex for the smallest $T$. We see that the
real parts of $\Phi (\omega )$ and $\Delta (\omega )$ are finite at $\omega
=0$ as should be the case in the superconducting state. The imaginary parts
of $\Phi (\omega )$ and $\Sigma (\omega )$ (and of $\Delta (\omega )$ and $%
Z(\omega )$) vanish at small frequencies and appear only above the threshold
frequency that is precisely $\Delta +\Delta _{s}$. Furthermore, all
variables have a complex internal structure at large frequencies. In the
next section we discuss the physical origin of the threshold at $\Delta
+\Delta _{s}$ and also show that one can  
extract $3\Delta $ from the derivative
of $\Sigma ^{\prime \prime }(\omega )$. 

Few words about the numbers. For $\lambda =2$,
 $\Delta \approx 0.3{\bar{\omega}}$ and $\Delta
_{s}\approx 0.2{\bar{\omega}}$, i.e., $\Delta$ and $\Delta_s$ are comparable to each other. For $\lambda \gg 1$ a numerical
solution of the Eliashberg equations leads to $\Delta \sim 2 T_{cr} \sim 0.35 {\bar \omega}$, and $%
\Delta _{s}\sim 0.25 {\bar \omega}/\lambda \ll \Delta$.

\section{Fingerprints of spin fermion pairing}

In this Section, we discuss the extent to which the ``fingerprints'' of
spin-mediated pairing can be extracted from experiments on materials that
are candidates for a magnetically-mediated superconductivity. Due to strong
spin-fermion coupling, there is unusually strong feedback from spin
excitations on fermions, specific to $d-$wave superconductors with a  magnetic
pairing interaction. The origin of this feedback is the emergence of a
propagating collective spin bosonic mode below $T_{cr}$. This mode is
present for any coupling strength, and its gap $\Delta _{s}$ is smaller than
the minimum energy $\sim 2\Delta $ that is necessary to break a Cooper pair.
In the vicinity of the antiferromagnetic phase, $\Delta _{s}\propto \xi
^{-1} $ where $\xi $ is the magnetic correlation length. We show that this
propagating spin mode changes the onset frequency for single particle
scattering, gives rise to the ``peak-dip-hump'' features in the
quasiparticle spectral function, the ``dip-peak'' features in tunneling SIS
and SIN conductances, and to singularities and fine structures in the
optical conductivity. In section 6, we apply these results to cuprate
superconductors and argue that (i) these features have been observed~\cite
{Norman97,shennat,Fedorov99,Kaminski00,fisher,zasad,basov,CSB99} (ii) ARPES~ 
\cite{Norman97,shennat,Fedorov99,Kaminski00}, tunneling~\cite{fisher,zasad},
and conductivity data~\cite{basov,CSB99} are consistent with each other, and
(iii) the value of $\Delta _{s}$ extracted from these various experiments
coincides with the resonance frequency measured directly in neutron
scattering experiments~\cite{neutrons,dai,neutrons2}.

\subsection{The physical origin of the effect}

The physical effect that accounts for dips and humps in the density of
states and spectral function of cuprates is not new and is known for
conventional $s-$wave superconductors as the Holstein effect \cite
{Scalapino69,holstein,varma}. 
\begin{figure}[tbp]
\epsfxsize=\columnwidth 
\begin{center}
\epsffile{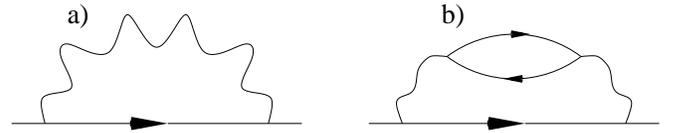}
\end{center}
\caption{a) The exchange diagram for a boson mediated interaction. The solid
line stands for a propagating fermion. The wiggly line is a phonon
propagator in the case of electron- phonon interaction, and a magnon line in
the case of a spin- fluctuation mediated interaction. b) The lowest order
diagram for the fermionic self energy due to a direct four fermion
interaction, also represented by a wiggly line. The figure is taken from ~ 
\protect\onlinecite{acs_finger}.}
\label{Figure1}
\end{figure}
Consider a clean $s-$wave superconductor, and suppose that the residual
interaction between fermions occurs via the exchange of an Einstein phonon.
Assume for simplicity that the fully renormalized electron phonon coupling
is some constant $g_{\mathrm{ep}}$, and that the phonon propagator $%
D(q,\omega )$ is independent of momentum $q$ and has a single pole at a
phonon frequency $\omega _{0}$ (the Holstein model) 
~\cite{holstein,varma,ssw}. Phonon exchange gives rise to a fermionic
self-energy (see Fig~\ref{Figure1}a) 
\begin{equation}
\Sigma (i\omega _{m})=g_{\mathrm{ep}}^{2}T\sum_{n}\int \frac{d^{d}k}{(2\pi
)^{d}}G_{{\bf{k}}}(i\omega _{n})D(i\omega _{m}-i\omega _{n})  \label{s}
\end{equation}
which is a convolution of $D(\omega )=1/(\omega _{0}^{2}-\left( \omega
+i 0^+ \right) ^{2})$ with the full fermionic propagator $G_{\bf k}(\omega )$,
which in a superconductor is given by Eq.\ref{GGorkov}: 
\begin{equation}
G_{{\bf{k}}}(\omega )=\frac{{\tilde{\Sigma}}(\omega )+\varepsilon _{%
{\bf{k}}}}{{\tilde{\Sigma}}(\omega )^{2}-\Phi ^{2}(\omega )-\varepsilon _{%
{\bf{ k}}}^{2}}.
\end{equation}
As before, $\Phi (\omega )$ is the pairing vertex, and $\varepsilon _{%
{\bf{k}}}$ is the band dispersion of the fermions. At $T=0$ both $\Sigma
^{\prime \prime }(\omega )$ and $\Phi ^{\prime \prime }(\omega )$ obviously
vanish for $\omega \leq \Delta $. This implies that the fermionic spectral
function $A_{{\bf{k}}}(\omega )=\left| G_{{\bf{k}}}^{\prime \prime
}(\omega )\right| /\pi $ for particles at the Fermi surface (${\bf{k}}=%
{\bf{k}}_{\mathrm{F}}$) has a $\delta -$function peak at $\omega =\Delta $%
, i.e. $\Delta $ is a sharp gap at $\ $zero temperature. The fermionic
density of states in a superconductor 
\begin{equation}
N(\omega )= N_0~ \mathrm{Im}\left[ \frac{{\tilde{\Sigma}}(\omega )}{(\Phi
^{2}(\omega )-{\tilde{\Sigma}}^{2}(\omega ))^{1/2}}\right]  \label{dos}
\end{equation}
vanishes for $\omega <\Delta $ and has a square-root singularity $N(\omega
)\propto (\omega -\Delta )^{-1/2}$ for frequencies above the gap 
($N_0$ is the normal state density of states).

The onset of the imaginary part of the self-energy, Eq.\ref{s}, can be
easily obtained by using the spectral representation for fermionic and
bosonic propagators in Eq.~\ref{s} and re-expressing the momentum
integration in terms of an integration over $\varepsilon _{{\bf{k}}}$. At $%
T=0$ we obtain 
\begin{equation}
\Sigma ^{\prime \prime }(\omega >0)\propto \int_{0}^{\omega }d\omega
^{\prime }N(\omega ^{\prime })D^{\prime \prime }(\omega -\omega ^{\prime })
\label{im}
\end{equation}
Since for positive frequencies, $D^{\prime \prime }(\omega )=(\pi
D_{0}/2\omega _{0})\delta (\omega -\omega _{0})$, the frequency integration
is elementary and yields 
\begin{equation}
\Sigma ^{\prime \prime }(\omega >0)\propto N(\omega -\omega _{0}).  \label{1}
\end{equation}
We see that the single particle scattering rate is directly proportional to
the density of states shifted by the phonon frequency. Clearly, the
imaginary part of the fermionic self-energy emerges only when $\omega $
exceeds the threshold 
\begin{equation}
\Omega _{\mathrm{t}}\equiv \Delta +\omega _{0},
\end{equation}
the sum of the superconducting gap and the phonon frequency. Right above
this threshold, ${\tilde{\Sigma}}^{\prime \prime }(\omega )\propto (\omega
-\Omega _{\mathrm{t}})^{-1/2}$. By the Kramers-Kronig relation, this
non-analyticity causes a square root divergence of ${\tilde{\Sigma}}^{\prime
}(\omega )$ at $\omega <\Omega _{\mathrm{t}}$. Combining the two results, we
find that near the threshold, ${\tilde{\Sigma}}(\omega )=A+C/\sqrt{\Omega _{%
\mathrm{t}}-\omega }$ where $A$ and $C$ are real numbers. By the same
reasoning, the pairing vertex $\Phi (\omega )$ also possesses a square-root
singularity at $\Omega _{\mathrm{t}}$. Near $\omega =\Omega _{\mathrm{t}}$, $%
\Phi (\omega )=B+C/\sqrt{\Omega _{\mathrm{t}}-\omega }$ with real $B$. Since 
$\Omega _{\mathrm{t}}>\Delta $, we have $A>B$.

\begin{figure}[tbp]
\epsfxsize=\columnwidth 
\begin{center}
\epsffile{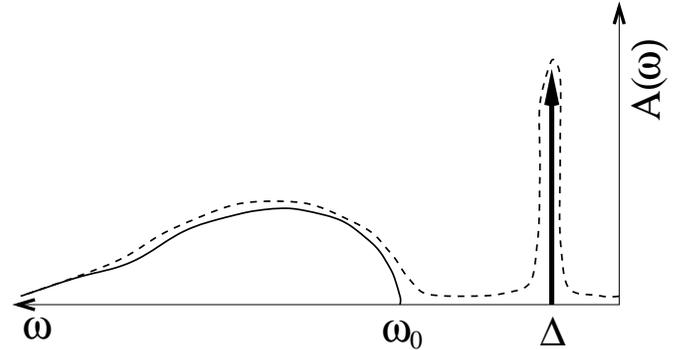}
\end{center}
\caption{The schematic form of the quasiparticle spectral function in an $s-$%
wave superconductor. Solid line -- $T=0$, dashed line -- at finite $T$. $%
\Omega _{\mathrm{t}}=\Delta +\protect\omega _{0}$ (from Ref. ~\protect\cite
{acs_finger}).}
\label{Figure2}
\end{figure}

The singularity in the fermionic self-energy gives rise to an extra dip-hump
structure of the fermionic spectral function at ${\bf{k}}={\bf{k}}_{F}$.
Below $\Omega _{\mathrm{t}}$, the spectral function is zero except for $%
\omega =\Delta $, where it has a $\delta -$functional peak. Immediately
above $\Omega _{\mathrm{t}}$, $\ A(\omega )\propto \mathrm{Im}({\tilde{\Sigma%
}}(\omega )/({\tilde{\Sigma}}^{2}(\omega )-\Phi ^{2}(\omega )))$ takes the
form $A(\omega )\propto (\omega -\Omega _{\mathrm{t}})^{1/2}$. At larger
frequencies, $A(\omega )$ passes through a maximum, and eventually vanishes.
Adding a small damping introduced by either impurities or finite
temperatures, one obtains the spectral function with a peak at $\omega
=\Delta $, a dip at $\omega \approx \Omega _{\mathrm{t}}$, and a hump at a
somewhat larger frequency. This behavior is shown schematically in Fig.~\ref
{Figure2}. 

The singularities in ${\tilde \Sigma} (\omega )$ and $\Phi (\omega )$ affect
other observables such as the fermionic DOS, optical conductivity, Raman
response, and the SIS tunneling dynamical conductance~\cite{varma,coffey}.

For a more complex phonon propagator, which depends on both frequency and
momentum, the singularities in the fermionic self-energy and other
observables are weaker and may only show up in the derivatives over
frequency~\cite{McMillan69,shaw}. 
Still, the opening of the new relaxational channel at $\Omega_{\mathrm{t}}$
gives rise to singularities in the electronic properties of a
phonon-mediated $s-$wave superconductor.

\subsection{Similarities and discrepancies between $d$- and $s-$wave \
superconductors}

As we already discussed, 
 for magnetically mediated $d-$wave superconductivity, spin fluctuations play
the role of phonons. Below $T_{c}$, spin excitations are propagating,
magnon-like modes with the gap $\Delta _{s}$. This $\Delta_s$ 
obviously plays the same role as $\omega _{0}$ for phonons, and hence we
expect that for spin-mediated pairing, 
the spectral function should display \ a peak-dip-hump structure
as well. Furthermore, 
we will demonstrate below that for observables such as the DOS,
Raman intensity and the optical conductivity, which measure the response
averaged over the Fermi surface, the angular dependence of the $d-$wave gap $%
\Delta (\theta )\propto \cos \left( 2\theta \right) $ softens the
singularities, but does not wash them out over a finite frequency range.
Indeed, we will find that the positions of the singularities are not
determined by some averaged gap amplitude but by the maximum value of the $%
d- $wave gap, $\Delta _{max}=\Delta $, i.e., the Holstein effect is still present for a $d-$wave superconductor.

Despite many similarities, the feedback effects for phonon-mediated $s-$wave
superconductors, and magnetically mediated $d-$wave superconductors are not
equivalent as we now demonstrate. The point is that for $s-$wave
superconductors, the exchange process shown in Fig.\ref{Figure1}a is not the
only possible source for the fermionic decay: there exists another process,
shown in Fig.\ref{Figure1}b, in which a fermion decays into three other
fermions. This process is due to a residual four-fermion interaction~\cite
{varma,coffey}. One can easily make sure that this second process also gives
rise to the fermionic decay when the external $\omega $ exceeds a minimum
energy of $3\Delta $, necessary to pull all three intermediate particles out
of the condensate of Cooper pairs. At the threshold, the fermionic spectral
function is non-analytic, much like that found at $\Delta +\omega _{0}$.
This implies that in $s$-wave superconductors, there are two physically
distinct singularities, at $\Delta +\omega _{0}$ and at $3\Delta $, which
come from \textit{different} processes and therefore are independent of each
other. Which of the two threshold frequencies is larger depends on the
strength of the coupling and on the shape of the phonon density of states.
At weak coupling, $\omega _{0}$ is exponentially larger than $\Delta $,
hence the $3\Delta $ threshold comes first. At strong coupling, $\omega _{0}$
and $\Delta $ are comparable, but calculations within the Eliashberg
formalism show that for real materials ( e.g. for lead or niobium), still $%
3\Delta <\Delta +\omega _{0}$.\cite{mahan}. This result is fully consistent
with the photoemission data for these materials\cite{C2000}.

For magnetically mediated $d$-wave superconductors the situation is
different. As we discussed in Section 2, in the one-band model for cuprates,
which we adopt, the underlying interaction is a Hubbard-type four-fermion
interaction. The introduction of a spin fluctuation as an extra degree of
freedom is just a way to account for the fact that there exists a particular
interaction channel, where the effective interaction between fermions is the
strongest due to the proximity to a magnetic instability. This implies that
the spin fluctuation propagator is made out of particle-hole bubbles like
those in Fig.\ref{Figure1}b. Then, to the lowest order in the interaction,
the fermionic self-energy is given by the diagram in Fig.\ref{Figure1}b.
Higher-order terms convert a particle-hole bubble in Fig.\ref{Figure1}b.
into a wiggly line, and transform this diagram into the one in Fig.\ref
{Figure1}a. Clearly then, inclusion of both diagrams would be double
counting, i.e., there is only a \textit{single} process which gives rise to
the threshold in the fermionic self-energy. Note also that the fact that the
diagram in Fig \ref{Figure1}b is a part of that in Fig.\ref{Figure1}a
implies that the development of a singularity in the spectral function at a
frequency different from $3\Delta $ cannot be due to effects outside the
spin-fermion model. Indeed, we will show that the model generates two
singularities: at $3\Delta $, and at $\Delta +\Delta _{s}<3\Delta $. The
fact that this is an internal effect, however, implies that $\Delta _{s}$ 
\textit{depends} on $\Delta $. The experimental verification of this
dependence 
can then be considered as a ``fingerprint'' of the spin-fluctuation
mechanism. Furthermore, as the singularities at $3\Delta $ and $\Delta
+\Delta _{s}$ are due to the same interaction, their relative intensity is
another gauge of the magnetic mechanism for the pairing. We will argue below
that some experiments on cuprates, particularly measurements of the optical
conductivity~\cite{CSB99}, allow one to detect both singularities, and that
their calculated relative intensity is consistent with the data.

We now discuss separately the behavior of the electronic spectral function,
the density of states, SIS tunneling, the Raman intensity and the optical
conductivity. To account for all features associated with $d-$wave pairing,
we will keep the momentum dependence of the fermionic self-energy and the
pairing vertex on momenta along the Fermi surface, although this dependence
is indeed weak near hot spots. For simplicity, we assume a circular Fermi
surface. In this situation, the $k$-dependence of the self-energy and the
pairing vertex reduces to the angular dependence, i.e. $\Sigma =\Sigma
(\theta ,\omega )$ and $\Phi =\Phi (\theta ,\omega )$

\subsection{The spectral function}

We first consider the spectral function $A_{{\bf{k}}}(\omega )=(1/\pi )|G_{%
{\bf{k}}}^{\prime \prime }(\omega )|$. In the superconducting state, for
quasiparticles near the Fermi surface 
\begin{equation}
A_{{\bf{k}}}(\omega >0)=\frac{1}{\pi }~\mathrm{Im}\left[ \frac{\omega +
\Sigma (\theta ,\omega )+\varepsilon _{{\bf{k}}}}{\left( \omega + \Sigma
(\theta ,\omega )\right) ^{2}-\Phi ^{2}(\theta ,\omega )-\varepsilon _{%
{\bf{k}}}^{2}}\right].  \label{sf}
\end{equation}
By definition, $A_{{\bf{k}}}(-\omega )=A_{{\bf{k}}}(\omega )$.

In a Fermi gas with $d$-wave pairing, $\Sigma (\theta ,\omega )=0$, and $%
\Phi (\theta ,\omega )=\Delta (\theta )\propto \cos \left( 2\theta \right) $%
. The spectral function then has a $\delta -$function peak at $\omega
=(\Delta ^{2}(\theta )+\varepsilon _{{\bf{k}}}^{2})^{1/2}$. It is obvious,
but essential for comparison with the strong coupling case, that the peak
disperses with ${\bf{k}}$ and that far away from the Fermi surface one
recovers normal state dispersion.

For strong coupling we consider the spectral function, $A_{{\bf{k}}%
}(\omega )$, for fermions located near hot spots, $\theta =\theta _{hs}$.
where the gap $\Delta (\theta )$ ( defined as a solution of ${\tilde{\Sigma}}%
^{\prime }(\omega =\Delta ,\theta _{hs})=\Phi ^{\prime }(\omega =\Delta
,\theta _{hs})$) is maximum. As discussed above, we expect the spectral
function to possess a peak at $\omega =\Delta $ and a singularity at $\omega
=\Omega _{\mathrm{t}}=\Delta +\Delta _{s}$. The behavior of $A(\omega )$
near the singularity is robust and can be obtained without a precise
knowledge of the frequency dependence of ${\tilde{\Sigma}}(\omega )$ and $%
\Phi (\omega )$. All we need to know is that near $\omega =\Delta $, ${%
\tilde{\Sigma}}^{2}(\omega )-\Phi ^{2}(\omega )\propto \omega -\Delta $.
Substituting this form into Eq.\ref{Elsi} and converting to the real axis
using the spectral representation, we obtain for $\omega =\Omega _{\mathrm{t}%
}+\epsilon $ 
\begin{equation}
{\tilde{\Sigma}}^{\prime \prime }(\omega )\propto \int_{0}^{\epsilon }~\frac{%
dx}{(x(\epsilon -x))^{1/2}}
\end{equation}
This integral is the same as in Eq.\ref{ya}, hence ${\tilde{\Sigma}}^{\prime
\prime }$ undergoes a finite jump at $\omega =\Omega _{\mathrm{t}}$, just as
the spin polarization operator does at $\omega =2\Delta $. By the
Kramers-Kronig relation, this jump gives rise to a logarithmic divergence of 
${\tilde{\Sigma}}^{\prime }$. The same singular behavior holds for the
pairing vertex $\Phi (\omega )$, with exactly the same prefactor in front of
the logarithm. The last result implies that ${\tilde{\Sigma}}(\omega )-\Phi
(\omega )$ is non-singular at $\omega =\Omega _{\mathrm{t}}$. Substituting
these results into Eq.\ref{sf}, we find that the spectral function $A(\omega
)$ behaves at $\omega >\Omega _{\mathrm{t}}$ as $1/\log ^{2}(\omega -\Omega
_{\mathrm{t}})$, i.e., almost discontinuously. Obviously, at a small but
finite $T$, the spectral function should have a dip very near $\omega
=\Omega _{\mathrm{t}}$, and a hump at a somewhat higher frequency.

\begin{figure}[tbp]
\epsfxsize=2.8in 
\epsfysize=3.4in
\begin{center}
\epsffile{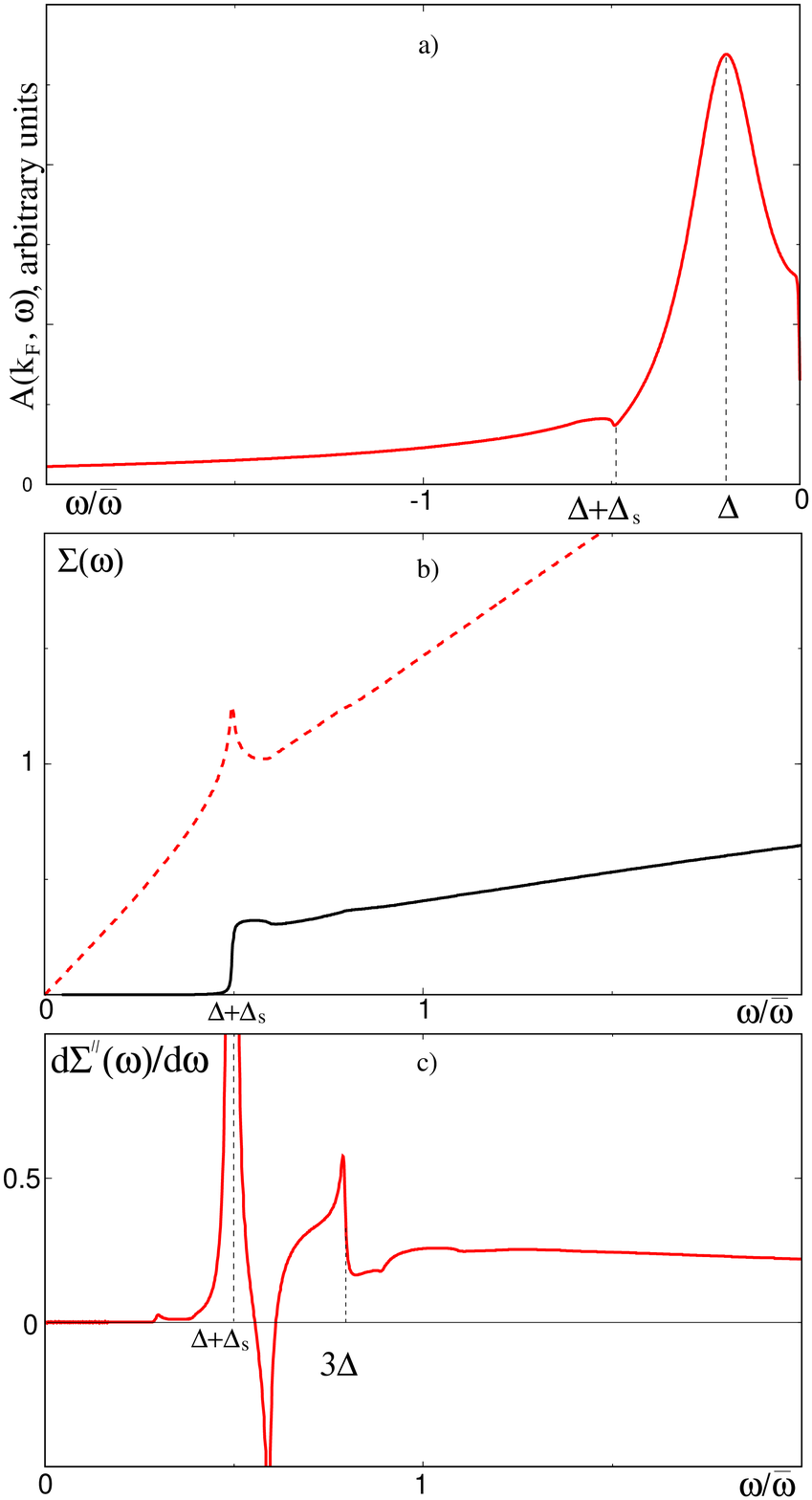}
\end{center}
\caption{Upper panel: the quasiparticle spectral function determined by
solving the coupled Eliashberg equations for $\protect\lambda =1$. The
peak-dip-hump structure of $A(\protect\omega)$ is clearly visible but not
dramatic. Middle panel: real and imaginary parts of the fermionic
self-energy (dashed and solid lines, respectively) . Lower panel -the
frequency derivative of $\Sigma^{\prime \prime} (\protect\omega)$. The extra
structure at $3\Delta$ is clearly visible. The figure is taken from Ref.~ 
\protect\onlinecite{acs_finger}.}
\label{Figure_specdens}
\end{figure}

In Fig.\ref{Figure_specdens} we present the result for $A(\omega )$ obtained
from a solution of the set of three coupled Eliashberg equations at 
 $T \ll T_{c}$~\cite{acs_finger}. This solution is consistent with our
analytical estimate. We clearly see that the fermionic spectral function has
a peak-dip-hump structure, and the peak-dip distance equals $\Delta _{s}$.
We also see in Fig.\ref{Figure_specdens} that the fermionic self-energy is
non-analytic at $\omega =3\Delta $. As we discussed above, this
 last non-analyticity  originates in the
non-analyticity of the dynamical spin susceptibility at $\omega =2\Delta $.

Another ``fingerprint'' of the spin-fluctuation scattering can be found by
studying the evolution of the spectral function as one moves away from the
Fermi surface. The argument here goes as follows: at strong coupling, where $%
\Delta \geq \omega _{\mathrm{sf}}$, probing the fermionic spectral function
at frequencies progressively larger than $\Delta $, one eventually probes
the normal state fermionic self-energy at $\omega \gg \omega _{\mathrm{sf}}$%
. Substituting the self energy Eq.\ref{SEns} into the fermionic propagator,
we find that up to $\omega \sim {\bar{\omega}}$, the spectral function in
the normal state does not have a quasiparticle peak at $\omega =\varepsilon
_{{\bf{k}}}$. Instead, it only displays a broad maximum at $\omega
=\varepsilon _{{\bf{k}}}^{2}/{\bar{\omega}}$. In other words, at $\omega _{%
\mathrm{sf}}<\omega <{\bar{\omega}}$, the spectral function in the normal
state displays non-Fermi liquid behavior with no quasiparticle peak (see
Fig.~\ref{Figure_NFL}). The absence of a quasiparticle peak in the normal
state implies that the sharp quasiparticle peak that we found at $\omega
=\Delta $ for momenta at the Fermi surface cannot simply disperse with $%
{\bf{k}}$, as it does for noninteracting fermions with a $d$-wave gap.
Specifically, the quasiparticle peak cannot move further in energy than $%
\Delta +\Delta _{s}$ since at larger frequencies, spin scattering rapidly
increases, and the fermionic spectral function should display roughly the
same non-Fermi-liquid behavior as in the normal state.

\begin{figure}[tbp]
\epsfxsize=\columnwidth 
\begin{center}
\epsffile{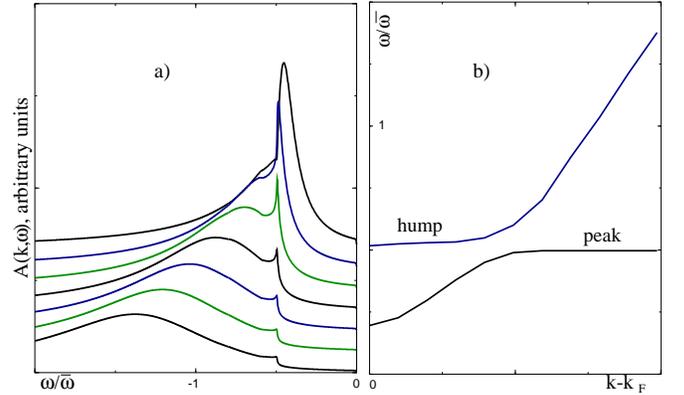}
\end{center}
\caption{a) Frequency dependence of the spectral function in the
superconducting state for different $\protect\epsilon_k$. The curve at the
bottom has a highest $\protect\epsilon_k$. No coherent quasiparticle peak
occurs for energies larger than $\Delta + \Delta_s$. Instead, the spectral
function displays a broad maximum, similar to that in the normal state.
(From Ref~\protect\onlinecite{acs_finger}.)}
\label{Figure_spec_sc_mom}
\end{figure}

In Fig.\ref{Figure_spec_sc_mom}a we present plots for the spectral function
as the momentum moves away from the Fermi surface. We see the behavior we
just described: the quasiparticle peak does not move further than $\Delta
+\Delta _{s}$. Instead, when ${\bf{k}}-{\bf{k}}_{\mathrm{F}}$ increases,
it gets pinned at $\Delta +\Delta _{s}$ and gradually looses its spectral
weight. At the same time, the hump disperses with ${\bf{k}}$ and for
frequencies larger than $\Delta +\Delta _{s}$ gradually transforms into a
broad maximum at $\omega =\varepsilon _{{\bf{k}}}^{2}/{\bar{\omega}}$. The
positions of the peak and the dip versus ${\bf{k}}-{\bf{k}}_{\mathrm{F}}$
are presented in Fig.\ref{Figure_spec_sc_mom}b.

\subsection{The density of states}

The quasiparticle density of states, $N(\omega )$, is the momentum integral
of the spectral function: 
\begin{equation}
N(\omega )=\int \frac{d^{2}{\bf{k}}}{4\pi ^{2}}~A_{{\bf{k}}}(\omega ).
\end{equation}
Substituting $A_{{\bf{k}}}(\omega )$ from Eq.\ref{sf} and integrating over 
$\varepsilon _{{\bf{k}}}$, one obtains 
\begin{equation}
N(\omega )\propto \mathrm{Im}~\int_{0}^{2\pi }d\theta ~\frac{\omega +\Sigma
(\theta ,\omega )}{(\Phi ^{2}(\theta ,\omega )-\left( \omega +\Sigma (\theta
,\omega )\right) ^{2})^{1/2}},  \label{eq}
\end{equation}
\begin{figure}[t]
\epsfxsize=2.8in 
\epsfysize=2.1in
\begin{center}
\epsffile{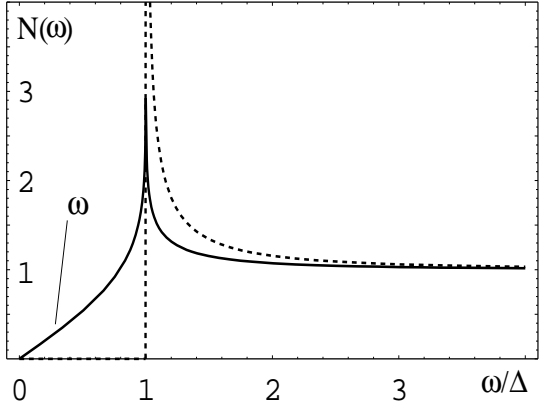}
\end{center}
\caption{Density of states of a noninteracting Fermi gas with $d$-wave gap
(solid line) and with $s$-wave gap (dashed line). (From Ref.~\protect\cite
{acs_finger}).}
\label{Figure_DOS_gas}
\end{figure}
We first consider $N(\omega )$ in a $d$-wave gas, and then discuss strong
coupling effects. In a $d-$wave gas, $\Sigma =0$ and $\Delta _{{\bf{k}}%
}=\Delta \cos \left( 2\theta \right) $. Integrating in Eq.\ref{eq} over $%
\theta $ we obtain~\cite{maki,acg} 
\begin{eqnarray}
N(\omega ) &=& N_0~ \mathrm{Re}~\left[ \frac{\omega }{2\pi }\int_{0}^{2\pi }\frac{%
d\theta }{\sqrt{\omega ^{2}-\Delta ^{2}\cos ^{2}(2\theta )}}\right] 
\nonumber \\
&=&\frac{2 N_0}{\pi } \cases{K(\Delta /\omega ) &for $~\omega >\Delta$; 
\cr (\omega /\Delta )K(\omega /\Delta )  & for $~\omega <\Delta$.\cr}
 \label{sin-gas}
\end{eqnarray}
where $K(x)$ is the elliptic integral of first kind. We see that $N(\omega
)\sim \omega $ for $\omega \ll \Delta $ and diverges logarithmic as $%
(1/\pi )\ln (8\Delta /|\Delta -\omega |)$ for $\omega \approx \Delta $. At
larger frequencies, $N(\omega )$ gradually decreases towards the frequency
independent, normal state value of the DOS, which we have normalized to
unity. The plot of $N(\omega )$ in a $d-$wave BCS superconductor is
presented in Fig.\ref{Figure_DOS_gas}.

For comparison, in an $s$-wave superconductor, the DOS vanishes at $\omega
<\Delta $ and diverges as $N(\omega) \propto 
(\omega -\Delta )^{-1/2}$ at $\omega \geq \Delta $%
. We see that a $d$-wave superconductor is different in that (i) the DOS is
finite down to the smallest frequencies, and (ii) the singularity at $\omega
=\Delta $ is weaker (logarithmic). Still, however, $N(\omega )$ is singular
only at a frequency which equals to the largest value of the $d-$wave gap.
This illustrates a point made earlier: the angular dependence of the $d$%
-wave gap reduces the strength of the singularity at $\omega = \Delta_{max} (\theta)$ , but does not wash it out
over a finite frequency range.

\begin{figure}[t]
\epsfxsize=2.8in 
\epsfysize=2.4in
\begin{center}
\epsffile{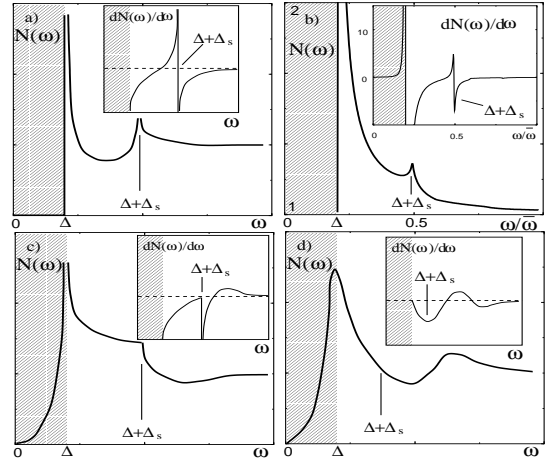}
\end{center}
\caption{(a) The behavior of the SIN tunneling conductance (i.e., DOS) in a
strongly coupled $d$-wave superconductor. Main pictures - $N(\protect\omega %
) $, insets - $dN(\protect\omega )/d\protect\omega $. (a) The schematic
behavior of the DOS for a flat gap. (b) The solution of the Eliashberg-type
equations for a flat gap. The shaded regions are the ones in which the flat
gap approximation is incorrect as the physics is dominated by nodal
quasiparticles. (c) The schematic behavior of $N(\protect\omega )$ for the
quadratic variation of the gap near its maxima. (d) The expected behavior of
the DOS in a real situation when singularities are softened out by finite $T$
or impurity scattering. The position of $\Delta +\Delta _{s}$ roughly
corresponds to a minimum of $dN(\protect\omega )/d\protect\omega $. The
figure is taken from Ref.~\protect\onlinecite{acs_finger}.}
\label{Figure_DOS_schematic}
\end{figure}

We now turn to strong coupling. We first demonstrate that the DOS possesses
extra peak-dip features, associated with the singularities in ${\tilde{\Sigma%
}}(\omega )$ and $\Phi (\omega )$ at $\omega =\Omega _{\mathrm{t}}$ which for spin-mediated pairing is 
\begin{equation}
\Omega _{\mathrm{t}}=\Delta +\Delta _{s}.  \label{omegatsf}
\end{equation}
An analytical approach proceeds as follows~\cite{acsin}. Consider first a
case when the gap is totally flat near a hot spot. At $\omega =\Omega _{%
\mathrm{t}}$, both ${\tilde{\Sigma}}(\omega )$ and $\Phi (\omega )$ diverge
logarithmically. On substituting these forms into Eq.~\ref{eq}, we find that $%
N(\omega )$ has a logarithmic singularity: 
\begin{equation}
N_{\mathrm{sing}}(\omega )\propto \left( \log {\frac{1}{|\omega -\Omega _{%
\mathrm{t}}|}}\right) ^{1/2}.
\end{equation}
This singularity gives rise to a strong divergence of $dN(\omega )/d\omega $
at $\omega =\Omega _{\mathrm{t}}$, a behavior schematically shown in Fig. (%
\ref{Figure_DOS_schematic}a). In part (b) of this figure we present the
result for $N(\omega )$ obtained by the solution of the Eliashberg-type Eqs.%
\ref{El1}-\ref{Elph}. A small but finite temperature was used to smear out
divergences. We recall that the Eliashberg set does not include the angular
dependence of the gap near hot spots, and hence the numerical result for the
DOS in Fig.~\ref{Figure_DOS_schematic}b should be compared with 
Fig.~\ref{Figure_DOS_schematic}a. We clearly see that $N(\omega )$ has a second peak
at $\omega =\Omega _{\mathrm{t}}$. This peak strongly affects the frequency
derivative of $N(\omega )$ which become singular near $\Omega _{\mathrm{t}}$.

The relatively small magnitude of the singularity in $N(\omega ) $ is a
consequence of the linearization of the fermionic dispersion near the Fermi
surface. For an actual $\epsilon _{{\bf{k}}}$ chosen to fit ARPES data~ 
\cite{Norman_1}, nonlinearities in the fermionic dispersion occur at energies
comparable to $\Omega_{\mathrm{t}}$. This is due to the fact that hot spots
are located close to $(0,\pi )$ and related points at which the Fermi
velocity vanishes. As a consequence, the momentum integration in the
spectral function should have a less pronounced smearing effect than found
in our calculations, and the frequency dependence of $N(\omega )$ should
more resemble that of $A(\omega )$ for momenta where the gap is at maximum.

For a momentum dependent gap, the behavior of fermions near hot spots is the
same as when the gap is flat, but now $\Omega_{\mathrm{t}}$ depends on $%
\theta $ as both $\Delta $ and $\Delta _{s}$ vary as one moves away from a
hot spot. The variation of $\Delta $ is obvious, the variation of $\Delta
_{s}$ is due to the fact that this frequency scales as $\Delta ^{1/2}$.
Since both $\Delta $ and $\Delta _{s}$ are maximal at a hot spot, we can
 model the momentum dependence by replacing 
\begin{equation}
\Omega_{\mathrm{t}}\rightarrow \Omega_{\mathrm{t}}-a{\tilde{\theta}}^{2}.
\end{equation}
where ${\tilde \theta} = \theta - \theta_{hs}$, and $a>0$. The singular
pieces of the self-energy and the pairing vertex then behave as $\log
(\Omega_{\mathrm{t}}-\omega -a\widetilde{\theta }^{2})^{-1}$. Substituting
these forms into Eq.\ref{eq} and using the fact that ${\tilde \Sigma}
(\omega )-\Phi (\omega )\approx const$ at $\omega \approx \Omega_{\mathrm{t}%
} $, we obtain 
\begin{equation}
N_{\mathrm{sing}}(\omega )\propto \mathrm{Re}\int d\widetilde{\theta }\left[
\log (\Omega_{\mathrm{t}}-\omega -a\widetilde{\theta }^{2})^{-1}\right]
^{-1/2}.
\end{equation}
A straightforward analysis of the integral shows that now $N(\omega )$
displays a one-sided non-analyticity at $\omega =\Omega_{\mathrm{t}}$: 
\begin{equation}
N_{\mathrm{sing}}(\omega )=-B\Theta (\omega -\Omega_{\mathrm{t}%
})\!\!\!~\left( \frac{\omega -\Omega_{\mathrm{t}}}{|\log (\omega -\Omega_{%
\mathrm{t}})|}\right) ^{1/2},  \label{nd}
\end{equation}
where $B>0$, and $\Theta (x)=1$ for $x>0$, and $\Theta (x)=0$ for $x<0$.
This non-analyticity gives rise to a cusp in $N(\omega )$ right above $%
\Omega_t$, and one-sided square-root divergence of the frequency
derivative of the DOS. This behavior is shown schematically in Fig. \ref
{Figure_DOS_schematic}c. Comparing this behavior with that shown in Fig. \ref
{Figure_DOS_schematic}a for a flat gap, we observe that the angular
dependence of the gap predominantly affects the form of $N(\omega )$ at $%
\omega \leq \Omega_{\mathrm{t}}$. At these frequencies, the angular
variation of the gap completely eliminates the singularity in $N(\omega )$.
At the same time, above $\Omega_{\mathrm{t}}$, the angular dependence of the
gap softens the singularity, but, still, the DOS sharply drops above $%
\Omega_{\mathrm{t}}$ in such a way that the derivative of the DOS diverges
on approaching $\Omega_{\mathrm{t}}$ from above. We see again that
 in a $d-$%
wave superconductor, the singularity in the DOS is softened by the angular
dependence of the gap, but still persists at a particular frequency related
to the maximum value of the gap. This point is essential as it enables us to
read off the maximum gap value directly from the experimental data without
any ''deconvolution'' \ of momentum averages.

For real materials, in which singularities are removed by e.g., impurity
scattering, $N(\omega )$ likely has a dip at $\omega \geq \Omega _{\mathrm{t}%
}$ and a hump at a larger frequency. This is schematically shown in Fig. (%
\ref{Figure_DOS_schematic})d. The location of $\Omega _{\mathrm{t}}$ is best
described as a point where the frequency derivative of the DOS passes
through a minimum. The singularity in $dN(\omega )/d\omega $ at $\Omega _{%
\mathrm{t}}$, and the dip-hump structure of $N(\omega )$ at $\omega \geq
\Omega _{\mathrm{t}}$ are additional ``fingerprints'' of the
spin-fluctuation mechanism in the single particle response.

\subsection{SIS tunneling}

Measurements of the dynamical conductance $dI/dV$ through a superconductor -
insulator - superconductor (SIS) junction offer another tool to search for
the fingerprints of the spin-fluctuation mechanism. The conductance through
this junction is the derivative over voltage of the convolution of the two
DOS~\cite{mahan}: $dI/dV\propto S(\omega )$ where 
\begin{equation}
S(\omega )=N^{-2}_0~\int_{0}^{\omega }d\Omega N(\omega -\Omega )~\partial _{\Omega
}N(\Omega ).  \label{sis}
\end{equation}

\begin{figure}[t]
\epsfxsize=2.8in 
\epsfysize=3.0in
\begin{center}
\epsffile{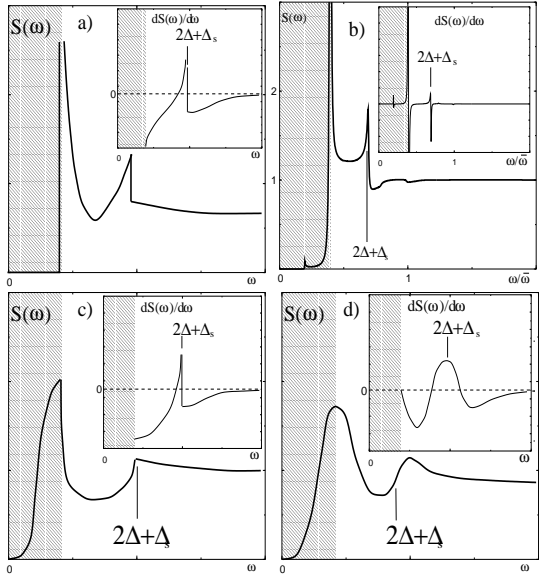}
\end{center}
\caption{ (a) The schematic behavior of the SIS tunneling conductance, $S(%
\protect\omega )$, in a strongly coupled $d$-wave superconductor. Main
pictures - $S(\protect\omega )$, insets - $dS(\protect\omega )/d\protect%
\omega $. (a) The schematic behavior of $S(\protect\omega )$ for a flat gap.
(b) The solution of the Eliashberg-type equations for a flat gap using the
DOS from the previous subsection. The shaded regions are the ones in which
the flat gap approximation is incorrect as the physics is dominated by nodal
quasiparticles. (c) The schematic behavior of $S(\protect\omega )$ for \ a
quadratic variation of the gap near its maxima. (d) The expected behavior of
the SIS conductance in a real situation when singularities are softened out
by finite $T$ or by impurity scattering. $2\Delta +\Delta _{s}$ roughly
corresponds to the maximum of $dS(\protect\omega )/d\protect\omega $. (From
Ref.~\protect\onlinecite{acs_finger}).}
\label{FIGURE_SIS}
\end{figure}

The DOS in a $d-$wave gas is given in Eq.~\ref{sin-gas}. Substituting this
form into Eq.~\ref{sis} and integrating over frequency we obtain the result
presented in Fig.\ref{FIGURE_SIS}. At small $\omega $, $S(\omega )$ is
quadratic in $\omega$~\cite{maki}. This is an obvious consequence of the
fact that the DOS is linear in $\omega $. At $\omega =2\Delta $, $S(\omega )$
undergoes a finite jump. This jump is related to the fact that near $2\Delta 
$, the integral over the two DOS includes the region around $\Omega =\Delta $
where both $N(\Omega )$ and $N(\omega -\Omega )$ are logarithmically
singular, and $\partial _{\Omega }N(\Omega )$ diverges as $1/(\Omega -\Delta
)$. The singular contribution to $S(\omega )$ from this region can be
evaluated analytically and yields 
\begin{equation}
S(\omega )=-\frac{1}{\pi ^{2}}P\int_{-\infty }^{\infty }\frac{dx~\log |x|}{%
x+\omega -2\Delta }=-\frac{1}{2}\mbox{sign}(\omega -2\Delta )
\label{sis-gas}
\end{equation}
Observe that the amplitude of the jump in the SIS conductance is a universal
number which does not depend on the value of $\Delta $. At larger
frequencies, $S(\omega )$ continuously goes down and eventually approaches a
value of $S(\omega \rightarrow \infty )=1$.

In the case of strong coupling one finds again that the quadratic behavior
at low frequencies and the discontinuity at $2\Delta $ survive at arbitrary
coupling. Indeed, the quadratic behavior at low $\omega $ is just a
consequence of the linearity of $N(\omega )$ at low frequencies. Therefore,
just as we did for the density of states we concentrate on  behavior
above $2\Delta $ that is sensitive to strong coupling effects.

\begin{figure}[t]
\epsfxsize=2.8in 
\epsfysize=1.7in
\begin{center}
\epsffile{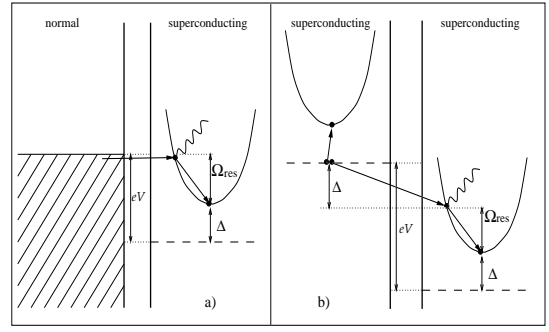}
\end{center}
\caption{The schematic diagram for the dip features in  SIN and SIS
tunneling conductances (figures a and b, respectively). For  SIN
tunneling, which measures the fermionic DOS, the electron which tunnels from a
normal metal can emit a propagating magnon if the voltage $eV=\Delta +\Delta
_{s}$. After emitting a magnon, the electron falls to the bottom of the
band. This leads to a sharp reduction of the current and produces a drop in $%
dI/dV$. For SIS tunneling, the physics is similar, but one first has to
break an electron pair, which costs energy $2\Delta $ (taken from Ref. 
\protect\onlinecite{acsin}). }
\label{sins}
\end{figure}

Consider first how the singularity in ${\tilde{\Sigma}}(\omega )$ at $\Omega
_{\mathrm{t}}$ affects $S(\omega )$. From a physical perspective, we would
expect a singularity in $S(\omega )$ at $\omega =\Delta +\Omega _{\mathrm{t}%
}=2\Delta +\Delta _{s}$. Indeed, to get a nonzero SIS conductance, one has
to first break a Cooper pair, which costs an energy of $2\Delta $. After a
pair is broken, one of the electrons becomes a quasiparticle in a
superconductor and takes the energy $\Delta $, while the other tunnels. If
the tunneling voltage equals $\Delta +\Omega _{\mathrm{t}}$, the electron
which tunnels through a barrier has energy $\Omega _{\mathrm{t}}$, and can
emit a spin excitation and fall to the bottom of the band (see Fig.~\ref
{sins}). This behavior is responsible for the drop in $dI/dV$ and is
schematically shown in Fig.~\ref{FIGURE_SIS}. 

Consider this effect in more detail~\cite{acsin,acs_finger}. We first note
that $\omega =\Delta +\Omega_{\mathrm{t}}$ is special for Eq.~\ref{sis}
because both $dN(\Omega )/d\Omega $ and $N(\omega -\Omega )$ diverge at the
same energy, $\Omega =\Omega_{\mathrm{t}}$. Substituting the general forms
of $N(\omega )$ near $\omega =\Omega_{\mathrm{t}}$ and $\omega =\Delta $, we
obtain after simple manipulations that for a flat gap, $S(\omega)$ has a
one-sided divergence at $\omega = \Omega_{\mathrm{t}} + \Delta = 2 \Delta +
\Delta_s$. 
\begin{equation}
S_{\mathrm{sing}} (\epsilon)\propto \frac{\Theta (-\epsilon )}{\sqrt{%
-\epsilon }}
\end{equation}
where $\epsilon =\omega -(\Omega_{\mathrm{t}}+\Delta)$. This obviously
causes a divergence of the frequency derivative of $S(\omega)$ (i.e., of $%
d^{2}I/dV^{2}$). This behavior is schematically shown in Fig. \ref
{FIGURE_SIS}a. In Fig. \ref{FIGURE_SIS}b we present the results for $%
S(\omega )$ obtained by integrating theoretical $N(\omega )$ from 
Fig.\ref{Figure_DOS_schematic}b. We clearly see
that $S(\omega )$ and its frequency derivative are singular at $\omega
=2\Delta + \Delta_s$, in agreement with the analytical prediction.

For a quadratic variation of the gap near the maxima, calculations similar
to those for the SIN tunneling yield the result that $S(\omega )$ is
continuous through $2\Delta +\Delta _{s}$, but its frequency derivative
diverges as 
\begin{equation}
\frac{dS(\omega )}{d\omega }\propto P\int_{0}^{\Delta }\frac{dx}{(x|\log
x|)^{1/2}\left( x-\epsilon \right) }~\sim \frac{\Theta (-\epsilon )}{%
|\epsilon \log |\epsilon ||^{1/2}}.  \label{sis2}
\end{equation}
The singularity in the derivative implies that near $\epsilon =0$ 
\begin{equation}
S(\epsilon )=S(0)-C~\Theta (-\epsilon )\left( \frac{-\epsilon }{|\log
(-\epsilon )|}\right) ^{1/2},
\end{equation}
where $C>0$. This behavior is schematically presented in Fig. \ref
{FIGURE_SIS}d. We again see that the angular dependence of the gap softens
the strength of the singularity, but the singularity remains confined to a
single frequency $\omega =2\Delta +\Delta _{s}$.

In real materials, the singularity in $S(\omega )$ is softened and
transforms into a dip slightly below $2\Delta +\Delta_s$, and a hump at a
frequency larger than $2\Delta + \Delta_s $. The frequency $2\Delta +
\Delta_s$ roughly corresponds to a maximum of the frequency derivative of
the SIS conductance.

\subsection{Optical conductivity and Raman response}

Further phenomena sensitive to $\Omega _{\mathrm{t}}$ are the optical
conductivity, $\sigma (\omega )$ and the Raman response, $R(\omega )$. 
Both are proportional to the fully renormalized particle-hole polarization
bubble, but with different signs attributed to the bubble composed of
anomalous propagators. Specifically, after integrating in the particle-hole
bubble over $\varepsilon _{{\bf{k}}}$, one obtains 
\begin{eqnarray}
R(\omega ) &=&\mathrm{Im}\int d\omega ^{\prime }d\theta V^{2}(\theta )\Pi
_{r}(\theta ,\omega ,\omega ^{\prime })  \nonumber \\
\sigma (\omega ) &=&\mathrm{Re}\left( \frac{i}{\omega +i\delta }~\int
d\omega ^{\prime }d\theta \Pi _{\sigma }(\theta ,\omega ,\omega ^{\prime
})\right) ,  \label{rs}
\end{eqnarray}
where $V(\theta )$ is a Raman vertex which depends on the scattering
geometry \cite{girsh}, and 
\begin{equation}
\Pi _{r,\sigma }(\theta ,\omega ,\omega ^{\prime })=~\frac{{\tilde{\Sigma}}%
_{+}{\tilde{\Sigma}}_{-}+\alpha \Phi _{+}\Phi _{-}+D_{+}D_{-}}{%
D_{+}D_{-}(D_{+}+D_{-})}  \label{pi}
\end{equation}
Here $\alpha =-1$ for $\Pi _{r}$, and $\alpha =1$ for $\Pi _{\sigma }$.
Also, ${\tilde{\Sigma}}_{\pm }={\tilde{\Sigma}}\left( \omega _{\pm }\right) $
and $\Phi _{\pm }=\Phi \left( \omega _{\pm }\right) $, where $\omega _{\pm
}=\omega ^{\prime }\pm \omega /2$. We also introduced $D_{\pm }=(\Phi _{\pm
}^{2}-{\tilde{\Sigma}}_{\pm }^{2})^{1/2}$. Note that ${\tilde{\Sigma}}$ and $%
\Phi $ depend on $\omega $ and $\theta $.

In a superconducting gas, the optical conductivity vanishes identically for
any nonzero frequency due to the absence of a physical scattering between
quasiparticles in a gas. The presence of a superconducting condensate,
however, gives rise to a $\delta $ functional term in $\sigma $ at $\omega
=0 $: $\sigma (\omega )=\pi \delta (\omega )\int d\theta d\omega ^{\prime
}\Pi _{\sigma }(\theta ,0,\omega ^{\prime })$. This behavior is typical for
any BCS superconductor~\cite{schrieffer}. The behavior of $\sigma (\omega )$
for a d-wave gas with additional impurities, causing inelastic scattering,
is more complex and has been discussed in Ref.\onlinecite{Quinlan96}.

The form of the Raman intensity depends on the scattering geometry. For the
scattering in the $B_{1g}$ channel, the Raman vertex has the same angular
dependence as the $d$-wave gap, i.e., $V(\theta )\propto \cos \left( 2\theta
\right)$~\cite{Devereaux,girsh}. Straightforward computations then show that
at low frequencies, $R(\omega )\propto \omega ^{3}$\cite{Devereaux}. For a
constant $V(\theta )$, we would have $R(\omega )\propto \omega $.

Near $\omega =2\Delta $, the $B_{1g}$ Raman intensity is singular. For this
frequency, both $D_{+}$ and $D_{-}$ vanish at $\omega ^{\prime }=0$ and $%
\theta =0$. This causes the integral for $R(\omega )$ to be divergent. The
singular contribution to $R(\omega )$ can be obtained analytically by
expanding in the integrand to leading order in $\omega ^{\prime }$ and in $%
\theta $. Using the spectral representation, we then obtain, for $\omega
=2\Delta +\delta $\cite{girsh} 
\begin{eqnarray}
R(\omega ) &=&\int_{0}^{\epsilon }d\Omega \int d{\tilde{\theta}}~\frac{1}{%
\sqrt{\Omega +a{\tilde{\theta}}^{2}}\sqrt{\delta -\Omega +a{\tilde{\theta}}%
^{2}}}  \nonumber \\
&&\times \frac{1}{(\sqrt{\Omega +a{\tilde{\theta}}^{2}}+\sqrt{\delta -\Omega
+a{\tilde{\theta}}^{2}})}  \label{gi}
\end{eqnarray}
where, as before, ${\tilde{\theta}}=\theta -\theta _{hs}$ For a flat band ($%
a=0$), $R(\omega )\propto Re[(\omega -2\Delta )^{-1/2}]$. For $a\neq 0$,
i.e., for a quadratic variation of the gap near its maximum, the 2d
integration in Eq.\ref{gi} is elementary, and yields 
$R(\omega )\propto \log |\omega -2 \Delta|$. 
At larger frequencies $R(\omega )$ gradually decreases.

\begin{figure}[tbp]
\epsfxsize=2.8in 
\epsfysize=1.7in
\begin{center}
\epsffile{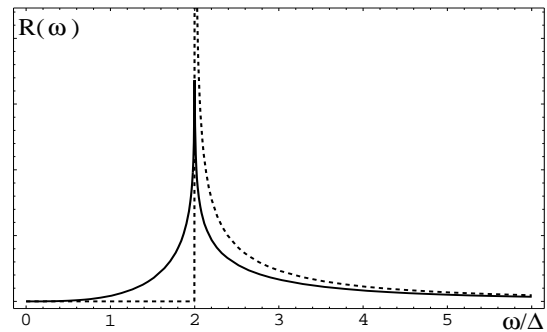}
\end{center}
\caption{The behavior of the Raman response in a BCS superconductor with a
flat gap (dashed line), and for a quadratic variation of the gap near its
maximum (solid line).}
\label{Figure_Raman_gas}
\end{figure}

The behavior of $R(\omega )$ in a d-wave gas is shown in Fig.\ref
{Figure_Raman_gas}. Observe that due to the interplay of numerical factors,
the logarithmic singularity shows up only in the near vicinity of $2\Delta $%
, while at somewhat larger $\omega $, the angular dependence of the gap
becomes irrelevant, and $R(\omega )$ behaves as $(\omega -2\Delta )^{1/2}$,
i.e. the same as for a flat gap~\cite{acg}. 

We now consider strong coupling effects. A nonzero fermionic self-energy
mostly affects the optical conductivity for the simple reason that it
becomes finite in the presence of spin scattering which can relax fermionic
momenta. For a momentum-independent gap, a finite conductivity emerges
above a sharp threshold. This threshold stems from the fact that at least
one of the two fermions in the conductivity bubble should have a finite ${%
\tilde \Sigma} ^{\prime \prime }$, i.e., its energy should be larger than $%
\Omega_{\mathrm{t}}$. Another fermion should be able to propagate, i.e., its
energy should be larger than $\Delta $. The combination of the two
requirements yields the threshold for $\sigma (\omega >0)$ at $2\Delta
+\Omega_{\mathrm{t}}$, i.e., at the same frequency where the SIS tunneling
conductance is singular. One can easily demonstrate that for a flat gap, the
conductivity behaves above the threshold as $\epsilon ^{1/2}/\log
^{2}\epsilon $, where $\epsilon =\omega -(\Delta +\Omega_{\mathrm{t}})=\omega
-(2\Delta +\Delta _{s})$. This singularity obviously causes a divergence of
the first derivative of the conductivity at $\epsilon =+0$.

\begin{figure}[tbp]
\epsfxsize=2.8in 
\epsfysize=1.7in
\begin{center}
\epsffile{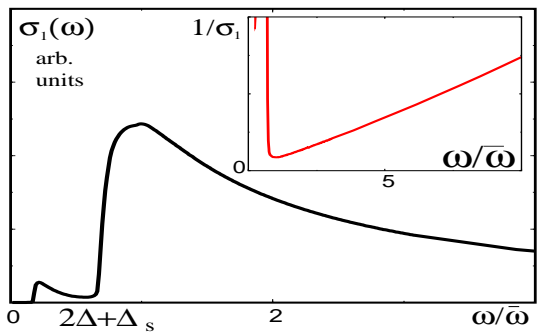}
\end{center}
\caption{The real part of the optical conductivity $\protect\sigma_1 (%
\protect\omega)$ at the lowest $T$ obtained using the self energy and the
pairing vertex from the solution of the Eliashberg equations for $\protect%
\lambda =1$. The onset of the optical response is $\protect\omega =2\Delta
+\Delta _{s}$. The contributions from nodal regions (not included in
calculations) yield a nonzero conductivity at all $\protect\omega$. Inset -
the behavior of the inverse conductivity vs frequency. (From Ref. ~ 
\protect\onlinecite{Abanov01}).}
\label{Figure_cond_strongc2}
\end{figure}

In Fig.\ref{Figure_cond_strongc2} we show the result for the conductivity
obtained by solving the set of coupled 
Eliashberg-type equations, Eqs.\ref{El1}-\ref{Elph}~\onlinecite{Abanov01,rob}.
 We see the expected singularity at $2\Delta
+\Delta _{s}$. The insert shows the behavior of the inverse conductivity $%
1/\sigma (\omega )$ Observe that $1/\sigma (\omega )$ is linear in $\omega $
over a rather wide frequency range~\cite{rob}.

For a true $d-$wave gap, the conductivity is finite for all frequencies
simply because the angular integration in Eq.~\ref{rs} involves the region
near the nodes, where ${\tilde \Sigma} ^{\prime \prime } $ is nonzero down
to the lowest frequencies. Still, the conductivity is singular at $\Omega_{%
\mathrm{t}}+\Delta $ as we now demonstrate. Indeed, as we already discussed,
at deviations from $\theta = \theta_{hs}$, where the gap is at maximum, both 
$\Delta $ and $\Delta _{s}$ decrease, hence $\Omega_{\mathrm{t}}(\theta
)=\Omega_{\mathrm{t}}-a{\tilde \theta}^{2}$, where ${\tilde \theta} = \theta
- \theta_{hs}$ and $a>0$. The singular pieces in ${\tilde \Sigma} (\omega)$
and $\Phi (\omega)$ then behave as $|\log (\Omega_{\mathrm{t}}-\omega
-a\theta ^{2})|$. Substituting these forms into the particle-hole bubble and
integrating over $\theta $, we find that the conductivity and its first
derivative are continuous at $\omega =2\Delta +\Delta _{s}$, but the second
derivative of the conductivity diverges as ${d^{2}\sigma }/{d\omega ^{2}}%
\propto 1/(|\epsilon |\log ^{2}\epsilon )$.
 We see that the singularity is weakened by the angular
dependence of the gap, but is still located exactly at $\Omega_{\mathrm{t}%
}+\Delta = 2\Delta +\Delta _{s}$.

The same reasoning can be applied to a region near $4\Delta$. The
singularity at $4\Delta$ is also weakened by the angular dependence of the
gap, but is not shifted and still should show up in the second derivative of
the conductivity.

For the Raman intensity, strong coupling effects are less relevant. First,
one can prove along the same lines as in previous subsections that the cubic
behavior at low frequencies for $B_{1g}$ scattering (and the linear behavior
for angular independent vertices), and the logarithmic singularity at $%
2\Delta $, are general properties of a $d-$wave superconductor, which
survive for all couplings. Thus, analogous to the density of states and the
SIS-tunneling spectrum, the Raman response below $2\Delta $ is not sensitive
to strong coupling effects. Second, near $\omega _{0}+\Delta $, singular
contributions which come from ${\tilde \Sigma} _{+}{\tilde \Sigma} _{-} $
and $\Phi _{+}\Phi _{-}$ terms in $\Pi _{r}$ in Eq.~\ref{rs} cancel each
other. As a result, for a flat gap, only the second derivative of $R(\omega
) $ diverges at $\Delta +\Omega_{\mathrm{t}}$. For a quadratic variation of
a gap near its maximum, the singularity is even weaker and shows up only in
the third derivative of $R(\omega )$. Obviously, this is a very weak effect,
and its experimental determination is difficult.

We now argue that measurements of the optical conductivity allow one not
only to verify the magnetic scenario, but also to determine both $\Delta
_{s} $ and $\Delta $ independently in the same experiment. In the magnetic
scenario, the fermionic self-energy is singular at two frequencies: at $%
\Omega _{\mathrm{t}}=\Delta +\Delta _{s}$, which is the onset frequency for
spin-fluctuation scattering near hot spots, and at $\omega =3\Delta $, where
fermionic damping near hot spots first emerges due to a direct four-fermion
interaction. Since in the spin-fluctuation mechanism, both singularities are
due to the same underlying interaction, their relative intensity can be
obtained within the model.

In general, the singularity at $3\Delta $ is much weaker at strong coupling,
and can be detected only in the analysis of the derivatives of the fermionic
self-energy. We recall that the singularity in ${\tilde{\Sigma}}(\omega )$
at $\Omega _{\mathrm{t}}$ gives rise to singularity in the conductivity at $%
\Delta +\Omega _{\mathrm{t}}$, while the $3\Delta $ singularity in ${\tilde{%
\Sigma}}(\omega )$ obviously causes a singularity in conductivity at $\omega
=4\Delta $. In addition, we expect a singularity in $\sigma (\omega )$ at $%
2\Omega _{\mathrm{t}}$, as at this frequency both fermions in the bubble
have a singular ${\tilde{\Sigma}}(\Omega _{\mathrm{t}})$.

\begin{figure}[tbp]
\epsfxsize=2.8in 
\epsfysize=1.7in
\begin{center}
\epsffile{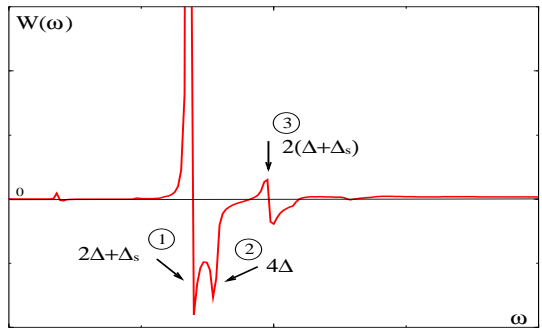}
\end{center}
\caption{ The calculated frequency dependence of $W(\protect\omega )=\frac{%
d^{2}}{d^{2}\protect\omega }[\protect\omega \mathrm{Re}[1/\protect\sigma (%
\protect\omega )]]$ at $T\rightarrow 0$. This quantity is a sensitive
measure of the fine structure in the optical response. The locations of the
extrema are: (1) $2\Delta +\Delta _{s}$, (2) $4\Delta $, (3) $2\Delta +2\Delta
_{s}$.  Observe that the maximum
shifts to a lower temperature, but the minimum remains at $2\Delta +\Delta
_{s}$. (From Ref. ~\protect\onlinecite{Abanov01}.)}
\label{Figure_d2sigdw2}
\end{figure}

For superconductors with pairing due to electron-phonon interaction the fine
structure of the optical conductivity has been analyzed by studying the
second derivative of conductivity via $W(\omega )=\frac{d^{2}}{d^{2}\omega }%
(\omega \mathrm{Re}\sigma ^{-1}(\omega ))$ which is proportional to $\alpha
^{2}(\omega )F(\omega )$ where $\alpha (\omega )$ is an effective
electron-phonon coupling, and $F(\omega )$ is a phonon DOS\cite{RMP-Carbotte}.
In Fig.\ref{Figure_d2sigdw2} we present the result of the strong coupling
calculations of $W(\omega )$~\cite{Abanov01}. There is a sharp maximum in $%
W(\omega )$ near $2\Delta +\Delta _{s}$, which is followed by a deep
minimum. This form is consistent with our analytical observation that for a
flat gap (which we used in our numerical analysis), the first derivative of
conductivity diverges at $\omega =2\Delta +\Delta _{s}$. At a finite $T$ (a
necessary attribute of a numerical solution), the singularity is smoothed,
and the divergence is transformed into a maximum. Accordingly, the second
derivative of the conductivity should have a maximum and a minimum near $%
2\Delta +\Delta _{s}$. The numerical analysis shows that the maximum shifts
to lower frequencies with increasing $T$, but the minimum moves very little
from $2\Delta +\Delta _{s}$, and is therefore a good measure of a magnetic
``fingerprint''.

Second, we note from Fig.\ref{Figure_d2sigdw2} that in addition to the
maximum and the minimum near $2\Delta +\Delta _{s}$, $W(\omega )$ has extra
extrema at $4\Delta $ and $2\Omega _{\mathrm{t}}=2\Delta +2\Delta _{s}$.
These are precisely the extra features that we expect: they are a primary
effect due to a singularity in ${\tilde{\Sigma}}(\omega )$ at $\omega
=3\Delta $ and a secondary effect due to a singularity in ${\tilde{\Sigma}}%
(\omega )$ at $\omega =\Omega _{\mathrm{t}}$. The experimental discovery of
these features will be a further argument in favor of spin-mediated pairing
and the applicability of the spin-fermion model.

\section{Comparison with the experiments on cuprates}

In this section we compare the theoretical results for the spin-fermion
model of the nearly antiferromagnetic Fermi liquid with the experimental
data for optimally doped members of the Bi$_{2}$Sr$_{2}$CaCu$_{2}$O$_{8}$
and YBa$_{2}$Cu$_{3}$O$_{7-y}$ families of cuprate superconductors. We make
the assumption that at this doping level, absent the influence of \ a
superstructure \ and imperfections that make the NMR lines in Bi$_{2}$Sr$%
_{2} $CaCu$_{2}$O$_{8}$ \ broad and difficult to interpret, \ the normal
state behavior of Bi$_{2}$Sr$_{2}$CaCu$_{2}$O$_{8}$ \ will resemble closely
that of YBa$_{2}$Cu$_{3}$O$_{7-y}$. This enables us to take the two input
parameters of the model from fits to NMR in the latter material. We then can
compare theory and experiment in the normal state and as $T\rightarrow 0$ in
the superconducting state. Finally, we discuss the general phase diagram of
\ the cuprates and the pseudogap physics of these materials.

\subsection{Parameters of the model}

The two input parameters of the theory are the coupling constant $\lambda $
and the overall energy scale ${\bar{\omega}}=4\lambda ^{2}\omega _{\mathrm{sf%
}}$. Alternatively, we can re-express $\lambda $ as $\lambda =3v_{F}\xi
^{-1}/(16\omega _{\mathrm{sf}})$ and use $v_{F}\xi ^{-1}$ and $\omega _{%
\mathrm{sf}}$ as inputs. The values of $\omega _{\mathrm{sf}}$ and $\xi $
can be extracted from the NMR measurements of the longitudinal and
transverse spin-lattice relaxation rates, and from neutron scattering data,
which measure $S({\bf{q}},\omega )\propto \omega
 /((1+({\bf{q}}-{\bf{Q}})^{2}\xi ^{2})^{2}+(\omega /\omega _{\mathrm{sf}})^{2})$. 
We will primarily rely on NMR data for near optimally doped
 YBa$_{2}$Cu$_{3}$O$_{6+\delta }$.
The NMR analysis\cite{Barzykin,nmr_extract} yields a moderately temperature
dependent $\omega _{\mathrm{sf}}$ and $\xi $ which take the values $\omega _{%
\mathrm{sf}}\sim 15-20\mathrm{meV}$ and $\xi \sim 2a$ in the vicinity of $T_{%
\mathrm{cr}}$, which for slightly overdoped materials will be close to $T_{c}
$. The neutron data \ from inelastic scattering (INS) experiments on the
normal state are more difficult to analyze because of the background which
increases the measured width of the neutron peak and because of the possible
influence of weak intrinsic inhomogeneities on a global probe such as INS.
The data show\cite{Bourges00} that the dynamical structure factor in the
normal state is indeed peaked at ${\bf{q}}={\bf{Q}}=(\pi /a,\pi /a)$,
and \ that the width of the peak increases with frequency and at $\omega =50%
\mathrm{meV}$ reaches $1.5$ of its value at $\omega =0$. A straightforward
fit to the theory yields $\omega _{\mathrm{sf}}\sim 35-40\mathrm{meV}$ and a
weakly temperature dependent $\xi \sim a$ which are, as expected, larger
than the $\omega _{\mathrm{sf}}$ and smaller than the $\xi $ values
extracted from NMR. We will be using $\omega _{\mathrm{sf}}\sim 20\mathrm{meV%
}$ and $\xi =2a$ for further estimates.

The value of the Fermi velocity can be obtained from the photoemission data
on Bi$_{2}$Sr$_{2}$CaCu$_{2}$O$_{8}$ at high frequencies, where the
self-energy corrections to the fermionic dispersion become relatively minor.
\ We note that because \ of problems related to the surface reconstruction \
in YBa$_{2}$Cu$_{3}$O$_{6+\delta }$ the vast majority of high quality
angular resolved photoemission spectroscopy (ARPES) \ experiments are
performed on Bi$_{2}$Sr$_{2}$CaCu$_{2}$O$_{8}$, the material where there are
much less reliable NMR experiments in part because of superstructure induced
line broadening. The three groups that report MDC (momentum distribution
curve) data for Bi$_{2}$Sr$_{2}$CaCu$_{2}$O$_{8}$ and momenta along the zone
diagonal \cite{Kaminski00,johnson1,bogdanov1} all agree that the value of the
bare Fermi velocity along the diagonal 
(determined at higher energies where mass renormalization is assumed to be small) is rather high: $2.5-3\mathrm{eV}$\AA , or $0.7-0.8%
\mathrm{eV}a$ where $a\simeq 3.8$\AA\ is the $\mathrm{Cu-Cu}$ distance. We
can use the $t-t^{\prime }$ tight binding model \ for the electronic
dispersion to relate this velocity with that at hot spots. Using the
experimental facts that the Fermi surface is located at ${\bf{k}}\approx
(0.4\pi /a,0.4\pi /a)$ for momenta along the zone diagonal and at ${\bf{k}}%
\approx (\pi /a,0.2\pi /a)$ for $k_{x}$ along the zone boundary, we find $%
t\sim 0.2-0.25\mathrm{eV}$, $t^{\prime }\approx -0.35t$ and $\mu \approx
-1.1t$. These numbers agree with those used in numerical studies \cite
{dagotto}. The hot spots are located at ${\bf{k}}_{\mathrm{hs}}=(0.16\pi
,0.84\pi )$ and symmetry related points, and the velocity at a hot spot is
then approximately half of that along zone diagonal. This yields $%
v_{F}\approx 0.35-0.4\mathrm{eV}a$.

Combining the results for $v_{F}$, $\xi $ and $\omega _{\mathrm{sf}}$, we
obtain $\lambda \sim 1.5-2$. This in turn yields ${\bar{\omega}}\sim 0.2-0.3%
\mathrm{eV}$. As an independent check of the internal consistency of these
estimates, we compare theoretical and experimental values of the resonance
spin frequency $\Delta _{s}$. 
As we said at the end of Section. 4, 
 $\Delta _{s}\simeq 0.2{\bar{\omega}}$ for $%
\lambda =2$. Substituting the value of ${\bar \omega}$, we obtain
$\Delta_s$  close to the experimental value of $40\mathrm{meV}$. 
  A smaller $\omega _{\mathrm{sf}}=15\mathrm{%
meV}$\textrm{\ }would require a slightly larger $\lambda $, but variations
of this magnitude are certainly beyond the quantitative accuracy of our
theory.

\begin{figure}[tbp]
\epsfxsize=\columnwidth 
\begin{center}
\epsffile{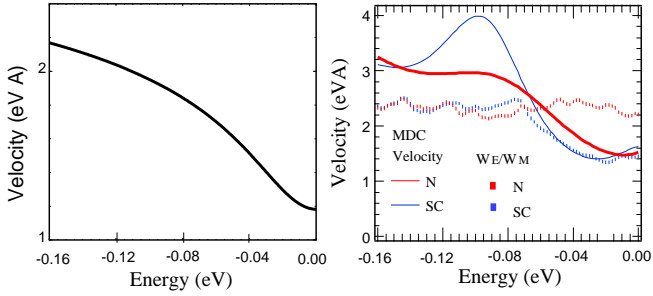}
\end{center}
\caption{a. The theoretical result for the effective velocity of the
quasiparticles $v_{F}^{\ast }=v_{F}/(1+\partial \Sigma ^{\prime }(\protect%
\omega )/\partial \protect\omega )$. For definiteness we used $\protect%
\omega _{sf}=20meV$, $\protect\lambda =1.7$ and bare velocity $v_{F}=3eVA$
along the diagonal. b. Experimental result for the effective velocity,
extracted from the MDC dispersion \protect\cite{Kaminski00} along the zone
diagonal. Observe the bump in the frequency dependence of the velocity at $%
70-80meV$ in the data and at about $3-4\protect\omega _{sf}$ in the theory.}
\label{velocity}
\end{figure}

Away from hot spots, the effective coupling decreases as $\lambda ({\bf{k}}%
)=\lambda /(1+(\delta k\xi )^{2})^{1/2}$ where $\delta k$ is the momentum
deviation from a hot spot along the Fermi surface. The largest $\delta k\xi $
is for ${\bf{k}}$ vectors along the zone diagonals. At optimal doping,
ARPES data yield $\delta k^{\mathrm{\max }}\sim 0.2\pi /a\approx 0.6/a$~\cite
{valla}. We see that $\lambda $ is reduced by at most $1.7$ as one moves
from hot spots to the zone diagonal. A prediction of the model is that $%
\omega _{\mathrm{sf}}(k)$ increases at deviations from hot spots. This
increase, however, should be at least partly compensated by the fact that $%
\omega _{\mathrm{sf}}\propto \sin \phi _{0}$, where $\phi _{0}$ is the angle
between Fermi velocities at ${\bf{k}}$ and ${\bf{k}}+{\bf{Q}}$, with \ 
$\phi _{0}\simeq $ $\frac{\pi }{2}$ in the vicinity of hot spots. $\phi _{0}$
tends to $\pi $ as ${\bf{k}}$ approaches the zone diagonal, and this 
\textit{reduces} $\omega _{\mathrm{sf}}$. In view of this competing effect
which we cannot fully control, we believe that the effective $\omega _{%
\mathrm{sf}}({\bf{k}})$ can best be obtained from the fit to the
photoemission data, particularly from the MDC measurements of the electronic
dispersion $\omega +\Sigma ^{\prime }(\omega )=\epsilon _{{\bf{k}}}$. In
Fig~\ref{velocity} we compare our $(1+\partial \Sigma ^{\prime }(\omega
)/\partial \omega )$ with the measured variation of the effective velocity $%
v_{\mathrm{F}}(\omega )$ of the electronic dispersion along zone diagonal~ 
\cite{Kaminski00}. We see that the theoretical dispersion has a bump at $%
\omega \sim 3\omega _{\mathrm{sf}}({\bf k}_{\mathrm{diag}})$. The experimental
curves look quite similar and show a bump at $\sim 70-80\mathrm{meV}$~\cite
{Kaminski00,johnson1,bogdanov1}. This yields
 $\omega _{\mathrm{sf}}({\bf k}_{\mathrm{%
diag}})\sim 25\mathrm{meV}$, a value only slightly larger than that near hot
spots.

Note in passing that although $\delta k\xi $ does not vary much when $k$
moves along the Fermi surface, the fact that the Fermi velocity is fairly
large implies that along the zone diagonal, $\epsilon _{{\bf{k}}_{\mathrm{F%
}}{\bf{+Q}}}$ is roughly $\sqrt{2}v_{\mathrm{F}}0.2\pi /a\approx 0.8eV$,
i.e., it is comparable to the bandwidth. This implies that the Fermi-surface
is very different from the near-perfect square that one would obtain for
only nearest neighbor hopping. Furthermore, the fact that the Fermi velocity
is large implies the physics at energies up to few hundred $\mathrm{meV}$ is
confined to the near vicinity of the Fermi surface, when one can safely
expand $\epsilon _{{\bf{k}}}$ to linear order in ${\bf{k-k}}_{\mathrm{F}%
} $. Finally, van-Hove singularities (which we neglected) do play some role 
\cite{onufr,katanin} but as $\epsilon _{\left( 0,\pi /a\right) }\approx
0.34t\sim 85\mathrm{meV} \gg \omega _{\mathrm{sf}}$, 
we expect that the van-Hove
singularity softens due to fermionic incoherence and should not
substantially affect the physics. The value of $\epsilon _{\left( 0,\pi
/a\right) }$ might however be affected by an additional bilayer splitting
which moves one of the bands closer to ${\bf{k}}=\left( 0,\pi /a\right) $.

Finally, in the analysis of the spin-fermion model we have neglected the
temperature dependence of the correlation length, and thus of $\omega _{%
\mathrm{sf}}$. Fits to NMR experiments on the near optimally doped member of
the YBa$_{2}$Cu$_{3}$O$_{6+\delta }$ family show that at $T_{\mathrm{cr}%
}\simeq T_{\mathrm{c}}$ both $\omega _{\mathrm{sf}}$ and $\xi $ display mean
field behavior with $\xi ^{-2}\simeq \xi _{0}^{-2}\left( 1+T/T_{0}\right)$
 and $\omega _{\mathrm{sf}}\xi ^{2}\simeq 70\mathrm{meV}$ almost independent on $T$. 
 
From a theoretical perspective, the leading temperature dependence of $\xi $ arises
from an interaction between spin-fluctuations and near the critical point in
two dimensions has the form $\xi ^{-2}(T)=\xi ^{-2}(T=0)+B\Gamma _{4}T\log T$
where $B=O(1)$, and $\Gamma _{4}$ is the effective four boson interaction
that is made out of fermions~\cite{Hertz,Millis93,ac2}. The $T\log T$ factor
is the universal contribution from the bosonic loop, confined to momenta
near ${\bf{Q}}$. The four-boson interaction has two contributions: one
comes from low-energy fermions and is universal; the other comes from
high-energy fermions and depends on the fermionic bandwidth $W$. One can
show \cite{ac2} that the temperature correction to $\xi $ involves only the
high-energy part of the interaction (this is what we labeled as $\Gamma _{4}$%
). The magnitude of $\Gamma _{4}$ can be easily estimated to be ${\bar{g}}%
^{2}/W^{3}$. Accordingly, the temperature correction to $\xi $ scales as $%
\frac{T}{{\bar{g}}}({\bar{g}}/W)^{3}$. As we have remarked, the theory is
universal as long as ${\bar{g}}\ll W$. In this limit, the temperature
dependence of $\xi $ is obviously small and can be neglected. This is what
we will do. Notice however that in the opposite limit, when ${\bar{g}}\gg W$%
, the full four-boson interaction differs from the lowest order term in ${%
\bar{g}}$ and might be estimated within an RPA-type summation. Estimates
show that in this limit, the full $\Gamma _{4}$ is fully determined within
the low-energy sector and scales as $O(1/J)$ where $J\sim W^{2}/{\bar{g}}$
is the magnetic exchange integral. This in turn yields a much stronger
temperature dependence of $\xi $: $\xi ^{-2}(T)-\xi ^{-2}(T=0)\sim (T/J)\log
T$. This result is similar to that obtained using a nonlinear $\sigma -$
model approach to near antiferromagnetism~\cite{subir_2,scs,lavagna}. The
agreement becomes obvious in the limit of a large spin-fermion interaction
(which, we recall, is the Hubbard $U$ if we derive the spin-fermion model
within the RPA); double occupancy is energetically unfavorable and the spin
susceptibility obeys the constraint $\int d^{2}qd\omega \chi ({\b q},i\omega
)\propto 1-x$. This is equivalent to imposing a constraint on the length of
the spin field in the nonlinear $\sigma $-model~\cite{subir_2,scs,lavagna}.

\subsection{The normal state}

In this section we compare the experimental and theoretical forms of the
fermionic spectral function and optical conductivity in the normal state. No
free parameters remain, since those which are needed to specify the model
completely have been taken from NMR and ARPES experiments. The discussion
will follow Ref. ~\onlinecite{Abanov01advphys}.

\subsubsection{The spectral function:}

The quasiparticle spectral function at various momenta is measured in angle
resolved photoemission experiments. In a sudden approximation (an electron,
hit by light, leaves the crystal without further interactions with other
electrons and without paying attention to selection rules for the optical
transition to its final state), the photoemission intensity is given by $I_{%
{\bf{k}}}(\omega )=A_{{\bf{k}}}(\omega )n_{F}(\omega )$ where $n_{F}$ is
the Fermi function and $A_{{\bf{k}}}(\Omega )=(1/\pi )\left| \mathrm{Im}G(%
{\bf{k}},\Omega )\right| $ is the spectral function.

\begin{figure}[tbp]
\epsfxsize=2.8in 
\epsfysize=3.0in
\begin{center}
\epsffile{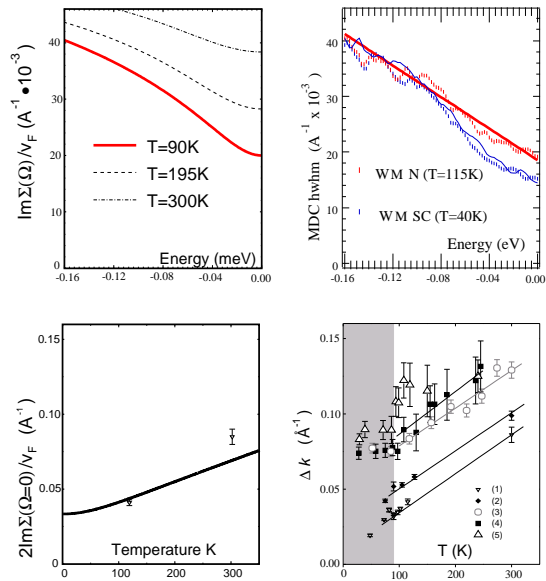}
\end{center}
\caption{A comparison between the theoretical results \ of the spin fermion
model and the photoemission MDC data. For the Lorentzian line-shape of the
MDC signal, observed in experiments, the MDC hwhm equals to $\Sigma ^{\prime
\prime }/v_{F}$. Upper panel - the results for the MDC hwhm vs frequency at
a given $T$. The experimental results are taken from\protect\onlinecite{Kaminski00}.
Lower panel - the MDC fwhm vs temperature at $\Omega =0$. The experimental
results (right figure and the points on the left figure) are taken from 
\protect\onlinecite{johnson1}. The figure is taken from \protect\cite
{Abanov01advphys}.}
\label{MDC_exp}
\end{figure}

We first use our form of the fermionic self-energy to fit MDC data which
measure the width of the photoemission peak as a function of ${\bf{k}}$ at
a given frequency. In Fig.~\ref{MDC_exp} we compare the theoretical results
for $\Delta k=\Sigma ^{\prime \prime }({\bf{k}},\Omega )/v_{\mathrm{F}}$
with the measured $\Delta k$ versus frequency at $T\sim 100\mathrm{K}$~\cite
{Kaminski00} and versus temperature at $\Omega \rightarrow 0$~\cite{johnson1}.
We used $\lambda =1.7$ and $\omega _{\mathrm{sf}}=20meV$. The slope of $%
\Delta k$ is chiefly controlled by $\lambda $. We obtain rather good
agreement with the data, both for the frequency and temperature dependence
of the self-energy. On the other hand, the magnitude of our $\Sigma ^{\prime
\prime }$ is smaller than that found in the experimental data. To account
for the values of $\Delta k$, we had to \textit{add} a constant of about $70%
\mathrm{meV}$ to $\Sigma ^{\prime \prime }$. The origin of this constant is
unclear and explaining it is presently a challenge to the theory.
 It may be the effect of elastic scattering by impurities\cite{va}, but 
the large
value of this constant implies that it is more likely the contribution from
scattering channels that we ignored. It is essential, however, that the
functional dependence of $\Sigma ^{\prime \prime }(\Omega ,T)$ can be
captured in \ the spin-fluctuation approach.

\begin{figure}[tbp]
\epsfxsize=\columnwidth 
\begin{center}
\epsffile{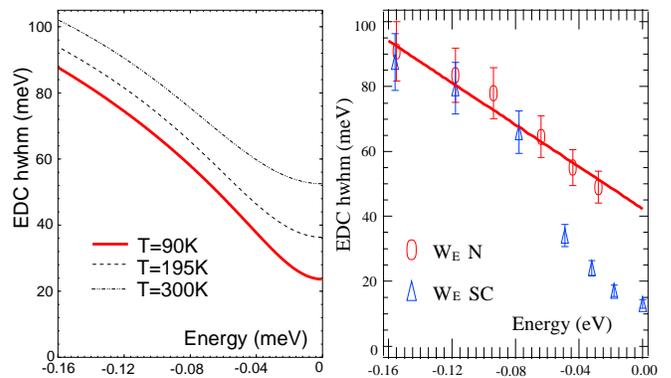}
\end{center}
\caption{A comparison of the theoretical result for $\Sigma ^{\prime \prime
}(\Omega )/(1+\Sigma ^{\prime }(\Omega )/\Omega )$ with the experimental
hwhm of the EDC dispersion from \protect\onlinecite{Kaminski00}. The figure is taken
from \protect\onlinecite{Abanov01advphys}. }
\label{EDC_exp}
\end{figure}

In Fig~\ref{EDC_exp} we present the results for the half width at half
maximum of the EDC (energy distribution curve) which measures fermionic $I_{%
{\bf{k}}}(\Omega )=A_{{\bf{k}}}(\Omega )n_{F}(\Omega )$ as a function of
frequency at a given ${\bf{k}}$. For a Lorentzian line-shape, the EDC hwhm
is given by $\Sigma ^{\prime \prime }(\Omega )/(1+\Sigma ^{\prime }(\Omega
)/\Omega )$. The data are taken at $T=115\mathrm{K}$~\cite{Kaminski00}. We see
that the theoretical slope agrees \ reasonably well with the experimental
one. The visible discrepancy is most likely associated with the fact that
the experimental line-shape is not a perfect Lorentzian.

\subsubsection{The optical conductivity}

\begin{figure}[tbp]
\epsfxsize=\columnwidth 
\begin{center}
\epsffile{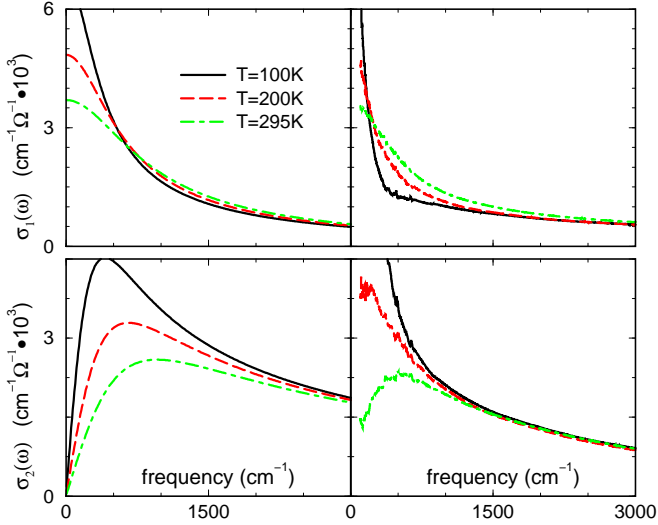}
\end{center}
\caption{The theoretical and experimental results for the real and imaginary
parts of optical conductivity. The data are from \protect\onlinecite{tu}. The
figure is taken from \protect\onlinecite{Abanov01advphys}. }
\label{conductivity_exp}
\end{figure}

In Fig.\ref{conductivity_exp} we compare the theoretical results for the
conductivity\cite{Abanov01} with the experimental data for $\sigma _{1}$ and 
$\sigma _{2}$ at different temperatures\cite{tu}. The theoretical results
are obtained using the same $\lambda =1.7$ and $\omega _{\mathrm{sf}}=20%
\mathrm{meV}$ as in the fit to the photoemission data along zone diagonals.
Changing $\lambda $ affects the ratio $\sigma _{2}/\sigma _{1}$ at high
frequencies, but does not change the functional forms of the conductivities 
\cite{Abanov01}. The value of the plasma frequency was adjusted to match the
d.c. conductivity and $\Sigma ^{\prime \prime }$ extracted from the MDC
photoemission data using $v_{F}\sim 3\mathrm{eV}$\AA . This adjustment
yields $\omega _{\mathrm{pl}}\sim 20000\mathrm{cm}^{-1}$. This value is
somewhat larger than  $\omega _{\mathrm{pl}%
}\sim 16000\mathrm{cm}^{-1}$ 
obtained experimentally by integrating $\sigma _{1}
$ up to about $2-2.5\mathrm{eV}$\cite{basov,tu,qui},
 however it agrees with the theoretical
result~ \cite{Abanov01} that the sum rule for $\sigma _{1}(\omega )$ is
exhausted only at extremely high frequencies of about $50{\bar{\omega}}$, that are much larger than $2\mathrm{eV}$. 
We see that theoretical calculations of $\sigma _{1}(\omega )$ and $\sigma
_{2}(\omega )$ capture the essential features of the measured forms of the
conductivities. In particular, the curves of $\sigma _{1}$ at different
temperatures cross such that at the lowest frequencies, the conductivity
decreases with $T$ while at larger frequencies it increases with $T$, a
behavior seen in Ref.\onlinecite{timusk,qui}. The imaginary part of conductivity
decreases with $T$ at any frequency, and the peak in $\sigma _{2}(\omega )$
increases in magnitude and shifts to lower $T$ with decreasing $T$~\cite
{tu,qui,orenst}. At frequencies above $1500\mathrm{cm}^{-1}$ both $\sigma
_{1}$ and $\sigma _{2}$ depend weakly on $T$ and are comparable in amplitude.

\begin{figure}[tbp]
\epsfxsize=\columnwidth 
\begin{center}
\epsffile{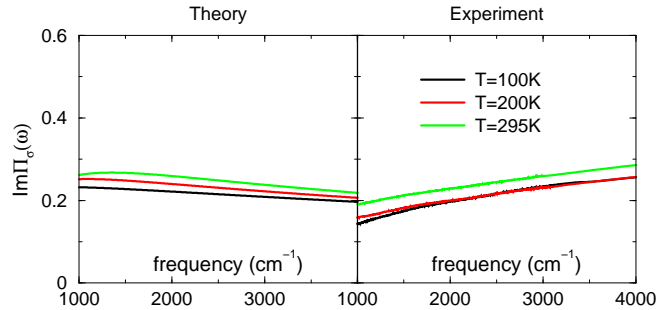}
\end{center}
\caption{The theoretical and experimental results for $\Pi_\protect\protect%
\sigma^{\prime \prime} (\protect\omega) = 4\protect\pi \protect\sigma_1 
\protect\omega/\protect\omega^2_{pl}$ (from \protect\onlinecite{Abanov01advphys}).
The data are from \protect\onlinecite{tu}. }
\label{pol}
\end{figure}

To make the comparison more quantitative, in Fig~\ref{pol} we present
experimental and theoretical results for the imaginary part of the full
particle-hole polarization bubble $\Pi _{\sigma }^{\prime \prime }(\omega
)=4\pi \sigma _{1}\omega /\omega _{\mathrm{pl}}^{2}$. Theoretically, at $T=0 
$, $\Pi _{\sigma }^{\prime \prime }(\omega )$ saturates at a value of about
0.2 \textit{independently} of $\lambda $ and remains almost independent of
frequency over a very wide frequency range~\cite{Abanov01}. The experimental
data also clearly show a near saturation of $\Pi _{\sigma }^{\prime \prime }$
at a value close to $0.2$.

\begin{figure}[tbp]
\epsfxsize=\columnwidth 
\begin{center}
\epsffile{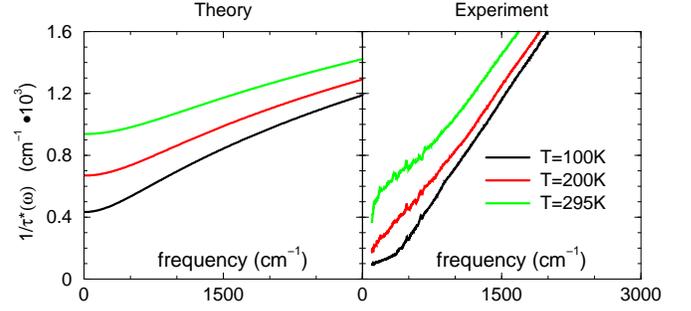}
\end{center}
\caption{The theoretical and experimental results for $1/\protect\tau^* = 
\protect\omega Re \protect\sigma/Im \protect\sigma$ (from \protect\cite
{Abanov01advphys}). The data are from \protect\onlinecite{tu}. }
\label{tau_star}
\end{figure}

The agreement between  theory and  experiment is, however, not a
perfect one. In Fig.~\ref{tau_star} we show theoretical and experimental
results for $1/\tau ^{\ast }=\omega \sigma _{1}/\sigma _{2}$. The advantage
of comparing $1/\tau ^{\ast }$ is that this quantity does not depend on the
unknown plasma frequency. We see that while both experimental and
theoretical curves are linear in frequency, the slopes are off roughly by a
factor of $3$. This discrepancy is possibly related to the fact that in the
spin-fermion model, $\Pi _{\sigma }^{\prime }(\omega )$ at high enough
frequencies is roughly 3 times larger than $\Pi _{\sigma }^{\prime \prime
}(\omega )$~\cite{Abanov01}, and hence $\sigma _{2}/\sigma _{1}\sim 3$,
whereas experimentally $\sigma _{1}$ and $\sigma _{2}$ are comparable in
magnitude at high frequencies. The discrepancy in $1/\tau ^{\ast }$
indicates that either the averaging over the Fermi surface, vertex
corrections inside a particle-hole bubble, or RPA-type corrections to the
conductivity\cite{ioffe_larkin} play some role. Still, Figs. \ref
{conductivity_exp} and \ref{pol} indicate that the general trends of the
behavior of the conductivities near optimal doping are reasonably well
captured within the spin-fluctuation approach.

\subsection{The superconducting state}

In this section, we apply our results from Section 5 to cuprates and examine
to what extent the ``fingerprints'' of spin-fluctuation pairing have been
detected in experiments on optimally doped high $T_{c}$ materials.

\subsubsection{The spin susceptibility}

The major prediction of the spin fermion model for the spin susceptibility in
 the superconducting state is the emergence of the resonance
 peak in $\chi''({\bf Q},\omega)$
at $\omega =\Delta_s$ if $\Delta_s < 2 \Delta$. The magnitude $\Delta_s$ is
fully determined within the theory and is chiefly set by the magnitude of the 
superconducting gap as well as the energy scale of magnetic fluctuations 
in the normal state,  $\omega_{\rm sf}$. For small doping concentration
 $\Delta_s \propto \xi^{-1}$ must decrease as one approaches the antiferromagnetic
 state. The resonance mode is confined to a small region in momentum space
 (where it  is of high intensity). For momenta away from ${\bf Q}$ and its close
 vicinity, magnetic excitations couple to gapless, nodal quasiparticles and become 
overdamped, eliminating the resonance mode.

\begin{figure}[tbp]
\epsfxsize=3.6in 
\epsfxsize=3.4in 
\epsfysize=4.5in
\begin{center}
\epsffile{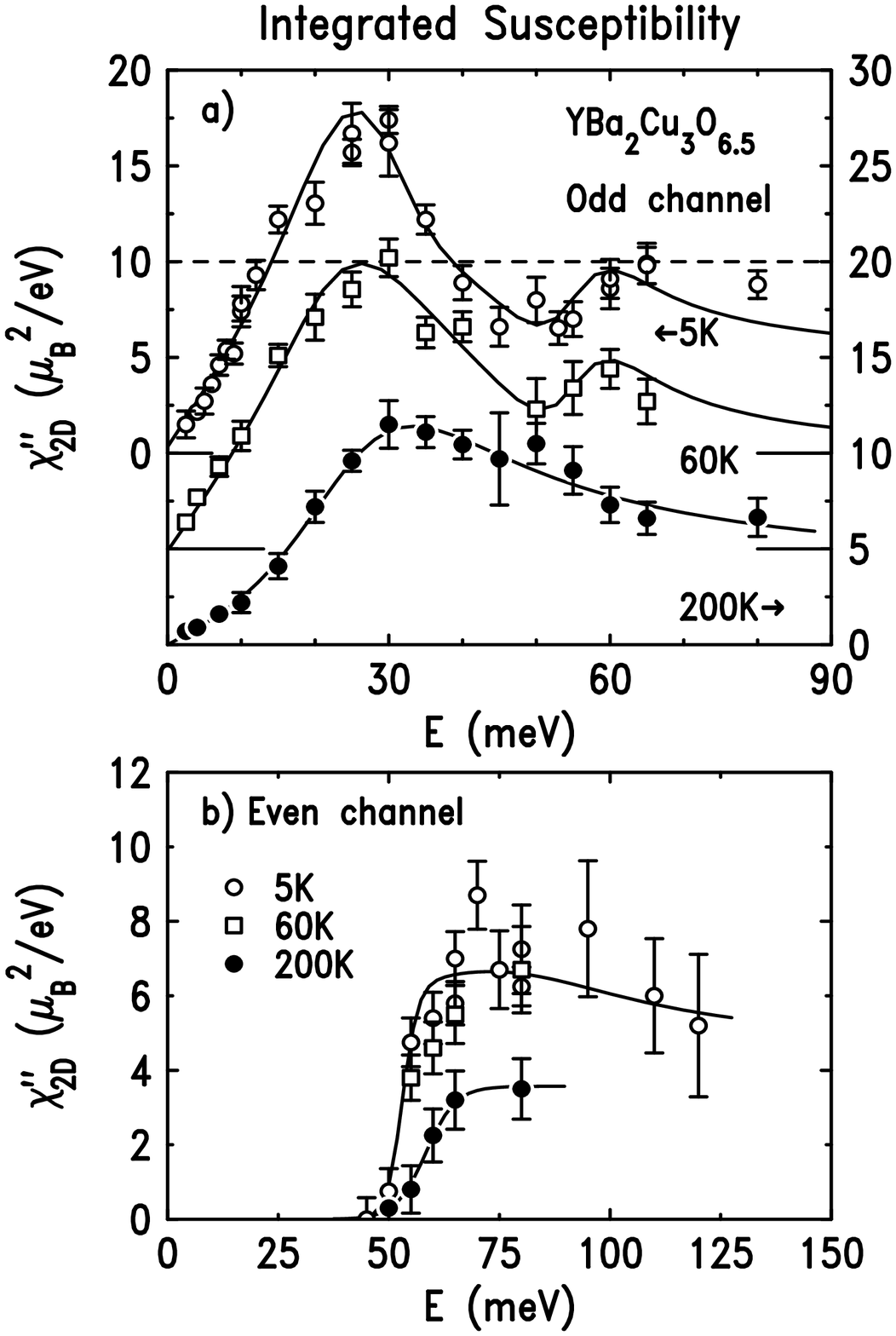}
\end{center}
\caption{Inelastic neutron scattering intensity for momentum 
${\bf{Q}}=\left( \protect\pi ,\protect\pi \right) $ as function of frequency for 
\textrm{YBa}$_{2}\mathrm{Cu}_{3}\mathrm{O}_{6.5}$. 
Data from Ref.~\protect\onlinecite{neutrons}.}
\label{Figure_resonance_exp}
\end{figure}

In Fig.\ref{Figure_resonance_exp} we show representative experimental data
for $\chi ^{\prime \prime }({\bf{Q}},\omega )$ showing the resonance peak
at $\omega \approx 41$ $\mathrm{meV}$ for optimally doped $\mathrm{YBa}_{2}%
\mathrm{Cu}_{3}\mathrm{O}_{6.9}$\cite{Bourges99}. As noted earlier, the
position of the peak is consistent with the prediction of the spin fermion
model. Similar behavior is found in $\mathrm{Bi}_{2}$Sr$_{2}$CaCu$_{2}$O$%
_{8} $\cite{neutrons2}; here the peak is at $43$ $\mathrm{meV}$. With
underdoping, the measured resonance energy goes down~\cite{neutrons,dai}. In
strongly underdoped $\mathrm{YBa}_{2}\mathrm{Cu}_{3}\mathrm{O}_{6.6}$, it is
approximately $25$ $\mathrm{meV}$\cite{neutrons}. The existence of the peak
and the downturn with underdoping agree with the predictions of the
spin-fluctuation theory. Further, the measured amplitude of $\ \chi ^{\prime
\prime }({\bf{Q}},\omega )$ decreases above the peak, but increases again
\ for $60-80$ $\mathrm{meV}$~\cite{dai,Bourges99}. This might indeed be a $%
2\Delta $ effect, which appears naturally within the model.

The full analysis of the resonance peak requires more care as (i) the peak
is only observed in two-layer materials, and only in the odd channel, (ii)
the momentum dispersion of the peak is more complex than that for magnons~ 
\cite{Bourges00}, (iii) the peak broadens with underdoping~\cite
{neutrons,dai}, and (iv) in underdoped materials, the peak emerges at the
onset of the pseudogap and only sharpens up at $T_{c}$~\cite{dai,Bourges99}.
\ All these features, already present on the level of a weak coupling~\
approach\cite{Ding,Norman_1,weakcoul_neutr}, \ have been explained within
the spin-fermion model\cite{oleg_1,Abanov01epl}. The broadening of the peak
was recently studied in detail in Ref.~\cite{lee_brink}. The explanation of
these effects, however, requires careful analysis of the details of the
electronic structure and is beyond the scope of this Chapter.

\begin{figure}[tbp]
\epsfxsize=3.4in 
\epsfysize=4.0in
\begin{center}
\epsffile{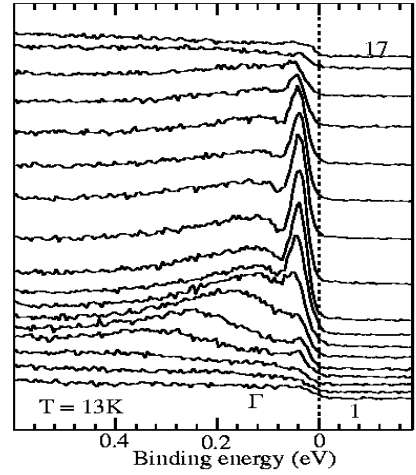}
\end{center}
\caption{ARPES spectrum  for near optimally doped \textrm{Bi}$2212$ for
momenta close to the hot spots. Data from
 Ref.{\protect\onlinecite{Norman97}}.}
\label{Figure_ARPES_exp}
\end{figure}
\subsubsection{The spectral function:}

\begin{figure}[tbp]
\epsfxsize=2.8in 
\epsfysize=2.8in
\begin{center}
\epsffile{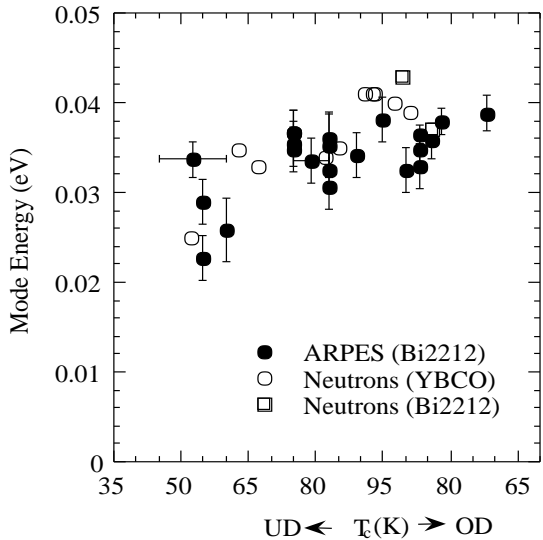}
\end{center}
\caption{The experimental peak-dip distance at various doping concentrations
compared with $\Delta _{s}$ extracted from neutron measurements. Data from
Ref.[30]. The theoretical result is presented in Fig.~\ref
{Figure_spec_sc_mom}.}
\label{Figure_Arpes_peak_dip}
\end{figure}

The predictions of our approach are a peak-dip 
structure of the spectral function, with a weakly dispersing peak at $\omega \approx \Delta$ and 
a peak-dip distance $\approx \Delta_s$. On the other hand  we expect a broad incoherent peak which 
disperses like $\epsilon_{\bf k}^2/\bar{\omega}$.
In Fig.\ref{Figure_ARPES_exp} we present ARPES data for near optimally doped 
$\mathrm{Bi}2212\mathrm{\ }$ with $T_{c}=87\mathrm{K}$ for momenta near a
hot spot~\cite{Norman97}. The intensity displays the predicted peak/dip/hump
structure. A sharp peak is located at $\sim 40\mathrm{meV}$, and the dip is
at $80\mathrm{meV}$ such that the peak-dip distance is $42\mathrm{meV}$\cite
{Norman97}. In the spin-fluctuation theory, the peak-dip distance is the
energy of the INS resonance peak frequency~\cite{Ding,ac}. The neutron
scattering data on $\mathrm{Bi}2212$ with nearly the same $T_{c}=91\mathrm{K}
$ yield~\cite{neutrons2} $\Delta _{s}=43\mathrm{meV}$, in excellent
agreement with this prediction. 
Furthermore, with underdoping, the peak-dip energy difference decreases and,
within error bars, remains equal to $\Delta _{s}$. This behavior is
illustrated in Fig.\ref{Figure_Arpes_peak_dip}. 

\begin{figure}[tbp]
\epsfxsize=2.8in 
\epsfysize=2.8in
\begin{center}
\epsffile{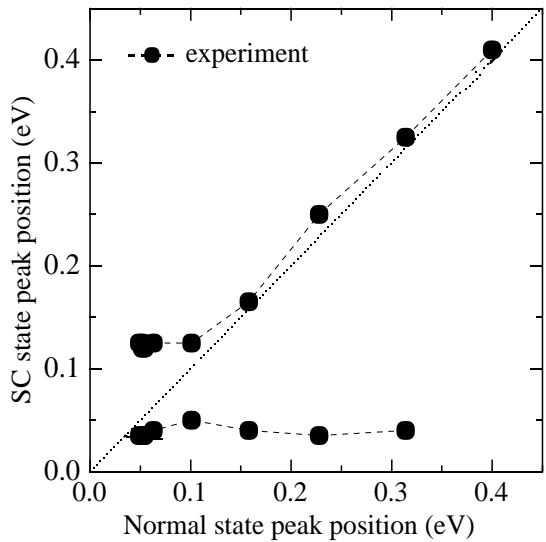}
\end{center}
\caption{The dependence of the experimental peak (flat curve) and hump
(dispersing curve) positions on the deviation from the Fermi surface. The
hump disperses with ${\bf{k}}-{\bf{k}}_{\mathrm{F}}$ (dotted line) and
eventually recovers the position of the broad maximum in the normal state,
while the peak position changes little with the deviation from ${\bf{k}}_{%
\mathrm{F}}$. Data from  Ref.{\protect\onlinecite{Norman97}}. The theoretical result is presented in Fig.~%
\ref{Figure_spec_sc_mom}.}
\label{Figure_ARPES_awaykF}
\end{figure}

In Fig.\ref{Figure_ARPES_awaykF} we present experimental results for the
variation of the peak and hump positions with the deviation from the Fermi
surface. These show that the hump disperses with ${\bf{k}}-{\bf{k}}_{%
\mathrm{F}}$ and eventually recovers the position of the broad maximum in
the normal state. At the same time, the peak shows little dispersion, and
does not move further in energy than $\Delta +\Delta _{s}$. Instead, the
amplitude of the peak dies off as ${\bf{k}}$ moves away from ${\bf{k}}_{%
\mathrm{F}}$. This behavior is again fully consistent with the theoretical
predictions~\cite{acsin,Abanov01advphys}. 

We regard the presence of the dip at $\Delta +\Delta _{s}$, and the absence
of the dispersion of the quasiparticle peak as two major ``fingerprints'' of
strong spin-fluctuation scattering in the spectral density of cuprate
superconductors.

\subsubsection{The density of states:}

The fermionic DOS $N(\omega )$ is proportional to the dynamical conductance $%
dI/dV$ through a superconductor-insulator-normal metal (SIN) measured at $%
\omega =eV~$ where $V$ is the applied voltage\cite{mahan}. The key prediction 
of our approach is the occurrence of a dip in the DOS at an energy $\approx \Delta_s$
away from the peak at $\omega =\Delta$. The drop in the
DOS at $\Omega _{\mathrm{t}}=\Delta +\Delta _{s}$ from Eq.\ref{omegatsf} can
be understood in terms of SIN conductance as follows: when the applied
voltage, $V$, equals $\Omega _{\mathrm{t}}/e$ an electron that tunnels from a normal
metal can emit a spin excitation and fall to the bottom of the band, losing
its group velocity. This loss leads to a sharp reduction of the current and
produces a drop in $dI/dV$. This process is shown schematically in Fig\ref
{sins}.

SIN tunneling experiments have been performed on $\mathrm{YBCO}$ and $%
\mathrm{Bi}2212$ materials~\cite{fisher}. 
Similar results have been recently obtained by 
Davis et al.~\cite{davis}. At low and moderate frequencies, the SIN conductance displays a
behavior which is generally expected in a $d-$wave superconductor, i.e., it
is linear in voltage for small voltages, and has a peak at $eV=\Delta $
where $\Delta $ is the maximum value of the $d-$wave gap~\cite{fisher,davis} 
The value of $\Delta$ extracted from tunneling agrees well with the maximum
value of the gap extracted from ARPES measurements~\cite
{Fedorov99,Kaminski00}. At frequencies larger than $\Delta $, the measured
SIN conductance  displays an extra dip-hump feature which become
visible at around optimal doping, and grows in amplitude with underdoping~ 
\cite{fisher}. At optimal doping, the distance between the peak at $\Delta $
and the dip is around $40\mathrm{meV}$. This is consistent with $\Delta _{s}$
extracted from neutron measurements.

\subsubsection{SIS tunneling:}
The major prediction of the spin-fermion model for the SIS tunneling conductance, $S(\omega )$,  is the emergence of a singularity at $\omega=2\Delta +\Delta_s$.
As mentioned above, this singularity  is likely softened due to thermal excitations 
or non-magnetic scattering processes and
transforms into a dip slightly below $2\Delta +\Delta_s$, and a hump at a
frequency larger than $2\Delta + \Delta_s $. 
\begin{figure}[tbp]
\epsfxsize=\columnwidth 
\begin{center}
\epsffile{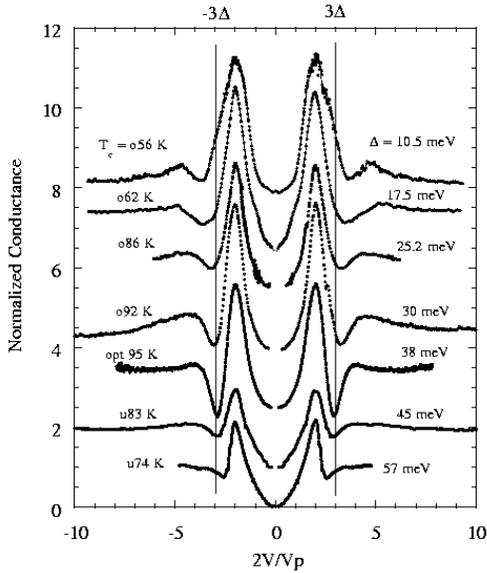}
\end{center}
\caption{SIS tunneling conductance normalized by $\Delta $ for $\mathrm{Bi}%
2212$ materials ranging from overdoped (top curves) to underdoped (bottom
curves) from Ref.{\protect\onlinecite{zasad}. The peak dip distance increases for
increasing doping and saturates at around }$3\Delta $ as expected in our
theory. The corresponding theoretical result is presented in Fig.~\ref
{FIGURE_SIS}.}
\label{Figure_SIS_Zasa}
\end{figure}
Recently, Zasadzinski \emph{et al.} obtained \ both new data and carefully
examined their previous SIS tunneling data for a set of $\mathrm{Bi}2212$
materials ranging from overdoped to underdoped\cite{zasad}. Their data,
presented in Fig.\ref{Figure_SIS_Zasa} show that in addition to the peak at $%
2\Delta $, the SIS conductance displays the dip and the hump at larger
frequencies. The distance between the peak and the dip (which approximately
equals $\Delta _{s}$ in the spin fluctuation model~\cite{ac,acs_finger}) is
close to $2\Delta $ in overdoped $\mathrm{Bi}2212$ materials, but goes down
with underdoping. Near optimal doping, this distance is around $40\mathrm{meV%
}$. For an underdoped, $T_{c}=74\mathrm{K}$, material, the peak-dip distance
is reduced to about $30\mathrm{meV}$. These results are in qualitative and
quantitative agreement with ARPES and neutron scattering data, as well as
with the theoretical estimates. The most important aspect is that with
underdoping, the experimentally measured peak-dip distance progressively
shifts down from $2\Delta $. This downturn deviation from $2\Delta $ is a
key feature of the spin-fluctuation mechanism. We regard the experimental
verification of this feature in the SIS tunneling data as an additional
strong argument in favor of the magnetic scenario for superconductivity.

\subsubsection{Optical and Raman response:}

Theoretical considerations show that optical measurements are much better
suited than Raman measurements~to search for the ``fingerprints'' of a
magnetic scenario\cite{acs_finger}. For the optical conductivity we predict a 
singular behavior at energies $2\Delta +\Delta _{s}$,  $4\Delta $,  $2\Delta +2\Delta
_{s}$, which can be amplified if one considers the second derivative of conductivity 
via $W(\omega )=\frac{d^{2}}{d^{2}\omega }(\omega \mathrm{Re}\sigma ^{-1}(\omega ))$.
Evidence for strong coupling effects in
the optical conductivity in superconducting cuprates has been reported in
Refs. \onlinecite{basov,CSB99,bonn,nuss}.
\begin{figure} 
\epsfxsize=2.8in 
\epsfysize=2.3in
\begin{center}
\epsffile{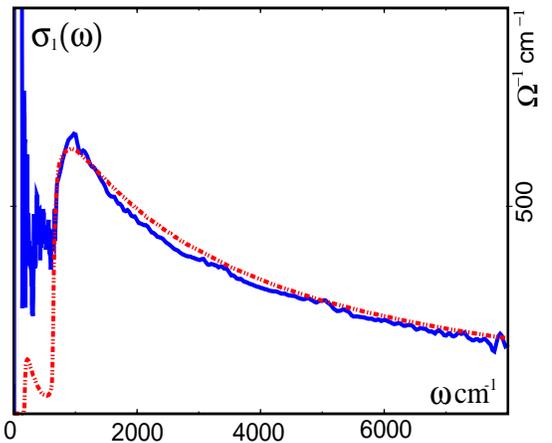}
\end{center}
\caption{A comparison between theoretical and experimental results
 for the optical conductivity (from Ref. ~\protect\onlinecite{Abanov01}). 
The experimental data are from Ref~\protect\onlinecite{basov}. }
\label{Figure_cond_strongc}
\end{figure}

We first discuss the form of $\sigma _{1}(\omega )$. In Fig. \ref{Figure_cond_strongc}
 we compare the theoretical result for $\sigma
_{1}(\omega )$~\cite{Abanov01,acs_finger} with the experimental data by
Puchkov \emph{et al.}\cite{basov} for optimally doped YBa$_{2}$Cu$_{3}$O$%
_{6+\delta }$ in the superconducting state. The parameters are the same as
in the normal state fits. As the theoretical formula does not include the
contributions from the nodes, the comparison is meaningful only for $\omega
>2\Delta $. We see that the frequency dependence of the conductivity at high
frequencies agrees well with the data. The measured conductivity drops at
about $100\mathrm{meV}$ in rough agreement with $2\Delta +\Delta _{s}$ which
for $\Delta \approx 30\mathrm{meV}$ and $\Delta _{s}\approx 40\mathrm{meV}$
is also around $100\mathrm{meV}$. The good agreement between theory and
experiment is also supportive of our argument that the momentum dependence
of the fermionic dynamics becomes irrelevant at high frequencies, and
fermions from all over the Fermi surface behave as if they were at hot spots.

\begin{figure}[tbp] 
\epsfxsize=2.8in 
\epsfysize=2.3in
\begin{center}
\epsffile{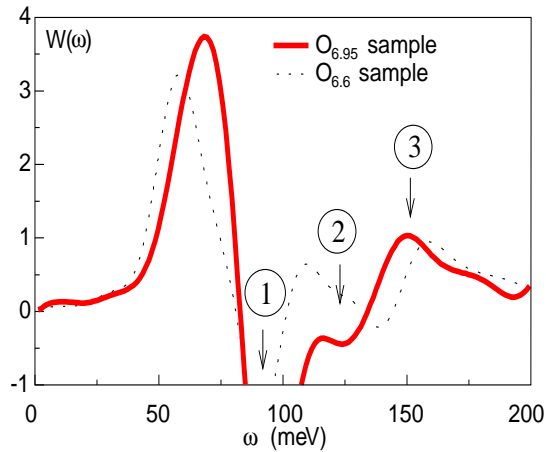} 
\end{center}
\caption{Experimental results for $W(\protect\omega )=\frac{d^{2}}{d^{2}
\protect\omega }(\protect\omega \mathrm{Re}\protect\sigma ^{-1}(\protect
\omega ))$ from Ref.\protect\onlinecite{CSB99}. The theoretical result is
presented in Fig. \ref{Figure_d2sigdw2}. The position of the deep minimum
agrees well with $2\Delta +\Delta _{s}$. The extrema at higher frequencies
are consistent with $4\Delta $ and $2\left( \Delta +\Delta _{s}\right) $
predicted by the theory.}
\label{Figure_d2sigdw2_exp}
\end{figure}

We next consider the singularities in the frequency dependence of the
conductivity in more detail and compare the theoretical and experimental
results for $W(\omega )=\frac{d^{2}}{d^{2}\omega }(\omega \mathrm{Re}\sigma
^{-1}(\omega ))$. The theoretical result for $W(\omega )$ is presented in
Fig.\ref{Figure_d2sigdw2} The experimental result for $W(\omega )$ in YBCO
is shown in Fig.\ref{Figure_d2sigdw2_exp}. We see that the theoretical and
experimental plots of $W(\omega )$ look rather similar, and the relative
intensities of the peaks are at least qualitatively consistent with the
theory. We identify (see explanations below) $2\Delta +\Delta _{s}$ with the
deep minimum in $W(\omega )$. This identification, that is consistent with
the analysis of $\sigma _{1}(\omega )$, yields $2\Delta +\Delta _{s}\approx
100\mathrm{meV}$. Identifying the extra extrema in the experimental $%
W(\omega )$ with $4\Delta $ and $2\Delta +2\Delta _{s}$, respectively, we
obtain $4\Delta \sim 130\mathrm{meV}$, and $2\Delta +2\Delta _{s}\sim 150%
\mathrm{meV}$. We see that three sets of data are consistent with each other
and yield $\Delta \sim 30\mathrm{meV}$ and $\Delta _{s}\sim 40-45\mathrm{meV}
$. The value of $\Delta $ is in good agreement with tunneling measurements~ 
\cite{Miyakawa99}, and $\Delta _{s}$ agrees well with the resonance
frequency extracted from neutron measurements~\cite{neutrons}. Indeed, the
analysis of a second derivative of a measured quantity is a very subtle
procedure. The good agreement between the theory and  experiment is
promising but has to be verified in further experimental studies. Still,
theoretical calculations~ \cite{Abanov01,acs_finger} clearly demonstrate the
presence and observability of these ''higher harmonics'' of the optical
response at $4\Delta $ and $2\Delta +2\Delta _{s}$.

Finally, we comment on the position of the $2\Delta +\Delta _{s}$ peak and
compare the results of Abanov et al~\cite{Abanov01,acs_finger} with those by
Carbotte \emph{et al.}\cite{CSB99}. Theoretically, at $T=0$ and in clean
limit, the maximum and minimum in $W(\omega )$ are located at the same
frequency. At a finite $T$, however, they quickly move apart (see Fig~\ref
{Figure_d2sigdw2}). 
%
%
Carbotte \emph{et al.}~\cite{CSB99} focused on the maximum in $W(\omega )$
and argued that it is located at $\Delta +\Delta _{s}$ instead of $2\Delta
+\Delta _{s}$. We see from Fig~\ref{Figure_d2sigdw2} that the maximum in $%
W(\omega )$ shifts to a lower frequency with increasing temperature and over
some $T$ range is located close to $\Delta +\Delta _{s}$. On the other hand,
the minimum in $W(\omega )$ moves very little with increasing $T$ and
virtually remains at the same frequency as at $T=0$. This result suggests
that the minimum in $W(\omega )$ is a more reliable feature for comparisons
with experiments. This conclusion is in agreement with recent conductivity
data on optimally doped $\mathrm{Bi}2212$~\cite{tu}. $W(\omega )$ extracted
from these data shows a strong downturn variation of the maximum in $W(\omega
) $ with increasing temperature, but the minimum in $W(\omega )$ is located
at around $110\mathrm{meV}$ for all temperatures.

\subsection{Experimental facts that we cannot yet describe}

There are several experimental results that we do not understand. First are
the results by Ando, Boebinger and collaborators\cite
{Ando95,Boebinger96,Ono00} on the behavior of the Lanthanum and Bismuthate
based superconductors in magnetic fields sufficiently strong to (almost)
destroy superconductivity. For doping levels close to the optimal one, they
found that the resistivity at low temperatures continues to be linear in $T$
with the same slope seen at higher temperatures. If the assumption that the
magnetic field destroys superconductivity but otherwise does not affect the
system properties is correct, this result poses a problem for the
spin-fluctuation model as the latter yields a linear in $T$ resistivity over
a wide range of temperatures, but only for $T$ larger than a fraction of $%
\omega _{\mathrm{sf}}$. To account for these data one might have to
invoke some kind of quantum-critical physics associated with the opening of
the pseudogap (see below).

Another experiment that is not yet  understood is the measurement of
the Hall angle, $\theta _{H}\equiv \rho _{xy}/\rho _{xx}$, which shows an
incredibly simple behavior, $\cot \theta _{H}\propto T^{2}$~\cite{anderson}
and also displays a particular frequency behavior~\cite{abr_varma,drue,ong}.
The orbital magnetoresistance $\Delta \rho /\rho $ also behaves in quite an
unusual way, violating Kohler's rule, according to which $\Delta \rho /\rho $
is a function of $H^{2}/\rho ^{2}$, independent of $T$, where $H$ is the
applied magnetic field. Some of this physics is already captured in the
semi-phenomenological calculations by Stojkovich and Pines \cite{branko};
however problems remain. In the description based on the spin-fermion model
the technical problem not yet solved is how to include in a controlled way
vertex corrections which are not small; in one of the vertices for the Hall
conductivity the momentum transfer is small. Some progress with these
calculations have been recently made by Katami and collaborators~\cite
{katami}. Another explanation of the Hall data has recently been proposed by Abrahams and Varma~\cite{abr_varma}.

Yet another unanswered question, already noted above, is the origin of a
large (almost $100\mathrm{meV}$), frequency and temperature independent
contribution to the self-energy that one has to invoke in order to fit
conductivity and ARPES data. It could, in principle, be due to inelastic
scattering by impurities, but its very large value makes this explanation
problematic.

Electronic Raman scattering reveals further puzzling behavior:
in all geometries one observes a frequency independent behavior over a very
large energy scale, frequently referred to as the positive background.
Moreover, the overall size of the background is very different in different
geometries~\cite{girsh,Devereaux}. 

There are also uncertainties associated with reconciling the
incommensurability of the magnetic response in the normal state of $214$
materials~\cite{Aeppli} with the commensurate peaks required to obtain a
consistent explanation of $^{17}$O and $^{63}$Cu NMR experiments, but these
are not likely to pose fundamental problems to the spin-fluctuation approach 
\cite{Barzykin,Zha}.

Finally, the claim of universality of the low-energy behavior relies heavily
on the existence of a quantum critical point at which the antiferromagnetic
correlation length diverges. In real materials there are indications that
the transition to antiferromagnetism is actually of first order. In this
situation, the theory we described is valid only if there still exists a
substantial region in parameter space where the system is critical before it
changes its behavior discontinuously. NMR and neutron scattering experiments
on optimally doped cuprates seem to support such behavior. Another reason
for concern is the role of disorder and inhomogeneities. Despite enormous
progress in sample fabrication, cuprates often tend to be very heterogeneous
materials. It has been established in several cases that these aspects are
actually intrinsic, forcing one to include effects due to inhomogeneities
and disorder into the theoretical description\cite{Schmalian_Wolynes}.

\subsection{Phase diagram}

In this section we discuss in detail the experimental phase diagram of
cuprate superconductors and comment on the origin of the pseudogap behavior
found for small charge carrier concentrations.

From a general perspective, the key to  understanding of cuprate
superconductors is identifying the nature of the protected behavior of the
novel states of matter encountered in the insulating, conducting, and
superconducting states as one varies doping and temperature, including the
possible existence of one or more quantum critical points. Consider first
the YBa$_{2}$Cu$_{3}$O$_{7-\delta }$ system on which the generic phase
diagram of Fig.\ref{PinesHouston} was based\cite{Barzykin}. A somewhat
similar diagram based on transport measurements was independently proposed
by Hwang \textit{et al.} \cite{Hwang94}, while one based on specific heat
and susceptibility measurements has been proposed recently by Tallon et al. 
\cite{Tallon}. 
As discussed in the Introduction, in addition to the $T_{c}$ line, there are
two crossover or phase transition lines in Fig.\ref{PinesHouston}. The upper
line $T=T_{\mathrm{cr}}$ is defined experimentally by a maximum in the
temperature dependent uniform magnetic susceptibility, $\chi _{0}$. It has
been further characterized\cite{Barzykin} as the temperature at which the
antiferromagnetic correlation length $\xi $ is of the order Cu-Cu lattice
spacing (Barzykin and Pines used a criterion $\xi (T_{\mathrm{cr}})=2a$).
The lower line $T=T^{\ast }$ may be defined experimentally as the
temperature at which the product of the copper spin-lattice relaxation time,$%
^{63}T_{1}$ and the temperature,$T$, reaches its minimum value. In the Bi$%
_{2}$Sr$_{2}$CaCu$_{4}$O$_{8}$ counterparts of the YBa$_{2}$Cu$_{3}$O$%
_{7-\delta }$ system, it corresponds to the temperature at which the leading
edge gap found in ARPES experiments for quasiparticles near $(\pi ,0) $
becomes fully open, effectively gapping that portion of the quasiparticle
Fermi surface. To a first approximation, on making use of the experimental
results for optimally and underdoped YBa$_{2}$Cu$_{3}$O$_{7-\delta }$
materials one finds that 
\begin{equation}
T^{\ast }\approx \frac{1}{3}T_{\mathrm{cr}}.
\end{equation}

The superconducting $T_{c}$ in Fig. \ref{PinesHouston} is obtained using the
empirical relation ~\cite{Tallon} 
\begin{equation}
x=0.16\pm 0.11\sqrt{1-\frac{Tc}{T_{c}^{\mathrm{max}}}}  \label{4.1}
\end{equation}
where $x$ is the doping level, and $T_{c}^{\mathrm{max}}$ is the maximal
transition temperature for a given class of materials. The location of $T_{%
\mathrm{cr}}$ can well be fitted by another empirical relation. 
\begin{equation}
T_{\mathrm{cr}}\approx 1250\,\mathrm{K}\left( 1-\frac{x}{x_{\mathrm{cr}}}%
\right) ,  \label{4.2}
\end{equation}
where $x_{\mathrm{cr}}\approx 0.19$. Similar expressions are found for the La%
$_{2-x}$Sr$_{x}$CuO$_{4}$ and Bi$_{2}$Sr$_{2}$CaCu$_{4}$O$_{8}$ materials.
This expression for $T_{\mathrm{cr}}$ is, in both its magnitude and doping
dependence, close to the pseudogap temperature obtained by Loram and his
collaborators\cite{Loram} from an analysis of specific heat experiments. A
remarkable result of this purely phenomenological analysis is that the
crossover temperature $T_{\mathrm{cr}}$ extrapolates at zero doping to the
known value of the antiferromagnetic super-exchange interaction $J$.

The fit to $T_{\mathrm{cr}}$ by Eq. \ref{4.2} raises the issue of whether $%
T^{\ast }$ and $T_{\mathrm{cr}}$ are independent of $T_{c}$ and would
extrapolate to the origin at a doping level $x=x_{\mathrm{cr}}$ if
superconductivity was absent. The system then would have an additional
quantum critical point at $x=x_{\mathrm{cr}}$ with a new kind of ordered
state for $x<x_{\mathrm{cr}}$. This issue is currently open and is a subject
of active research. Support for a phase diagram with an additional quantum
critical point at $x=x_{\mathrm{cr}}$ comes from the work of Loram , Tallon,
and their collaborators~\cite{Loram,Tallon}, who have proposed such behavior
based on a detailed analysis of their specific heat experiments on
underdoped and overdoped systems. Moreover, as Loram, Tanner, Panagopoulos
and others have emphasized~ \cite{Loram,Tallon,Panagopolis}, in the
superconducting state of the low-doping side of $T_{\rm cr }$ one has ''weak''
superconducting behavior, with a superfluid density $\rho _{s}$ decreasing
with decreasing doping, while on the high doping side one has a
``conventional'' superconductivity, and a value of $\rho _{s}$ that is
nearly independent of the doping concentration. Further support for the idea
of an additional quantum-critical point comes from the well established fact
that optimally doped cuprates are the ones for which the extension of the
linear resistivity to $T=0$ yields very small residual resistivity, and from the
experiments of Refs.\onlinecite{Ando95,Boebinger96} which, we recall, show that in
the absence of superconductivity the linear temperature dependence of the
resistivity extends to lower $T$ indicating that at some doping the
resistivity can be linear down to $T=0$. As
 Laughlin \emph{et al. }\cite{Laughlin} have
emphasized, the presence of a quantum critical point with a large domain of
influence, together with superconductivity, serves to conceal the nature of
the non-superconducting ground states on either side of the quantum critical
point. One might hope that ARPES experiments near optimal doping would distinguish between a quantum critical behavior with quantum critical point at 
around optimal doping and a spin fermion scenario with antiferromagnetic quantum critical point at considerably smaller doping concentration. However, a recent analysis of Haslinger {\em et al.}\cite{rob2} showed that fits to
 current experiments with either model is possible and requires in both cases
the introduction of a large temperature and frequency independent scattering rate, as noted earlier.

The variety of experimental results for the pseudogap allows one to
understand it phenomenologically, without invoking a particular microscopic
mechanism. First, as $T^{\ast }$ and $T_{\mathrm{cr}}$ scale with each
other, it is natural to attribute both $T^{\ast }$ and $T_{\mathrm{cr}}$ to
different aspects of the same physical phenomenon which begins at $T_{%
\mathrm{cr}}$ and gains full strength at $T^{\ast }$. This idea is fully
consistent with NMR data which show the onset of changes in $T_{1}T$ at $%
T_{cr}$, which eventually give rise to a sign change of the temperature
derivative of $^{63}T_{1}T$ at $T^{\ast }$. Second, ARPES data on the
leading edge gap clearly demonstrate that the pseudogap physics below $%
T^{\ast }$ is associated with the redistribution of the spectral weight for
hot quasiparticles; quasiparticles near the nodes are almost unaffected by
the development of the pseudogap. In the ARPES literature, this effect is
described as a progressive development of the arcs of the Fermi surface
centered around nodal points. The evolution of the full Fermi surface into
the arcs begins at around $T^{\ast }$, and at $T_{c}$ the whole Fermi
surface becomes gapped. The ``gapping'' of hot fermions obviously affects
NMR relaxation rates dominated by momenta near ${\bf{Q}}$ (such as $%
^{63}Cu $ $T_{1}T$) as a spectral weight transfer would lead to a reduction
of a decay rate of a spin fluctuation into a particle-hole pair~\cite
{Chubukov95,Schmalian98}. NMR experiments by Curro \emph{et al.}\cite
{Curro00} and Haase \emph{et al.} \cite{Haase01} show that this is indeed
the case. The gapping of hot quasiparticles should also lead to a
temperature-dependent reduction in the uniform magnetic
 susceptibility \cite{Alloul}.

The phenomenological description is of course not enough as it leaves open
the key question, namely what causes the spectral weight transfer for hot
quasiparticles. We now discuss how the experimental phase diagram fits into
the spin-fluctuation scenario.

First of all a general phase diagram based on spin fluctuation approach
should distinguish between weakly and strongly antiferromagnetic materials.
Weakly antiferromagnetic materials are those to the right of $T_{\mathrm{cr}%
} $ for which the dimensionless coupling constant $\lambda $ is smaller than
unity, which corresponds to a correlation length smaller than a few lattice
constants. For these materials, the normal state is a renormalized Fermi
liquid, the nearly antiferromagnetic Fermi liquid, and $T_{c}$ signals a
transition to a BCS-like superconducting state with a $d_{x^{2}-y^{2}}$
order parameter. For strongly antiferromagnetic materials, on the other
hand, the dimensionless coupling constant $\lambda \geq 1$ and $\xi \geq 2a$%
. In this situation, the normal state behavior deviates from a Fermi liquid
already at comparatively small $\omega $ and $T$ although at the lowest
frequencies the system still would display a Fermi liquid behavior if indeed
one could extend the normal state down to $T=0$. For these systems, we also
know that the pairing instability temperature $T_{\mathrm{cr}}$ increases
with decreasing doping and for large enough coupling saturates at a value comparable to
the magnetic $J$. Applying this to the experimental phase diagram, we see
that optimally doped materials are at the borderline between being weakly
and strongly antiferromagnetic: on the one hand the dimensionless coupling
is already not small, on the other hand, the pseudogap phase extends at best
over a $T$ range which is only a fraction of $T_{\rm cr}$.

As we have noted above, these results of the spin-fermion model make it a strong 
candidate for the microscopic description of the pseudogap phase: $T_{\rm cr}$ saturates   $T_{\mathrm{cr}}$ saturates at a finite value at the
magnetic transition;  for $\lambda \geq 1$, the pairing involves
non-Fermi liquid fermions;  at $T=0$ there are two distinct energy
scales in the problem, a fermionic gap $\Delta \propto T_{\mathrm{cr}}$ and
a bosonic gap $\Delta _{s}\propto T_{\mathrm{cr}}\lambda ^{-1}\ll T_{\mathrm{%
cr}}$.  The central issue is whether the pairing of
incoherent fermions only gradually changes the fermionic self-energy, or \
whether it creates a feedback on fermions which immediately gives rise to a
coherent quasiparticle behavior at the lowest frequencies, as happens in
dirty superconductors where $\Sigma (\omega )=i\gamma $ in the normal state
transforms below $T_{c}$ into a mass renormalization at the smallest $\omega 
$ 
\begin{equation}
\Sigma (\omega )=i\gamma \frac{\omega }{(\omega ^{2}-\Delta ^{2})^{1/2}}%
\simeq \gamma \frac{\omega }{\Delta }+O\left( \omega ^{3}\right) .~
\end{equation}
If the feedback is gradual, then the pairing creates bound states of
incoherent fermions with $S=0$. In this situation, there is a reduction in
the density of states below $T_{\mathrm{cr}}$, but a full superconducting
gap does not develop until a smaller temperature, $T_{c}$. A simple toy
model which describes this physics would be one~\cite{Chubukov95} in which
fermions in the normal state display a quantum critical behavior with ${%
\tilde{\Sigma}}(\omega )\approx \Sigma (\omega )=(i\omega {\bar{\omega}}%
)^{1/2}$, and pairing creates a nonzero flat 
pairing vertex $\Phi $ but does not
affect $\Sigma (\omega )$. In this situation, the fermionic propagator
acquires a gap at a finite but imaginary frequency: 
\begin{equation}
G_{{\bf k}}(\omega )\propto \frac{\sqrt{i\omega {\bar{\omega}}}+\epsilon _{\bf k}}{%
i\omega -E_{\bf k}}  \label{ch1}
\end{equation}
where 
\begin{equation}
E_{\bf k}=(\Phi ^{2}+\epsilon _{\bf k}^{2})/{\bar{\omega}.}
\end{equation}
The spectral function at ${\bf{k}}={\bf{k}}_{F}$ and the fermionic DOS
both have broad maxima at $\omega =\Delta =\Phi ^{2}/{\bar{\omega}}$, but
the spectral weight is finite at any finite $\omega $, although reduced at
low frequencies. Within this model, the transition to the true
superconducting state can be understood as a rotation of the pole in \ Eq. (%
\ref{ch1}) from the imaginary to the real frequency axis such that at the
lowest frequencies the $i\omega $ term becomes purely real. The frequency up
to which $N(\omega )=0$ then give an estimate for the actual $T_{c}$.

It is unclear to what extent the results of this toy model reflect the
physics of the spin-fermion model below $T_{\mathrm{cr}}$. Without
elaborating on this subject of current research we mention that as long as
an Eliashberg approach is justified, phase fluctuations of the
superconducting order parameter cannot substantially reduce $T_{c}$ compared
to $T_{\mathrm{cr}}$. Behavior different from that in dirty superconductors
could emerge only if longitudinal fluctuations of the superconducting order
parameter are soft and \ able to destroy the superconducting coherence at $%
T_{c}\ll T_{\mathrm{cr}}$. We have both numerical~\cite{Abanov01epl} and
analytical~ \cite{altshuler} evidence that such degeneracy does exist in the
limit $\lambda =\infty $. Still, this subject is far from being fully
understood and clearly requires further study.

Another subtle issue is whether the spin-fermion model displays quantum
critical behavior at $x=x_{cr}$. Physically, this would imply that the
pairing of incoherent fermions at $T_{\mathrm{cr}}$ and the pairing of
coherent fermions at $T_{c}$ are uncorrelated phenomena - the first gives
rise only to the pseudogap, while the latter yields BCS-type
superconductivity. Since the pairing of incoherent fermions is not a
perturbative phenomenon and requires the interaction to exceed a threshold
value~\cite{ACF}, the pairing of incoherent fermions involves only
quasiparticles in some \textit{finite} region around a hot spot and would
form a dome on top of a magnetic quantum critical point and vanish at a
finite $x$. It is not clear how well one can separate coherent and
incoherent pairings.

Another issue related to the possible explanation of the pseudogap within
the spin-fermion model is whether one can smoothly interpolate between the
limit ${\bar{g}}\ll W$ ($W$ is the fermionic bandwidth), where one can
perform calculations in a controlled fashion, and ${\bar{g}}\gg W$ where
Mott physics become relevant. In essence the issue is whether or not there
is a qualitative difference between limits in which the effective
interaction is either much larger or much smaller than the fermionic
bandwidth. In the latter case, it is appealing to conjecture that the
pseudogap is associated with the fact that it is difficult for hot
quasiparticles to be both itinerant and localized. $T_{\mathrm{cr}}$ then
would mark the onset of insulating behavior associated with such
localization, and the pseudogap phase would represent a kind of partial Mott
insulator. A sign that these two limits may describe some aspects of the
basic physics similarly is that, as shown earlier, $T_{\mathrm{cr}}$ scales
with ${\bar{g}}$ for ${\bar{g}}<W$, but crosses over to $W^{2}{/}\overline{{g%
}}\propto J$ for ${\bar{g}}>W$. At the same time, the limit ${\bar{g}}>W$ is
probably more rich than the small ${\bar{g}}$ limit as the physics
associated with the localization in the Mott insulator is not included into
our analysis. We speculate that due to this Mott physics, bound singlet
pairs of fermions that emerge below $T_{\mathrm{cr}}$ could order for
example in columnar fashion as suggested by Sachdev and collaborators\cite{subir} 
who arrived at a columnar phase by studying weakly doped Mott
antiferromagnets. This ordering in turn would imply that $T_{\mathrm{cr}}$
is a true phase transition line below which $Z_{4}$ symmetry is broken.
Columnar ordering also opens a link between our approach and the approaches
which depart from Mott insulator at half-filling. In particular, columnar
ordering of bound electron pairs naturally leads to stripe physics\cite
{subir}. An alternative possibility is that singlet pairs remain spatially
disordered\cite{rvb,rvb2}. In any event, the role of  localization effects
certainly increases as the system approaches half-filling. Whether they
remain strong near optimal doping in the normal state is a subject of
debate, but still, strong localization effects should reduce the number of
low energy carriers and therefore change the volume of the Fermi surface or
increase their mass. ARPES experiments on the other hand indicate that in
the normal phase, the Fermi surface is large and obeys Luttinger's theorem
without dramatic mass renormalizations. We therefore believe that near
optimal doping localization effects are at best moderate.

We conclude this discussion of the phase diagram by mentioning two
alternative scenarios for the pseudogap and anomalous normal state
properties. The first scenario, pioneered by P.W. Anderson\cite{rvb}, 
X.-G. Wen and P. A. Lee\cite{WenLee} and others\cite{rvb2,rvb3,rvb4,rvb5,rvb6,rvb7,rvb8,rvb9} and later modified chiefly by M.P.A. Fisher 
and T. Senthil\cite{Senthil}, assumes
spin-charge separation at half-filling and explains the whole phase diagram
as a result of \ weak doping of a Mott insulator.  A second scenario, on
the contrary, assumes that one can understand the phase diagram within
mean-field theory; as Chakravarty \emph{et al.}\cite{Chakravartyddw} have
proposed the pseudogap might then be a new protected state of matter that is
the result of the breaking of a hidden symmetry.

\section{Conclusions}

In this Chapter we have demonstrated that superconducting pairing mediated
by the exchange of spin fluctuations is a viable alternative to conventional
phonon-mediated pairing. We discussed in detail the normal state properties,
\ the pairing instability and the superconducting behavior of a material
near an antiferromagnetic instability, when the dominant interaction between
quasiparticles is of electronic origin and, at energies much smaller than
the fermionic bandwidth, can be viewed as being due to the emission and
absorption of a collective, soft spin degree of freedom. We argued that the
spin-fluctuation exchange yields an attraction in the $d_{x^{2}-y^{2}}$
channel in agreement with what nearly all researchers now believe is the
pairing symmetry in the cuprates.

We demonstrated that the physics is qualitatively different depending on
whether or not Fermi surface geometry allows a process in which a collective
mode decays into a particle and a hole. For this to be possible, the Fermi
surface should contain hot spots. We focused on the case in which the Fermi
surface does contain hot spots (as the photoemission experiments in cuprates
indicate) and showed that spin fluctuations are then overdamped and that
their diffusive dynamics should be analyzed in a consistent manner with the
low energy dynamics of the fermions. We further argued that contrary to
naive expectations, this case is better for $d-$wave pairing than one in
which spin fluctuations are propagating, magnon-like quasiparticles.

We showed that the low-energy theory for fermions interacting with
overdamped collective spin excitations is universal, independent of the
details of the underlying lattice Hamiltonian and is characterized by only
two input parameters: the dimensionless coupling constant, $\lambda $, and
an overall energy scale, $\overline{\omega }$, proportional to the effective
spin-fermion interaction $\overline{g}$. 
 In so doing, we have developed the microscopic justification for the NAFL. 
The coupling constant, $\lambda $, scales with the magnetic correlation
length, so that close enough to a magnetic transition, the system falls into
a strong coupling regime. A universal description is valid if $\overline{g}$
is smaller than the fermionic bandwidth $W$ (that would correspond to a weak
coupling limit if the system was far away from a magnetic instability). In
the opposite limit which we did not discuss in detail, lattice effects
become important, and the universality is lost.

At sufficiently low temperatures and energies, the nearly 
 antiferromagnetic Fermi liquid (NAFL) is a Fermi liquid quantum  protectorate, according  to Landau's criterion--that one can obtain a
 one-to-one correspondence between the low-lying states of a Fermi gas
 and the Fermi liquid, as though the particle interaction was turned
 on adiabatically~\cite{PinesNoz}.
   However, because of the closeness to an 
 antiferromagnetic instability,
  it is an unconventional 
 Fermi liquid, in that the characteristic energy above which this
 description is no longer valid is not the Fermi energy, but is the
 much much lower spin-fluctuation energy $\omega_{\rm sf} = {\bar \omega}/(4\lambda^2)$ that is
 typically  two orders of magnitude smaller than the Fermi energy.
 For energies (or temperatures) between $\omega_{\rm sf}$ and the Fermi 
 energy, we have seen using the spin-fermion model that 
 the system behavior is again universal and depends only on a very limited 
 number of phenomenological parameters. 
 So in this sense, the behavior of an NAFL at energies and
temperatures above $\omega_{\rm sf}$ is also protected, and one finds, in the
 NAFL, two distinct protected states of matter, depending on the energy
 or temperature one encounters.

We compared in detail the spin fluctuation approach with the Eliashberg
approach to phonon superconductors and showed that despite the absence of
the small electron to ionic mass ratio that justified Eliashberg theory for
phonons, an Eliashberg-type approach to the spin-fermion model is still
justified, but only at strong coupling $\lambda \geq 1$.

We showed that at large $\lambda $, there are two  distinct energy scales
for the normal state problem: ${\bar{\omega}}=A\overline{g}$, $A=O(1)$,
which at ${\overline{g}}\sim W$ is of the order 
of (even though numerically smaller by about an order of magnitude than) 
the Fermi energy, and a
much smaller $\omega _{\mathrm{sf}}={\bar{\omega}}/(4\lambda ^{2})\ll {\bar{%
\omega}}$. Conventional Fermi liquid behavior with $\Sigma ^{\prime \prime
}\propto \omega ^{2}+(\pi T)^{2}$ and an almost temperature independent
static spin susceptibility exists only at frequencies and temperatures
smaller that $\omega _{\mathrm{sf}}$. At frequencies between $\omega _{%
\mathrm{sf}}$ and ${\overline{\omega }}$, the system crosses over into a
regime in which spin susceptibility is diffusive and $\Sigma ^{\prime \prime
}\propto \omega ^{\alpha }$. We found that $\alpha =\frac{1}{2}$, while the
behavior in the crossover region from Fermi liquid to non Fermi liquid
behavior resembles a  linear in frequency dependence. In the diffusive
regime, both spin and fermionic propagators become independent of $\lambda $%
, i.e., the system displays magnetic quantum-critical behavior. Finally, at $%
\omega >\overline{\omega }$, the self-energy gets smaller than the bare $%
\omega $, although $\Sigma ^{\prime \prime }\propto \omega ^{1/2}$ still
holds.

We next argued that the existence of the two distinct energy scales has a
strong impact on the pairing problem: the pairing instability temperature $%
T_{\mathrm{cr}}$ is predominantly determined by incoherent, non-Fermi liquid
fermions with  energies between $\omega _{\mathrm{sf}}$ and $\overline{%
\omega }$. We then considered the superconducting state, found the pairing
gap $\Delta \propto T_{\mathrm{cr}}$ ($2\Delta /T_{\mathrm{cr}}\sim 4$ for $%
\lambda \gg 1$), and argued that one of the signatures of the
spin-fluctuation mechanism is the change of the low-frequency spin dynamics
from   relaxational to propagating  due to the feedback from the pairing.
This give rise to the emergence of the resonance peak in the spin propagator
at a frequency $\Delta _{s}$ which at large $\lambda $ scales as 
$\Delta _{s}\sim 1.5 T_{\mathrm{cr}}/\lambda$. For smaller $%
\lambda $, the resonance frequency goes up but always remain smaller than $%
2\Delta $. This restriction (not found for phonon superconductors) is a
consequence of feedback between fermionic and bosonic dynamics.

We discussed how the emergence of the resonance mode affects fermionic
properties and identified several ``fingerprints'' of the strong coupling
spin-fluctuation scenario. These include, but are not limited to, the
peak/dip/hump in the spectral function, the near-absence of the dispersion
of the quasiparticle peak, the peak/dip features in the SIN and SIS
tunneling conductances, and in the second derivative of the optical
conductivity.

Finally, we compared the theoretical results with the normal state data for
near optimally-doped cuprates, and found that a large number of experimental
results can be understood within the spin-fluctuation scenario. In
particular, we argued that the tunneling, photoemission and optical data for
the superconducting state display the expected ``fingerprints'' of the
spin-fluctuation mechanism for superconductivity in the cuprates.

The spin-fluctuation theory should be equally applicable to quasi two
dimensional organic superconductors or compounds of the family \textrm{CeXIn}%
$_{5}$ with \textrm{X=Co, Rh, Ir}, if the Fermi surface of these systems
possesses  hot spots. A detailed experimental investigation of spectroscopic
properties of these systems in the superconducting state will provide
important clues on whether they are indeed magnetically mediated
superconductors.

Despite the successful identification of the ''fingerprints'' of spin
mediated pairing in the cuprates, we listed in Section 6.4 some  properties
of these materials which we  have not explained within the spin-fluctuation
scenario. Perhaps the most important remaining question is to what extent
the spin-fluctuation theory explains the pseudogap physics in underdoped
cuprates. The existence of the two distinct energy scales in the
spin-fermion model and the presence of fermionic incoherence in the normal
state make this model a good candidate model for the pseudogap. The
plausible argument here is that while the pairing of the incoherent fermions
will create spin singlets, fermions will still behave incoherently and hence
not superconduct until a smaller $T_{c}$ is reached when feedback from the
pairing restores fermionic coherence. Particularly relevant here is whether
or not there exists an additional critical point on the phase diagram at
around $x=0.19$. This critical point emerges in the spin-fluctuation theory
if the pairing of incoherent fermions from frequencies above $\omega _{sf}$
and the pairing of coherent fermions from frequencies below $\omega _{%
\mathrm{sf}}$ are separate phenomena (the first then definitely gives rise
to a pseudogap physics while the second yields a true superconductivity).
The pairing of incoherent fermions is a threshold phenomenon and thus occurs
at $T_{\mathrm{cr}}(x)$ . We still however need to understand whether
incoherent pairing and coherent pairing can be totally separated from each
other.

Another unresolved issue is to what extent the fact that the parent
compounds of the cuprates ($\mathrm{La}_{2}\mathrm{CuO}_{4}$ or $\mathrm{YBa}%
_{2}\mathrm{Cu}_{3}\mathrm{O}_{6}$) are Mott insulators and nearest-neighbor
Heisenberg antiferromagnets affects the behavior of doped materials. For the
spin-fermion model, this question could be reformulated as whether the limit 
${\bar{g}}\ll W$, in which we can separate low and high energies and perform
controlled calculations, and the limit ${\bar{g}}\gg W$, are only
quantitatively different or are qualitatively different. An encouraging sign
that the two limits may not be very different is the result that $T_{\mathrm{%
cr}}$ smoothly interpolates between $O({\bar{g}})$ for ${\bar{g}}\ll W$ to $%
O(W^{2}/{\bar{g}})$ for ${\bar{g}}\gg W$, in the latter case $T_{\mathrm{cr}%
} $ becomes of order of a magnetic exchange integral $J$. On the other hand,
the localization effects which accompany a transition to a Mott insulator
are not included in the theory presented here. In particular, we cannot
predict what would happen with singlet pairs below $T_{\mathrm{cr}}$, if it
indeed is the onset of the pseudogap - whether they remain disordered or
form columnar stripes as Sachdev and collaborators\cite{subir} suggested.

The enduring presence and  richness of unsolved problems in the
 field of  unconventional
superconductivity  makes us optimistic that we will continue to see
unexpected experimental observations and  new, creative theoretical
concepts in this  field of research. We have enjoyed over a number
of years our respective collaborations on the spin fluctuation approach
because of its clarity and ability to make falsifiable predictions for
experiment. We hope that the readers of this Chapter will share our
excitement  for this approach to magnetically mediated superconductivity.

\section{Acknowledgments}

It is our pleasure to thank our immediate collaborators Artem Abanov, Victor
Barzykin, Alexander Finkelstein, Rob Haslinger, 
 Philippe Monthoux and Branko Stojkovic for
stimulating discussions on essentially all aspects of the spin fluctuation
approach presented here. We are also thankful to E. Abrahams, A. A.
Abrikosov, G. Aeppli, 
B. L. Altshuler, N. Andrei, A. Balatsky, D. Basov, K. H. Bennemann, G.
Blumberg, J. Brinckmann,  J.C. Campuzano, P. Chandra, 
P. Coleman, E. Dagotto, J. C. Davis, E. Demler, 
M. P. A. Fisher,  A. Georges, L.P. Gor`kov,
M. Grilli, L. Ioffe, P. Johnson, R. Joynt, A. Kaminski, B.
Keimer, D. Khveschenko, S. Kivelson, G. Kotliar, R. Laughlin, 
 M. Lavagna, P. A. Lee,  G. Lonzarich,
H. von L\"{o}hneysen, A. Millis, H. Monien, 
 D. K. Morr, M. Norman, C. P\'{e}pin, A.
Rosch, S. Sachdev, D. Scalapino, J. R. Schrieffer, T. Senthil,  Q. Si,
 C. P. Slichter, S. Sondhi, Z. Tesanovic,
 O. Tchernyshyov, A.Tsvelik, J. Tu, C. Varma, P. G. Wolynes, A. Yazdani,
 J. Zasadzinski and  S.-C. Zhang
for useful conversations. We further wish to thank Ar. Abanov, D. Basov, J.
C. Davis and J. Tu for sharing their unpublished results with us.

The research which led to this review was supported in part by 
 NSF DMR-9979749 ( A. Ch.), by the Institute
for Complex Adaptive Matter, an independent unit of the University of
California, by Project DR 200153 on \emph{Emergent Properties of Correlated
Electron Superconductors} and general DOE support to Los Alamos National
Laboratory (D. P.), and by the Ames
Laboratory, operated for the U.S. Department of Energy by Iowa State
University under contract No. W-7405-Eng-82 \ (J.S).

%
\end{multicols}{2}  
\end{document}